\documentclass[12pt,english]{article}

\newcommand{\CO}{O} 

\usepackage[latin9]{inputenc}
\usepackage{geometry}
\usepackage{verbatim}
\usepackage{prettyref}
\usepackage{amsmath}
\usepackage{amssymb}
\usepackage{graphicx}

\usepackage{mathrsfs}

 \usepackage{slashed}

\usepackage{dsfont}

\usepackage{color}
\definecolor{darkgreen}{rgb}{0,0.5,0}
\definecolor{darkblue}{rgb}{0,0,0.6}
\definecolor{purple}{rgb}{0.4,.2,0.7}
\usepackage[colorlinks=true,citecolor=darkgreen,linkcolor=purple,urlcolor=purple]{hyperref}

\numberwithin{equation}{section}
\numberwithin{figure}{section}
\numberwithin{table}{section}

\def\CG{{\cal G}} 
\def\CH{{\cal H}}

\def\CN{{\cal N}}
\def\CD{{\cal D}}

\def\CA{{\cal A}}
\def\CB{{\cal B}}
\def\CR{{\cal R}}

\def\CCH{{\mathscr H}}

\def\tH{{\rm H}}
\def\tQ{{\tt \it Q}}
\def\tB{{\tt \it B}}

\def\IR{{\mathbb R}}
\def\IZ{{\mathbb Z}}

\def\llang{\langle}
\def\rrang{\rangle_{\rm Sp(N)}}

\DeclareMathOperator{\Tr}{Tr}

\def\N{N}

\DeclareFontShape{OT1}{cmr}{mx}{n}{<->cmr10}{}

\begin{document}

\fontseries{mx}\selectfont

\begin{center}
{\LARGE  Higher Spin de Sitter Hilbert Space}  
\end{center}
\vskip5mm
\begin{center}
 Dionysios Anninos$^1$, Frederik Denef$^{\, 2,3}$, Ruben Monten$^4$, Zimo Sun$^2$
\end{center}
\vskip5mm
{\it{\footnotesize $^1$ $\Delta$ Institute for Theoretical Physics $\&$ Institute for Theoretical Physics Amsterdam, University of Amsterdam, Science Park 904, 1098 XH Amsterdam, The Netherlands} \\
\it{\footnotesize $^2$ Department of Physics, Columbia University, 538 West 120th Street, New York, New York 10027, U.S. }\\
\it{\footnotesize $^3$ Institute for Theoretical Physics, KU Leuven, Belgium
} \\
\it{\footnotesize $^4$ Institut de Physique Th{\'e}orique, CEA Saclay, 91191 Gif-sur-Yvette, France \\}}

\vskip10mm
{\bf \center Abstract\\}
\vskip8mm
{
{We propose a complete microscopic definition of the Hilbert space of minimal higher spin de Sitter quantum gravity and its Hartle-Hawking vacuum state. The fundamental degrees of freedom are $2N$ bosonic fields living on the future conformal boundary, where $N$ is proportional to the de Sitter horizon entropy. The vacuum state is normalizable. The model agrees in perturbation theory with expectations from a previously proposed dS-CFT description in terms of a fermionic  Sp(N) model, but it goes beyond this, both in its conceptual scope and in its computational power. In particular it resolves the apparent pathologies affecting the Sp(N) model, and it provides an exact formula for late time vacuum correlation functions. We illustrate this by computing probabilities for arbitrarily large field excursions, and by giving fully explicit examples of vacuum 3- and 4-point functions. We discuss bulk reconstruction and show the perturbative bulk QFT canonical commutations relations can be reproduced from the fundamental operator algebra, but only up to a minimal error term $\sim e^{-\mathcal{O}(N)}$, and only if the operators are coarse grained in such a way that the number of accessible  ``pixels'' is less than $\mathcal{O}(N)$. Independent of this, we show that upon gauging the higher spin symmetry group, one is left with $2N$ physical degrees of freedom, and that all gauge invariant quantities can be computed by a $2N \times 2N$ matrix model. This suggests a concrete realization of the idea of cosmological complementarity.  
}



\newpage

\tableofcontents

\def\nn{\nonumber}

\newpage

\section{Introduction and summary}

Finding a precise and complete theory of quantum gravity has been a longstanding problem. For certain systems living in an infinitely deep gravitational potential well, shaped by a negative vacuum energy density, this problem has been solved: the Hilbert space and operator algebra of these theories are those of a conformal field theory living on the boundary of the well \cite{Maldacena:1997re}. 
This discovery, known as the AdS-CFT correspondence, has had a profound impact on theoretical research in quantum gravity and quantum field theory over the past twenty years.

However, the universe we find ourselves in does not remotely resemble such spacetimes. Rather than being trapped together in an infinitely deep gravitational well, galaxies surrounding us recede at ever increasing speeds, pushed apart by a small but positive vacuum energy density. Extrapolated to the far future, the geometry of our spacetime is neither asymptotically flat nor asymptotically anti de Sitter, but asymptotically de Sitter. Furthermore, the primordial universe that spawned us all also appears to be well-approximated by a dS-like geometry, albeit one with a much larger vacuum energy density. 

Despite its evident importance, to this date, no precise, complete definition exists of any theory of quantum gravity in a four-dimensional universe with positive vacuum energy density, even when disregarding all other observational constraints such as the properties of particles beyond the graviton. Although low energy effective field theory approaches to this problem are perfectly adequate for many purposes, they also lead to many deep problems and conceptual paradoxes \cite{Linde:1994gy,Dyson:2002pf,Page:2006dt,Bousso:2006xc,DeSimone:2008if,Bousso:2010yn,Borde:2001nh,Guth:2007ng} as well as formidable {technical} challenges \cite{Douglas:2006es,Denef:2008wq,Denef:2006ad,Banks:2010tj,Sethi:2017phn}, and to longstanding disagreements on how to resolve them \cite{Hawking:1984hk,Duff:1989ah}. It is unlikely that definitive progress will be made on these issues in the absence of a theoretical framework on par with AdS-CFT. 

{There have been several efforts to go beyond four-dimensional low energy effective field theory, towards a fundamental theory of quantum gravity in universes with a positive vacuum energy density. These include, but are certainly not limited to, string theory constructions of metastable de Sitter vacua \cite{Silverstein:2001xn,Bousso:2000xa,Kachru:2003aw,Dong:2010pm}, holographic considerations of the de Sitter observer's static region \cite{Gibbons:1977mu,Banks:2002wr,Banks:2005bm,Goheer:2002vf,Parikh:2004wh,Alishahiha:2004md,Anninos:2011af,Anninos:2017hhn,Verlinde:2016toy,Neiman:2017zdr}, more general holographic considerations of the landscape \cite{Freivogel:2009rf,Susskind:2007pv,Garriga:2008ks,Freivogel:2006xu,Maltz:2016max}, and the dS-CFT correspondence \cite{Strominger:2001pn,Witten:2001kn,Maldacena:2002vr}. For an overview, see e.g.\ \cite{Spradlin:2001pw,Anninos:2012qw,Bousso:2007gp}. However, these efforts fall short of providing concrete models with a precise, microscopic description of the fundamental degrees of freedom, Hilbert space and operator content, capable in principle of answering all physically sensible questions to any desired precision.}

{In this paper, we propose such a model. The perturbative low energy bulk field content of this theory includes a scalar, the graviton and an infinite tower of interacting massless higher spin fields, whose classical  dynamics is governed by Vasiliev's minimal higher spin gravity equations of motion \cite{Vasiliev:1990en,Vasiliev:2003ev}. More specifically, we propose a precise microscopic definition of the Hilbert space of this theory, its operator algebra and its Hartle-Hawking vacuum state. We show that our construction is consistent with perturbative bulk field theory expectations within the realm of their applicability, including cosmological vacuum correlation functions. We demonstrate that the theory is furthermore capable of reaching deep into the nonperturbative regime, by computing the probability of arbitrarily large field excursions. We show that the perturbative bulk QFT Heisenberg algebra, a prerequisite for any attempt at reconstructing standard perturbative bulk quantum field theory, can be reproduced from the fundamental operator algebra, but only up to a minimal error term $\sim e^{-cN}$ where $N \sim \ell^2_{\rm dS}/G_{\rm Newt} \sim S_{\rm dS}$, and only if the QFT is coarse grained and limited to access a maximal number of ``pixels'' of order $N$. In the same spirit, we argue that the computation of any gauge invariant observable in the theory can be reduced to a finite dimensional $2N \times 2N$ matrix integral. Consistent with this, we show that the physical Hilbert space of gauge-invariant $n$-particle states is finite-dimensional for any given $n$. Although much work remains to be done, there seem to be no insuperable obstacles to a precise identification and microscopic derivation of the de Sitter entropy $S_{\rm dS}$ within this framework.} 

In what follows we will give an overview of the basic formal elements of our construction, leaving the finer points and applications to the bulk of the text. {The subdivision in sections of this summary follows the subdivision in sections of the remainder of the paper.}


\subsection{Preliminaries and review} \label{sec:introSpN}

Our construction {provides a complete Hilbert space framework for the dS-CFT idea as envisioned in  \cite{Strominger:2001pn,Witten:2001kn,Maldacena:2002vr}, and more specifically for the concrete proposal of \cite{Anninos:2011ui}.} As we review in section \ref{sec:prelandrev}, the latter can be phrased roughly as the statement that the Hartle-Hawking wave function of the minimal higher spin dS$_{3+1}$ universe is given (up to contact terms) by the generating function of correlation functions of single-trace primaries in a theory of $N$ free {\it anti}-commuting scalars $\chi^a_x$, $a=1,\ldots,N$, $x \in \IR^3$:
\begin{align} \label{psiHHintro}
 \psi_{\rm HH}(\CB) \, = \, \frac{1}{Z_0} \int d\chi \, e^{-\frac{1}{2} \int \chi \CD \chi + : \chi \CB \chi  :} = \det\bigl(1+\CD^{-1} \CB\bigr)^{\frac{N}{2}} \, e^{-\frac{N}{2} \Tr(\CD^{-1} \CB)} \, .
\end{align}
Here $\CD$ is minus the Laplacian, $\CB^{xy}$ represents a general source \cite{Das:2003vw,Douglas:2010rc} coupling to bilinears $\chi_x \chi_y$, and the trace term on the right hand side appears due to the normal ordering $:\!\chi\chi\!:$ of the bilinears (defined in the usual way by subtracting a $c$-number such that the 1-point function vanishes). The bilinear contractions are defined as $\chi \chi \equiv \epsilon_{ab} \chi^a \chi^b$, where $\epsilon_{ab}$ is a constant antisymmetric matrix. Thus the theory has an Sp(N) symmetry, and one considers only source deformations in the singlet sector of Sp(N). This Sp(N) model can be viewed as effectively implementing an analytic continuation $N \to -N$ of the bosonic O(N) model known to be dual to minimal Vasiliev gravity in AdS$_{3+1}$ \cite{Klebanov:2002ja,Giombi:2009wh}, in effect flipping the sign of the cosmological constant and continuing AdS $\to$ dS \cite{Anninos:2011ui}. 

In slightly more detail, a suitable decomposition of the source $\CB$ into local differential operators,
\begin{align} \label{Bxyexpintro}
  \CB^{xy} = \sum_{s \in 2\IZ} \int d^3 z \, b^{i_1 \cdots i_s}(z) \, \CD^{xy}_{z,i_1 \cdots i_s}  \, \equiv \, b^I \CD^{xy}_I \,  ,
\end{align}
decomposes the source terms accordingly into the standard traceless conserved local (even) spin-$s$ currents $O_{i_1 \cdots i_s}(z) = \chi \partial_{i_1} \cdots \partial_{i_s} \chi + \cdots$:
\begin{align} \label{bilinearcurrentsintro}
 :\! \chi \CB \chi \!: = b^I O_I \, , \qquad O_I \equiv :\!\chi_x \CD^{xy}_I \chi_y \! : \, .
\end{align}
Here we introduced a convenient shorthand index notation $I=(z,i_1 \cdots i_s)$ labeling both spatial points $z$ and tensor indices $(i_1 \cdots i_s)$. Contracted indices are understood to be summed and integrated over. For example for $s=0$, we have $\CD_z^{xy} = \delta_z^x \delta_z^y$ and $O_z = :\! \chi_z \chi_z \!:$.
The source components $b^I$ are interpreted in the bulk as boundary values of bulk higher spin fields; for example $b^{z,ij}$ is the conformal boundary value of the spatial metric fluctuation field $h_{ij}(z)$. Taking derivatives with respect to the sources $b^I$ generates the correlation functions of the $O_I$. In particular the 2-point function is
\begin{align} \label{GIJdefintro}
 \partial_{b^I} \partial_{b^J} \psi_{\rm HH}(b)|_{b=0} = \frac{1}{4} \langle O_I O_J \rangle_{\rm Sp(N)} = - \frac{N}{2} G_{IJ} \, , \qquad G_{IJ} = \Tr(\CD^{-1} \CD_I \CD^{-1} \CD_J) \, ,
\end{align}
so for example for the scalar-scalar case we get $G_{zz'} = (\CD^{-1})_{zz'} (\CD^{-1})_{z'z} = \frac{1}{(4\pi|z-z'|)^2}$.

\subsection{Problems}

The problem with this story is that it at most half a story. We explain this  in some detail in section \ref{sec:missingHilbertspace}. The main issue is that just specifying a wave function without specifying the Hilbert space to which it belongs is physically meaningless. In particular, without an inner product, probabilities, expectation values and vacuum correlation functions cannot be computed. Naively postulating an inner product with a flat measure $d\CB$, or equivalently $d\CH$ with $\CH = \CD + \CB$, turns out to be inconsistent for a number of reasons, and indeed ignoring this immediately leads to fatal normalizability problems in the nonperturbative regime \cite{Anninos:2012ft,Banerjee:2013mca}. A better motivated inner product is obtained by requiring invariance of the measure $[d\CH]$ under  transformations induced by linear field redefinitions $\chi_x \to {R_x}^y \chi_y$, i.e.\ $\CH \to R^T \CH R$. Moreover this  naturally allows restricting the integration domain to positive definite $\CH$,  analogous to the restriction of the integration domain in a Euclidean gravity path integral to positive definite metrics. This gives a sensible measure with a normalizable $\psi(\CH)$ in discretized toy models with $K$ lattice points, as long as $K \leq 2N$, but it breaks down when $K > 2N$, including in particular in the continuum limit $K=\infty$. All of these problems can be traced down to an even more fundamental problem: there is no reason to assume the sources $b^I$ constitute a linearly independent set of operators in the fundamental description of the Hilbert space. In fact, given that the Sp(N) model itself has only $N$ degrees of freedom per spatial point, while the  sources $b^I$, purportedly canonically conjugate to the Sp(N) bilinears $O_I$, have infinitely many degrees of freedom per spatial point, makes it fairly obvious that this is unlikely to be correct. This  leads to the question: what are the proper degrees of freedom in the fundamental description of the Hilbert space?

\subsection{Proposal}

\def\CCH{{\mathscr H}}

The answer turns out to be staggeringly simple. We argue in section \ref{sec:Proposal} that the proper fundamental degrees of freedom are $2N$ {\it bosonic} fields $Q_x^\alpha$, $\alpha=1,\ldots,2N$, $x \in \IR^3$. We define a Hilbert space $\CCH$ consisting of O(2N)-invariant wave functions $\psi(Q)$ with inner product
\begin{align} \label{innerproddefintro}
 \langle \psi_1|\psi_2\rangle = \int dQ \, \psi_1(Q)^* \psi_2(Q) \, ,
\end{align}
where $dQ$ is the flat measure. The vacuum wave function is 
\begin{align} \label{psi0defintro}
 \psi_0(Q) = c \, e^{-\frac{1}{2} \int Q \CD Q} \, ,
\end{align}
with $c$ chosen such that $\langle \psi_0|\psi_0\rangle = 1$. The true physical Hilbert space $\CCH_{\rm phys}$ is obtained as the subspace of $\CCH$ invariant under global higher spin transformations. We will have more to say about this below in section \ref{sec:Hphys}; for now we stick to $\CCH$. The relation to the original source fields $\CB$ appearing in (\ref{psiHHintro}) is given by
\begin{align} \label{BQQrelintro}
 \CB^{xy} = \CD^{xx'} B_{x'y'} \CD^{y'y} \, , \qquad B_{xy} \, \equiv \, \frac{1}{N} :\! Q_x Q_y \! :  = \frac{1}{N} \, Q_x Q_y - (\CD^{-1})_{xy} \, .
\end{align}
Thus, given the bulk interpretation of the sources $\CB^{xy}$, we may read off the bulk interpretation of the $QQ$ bilinears $B_{xy}$. 
In terms of the expansion (\ref{Bxyexpintro}), this means that if we define the analog of the bilinear currents (\ref{bilinearcurrentsintro}),
\begin{align}
 B_I \, \equiv \, \Tr(\CD_I B) =  \frac{1}{N} \! :\! Q_x \CD_I^{xy} Q_y \!: \, ,
\end{align}
we get the relation $B_I = \Tr(\CD_I \CD^{-1} \CB \CD^{-1}) = G_{IJ} b^J$, with  $G_{IJ}$ defined in (\ref{GIJdefintro}), or
\begin{align} \label{bIBJidintro}
 b^I = G^{IJ} B_J \, , 
\end{align}
where $G^{IJ}$ is the inverse of $G_{IJ}$. Since by (\ref{GIJdefintro}), $G_{IJ}$ has the interpretation of a CFT 2-point function, this means the source fields $b^I$ and the $QQ$ bilinears $B_I$ are related to each other by a CFT {\it shadow transform} \cite{Ferrara:1972uq,SimmonsDuffin:2012uy}, reviewed in appendix  \ref{shadowapp}.  Cosmological vacuum correlation functions of the $b^I$ can thus be extracted by first computing the correlation functions of the $QQ$ bilinears $B_I$ in the Gaussian theory with weight $e^{-\int Q \CD Q}$ and then shadow transforming the result:
\begin{align} \label{Qmodelcorrelations}
 \langle b^{I_1} \cdots b^{I_n} \rangle &= G^{I_1 J_1} \cdots G^{I_n J_n} \langle \psi_0|B_{J_1} \cdots B_{J_n} |\psi_0 \rangle \, \nn \\
 &=\frac{c^2}{N^n} \, G^{I_1 J_1} \cdots G^{I_n J_n} \int dQ \,  e^{-\int Q \CD Q} \, :\!(Q \CD_{J_1} Q) \!: \, \cdots \, :\!(Q \CD_{J_n} Q) \!: \, .
\end{align}
Our arguments rest on two key observations:
\begin{enumerate}
 \item To leading order at large $N$ (i.e.\ at tree level from the bulk point of view), the vacuum correlation functions computed by the $Q$-model as in (\ref{Qmodelcorrelations}) coincide with the vacuum correlation functions computed using $\psi_{\rm HH}(\CB)$ defined by the Sp(N) model as in (\ref{psiHHintro}). This is established by comparing the generating functions of vacuum correlations functions to leading order in a large $N$-saddle point approximation, which at least formally can be done without knowing the measure $[d\CB]$.
 \item In discretized toy models with $K \leq 2N$ lattice points, we can pick a natural O(2N)-invariant basis $|\CH\rangle$ for $\CCH$, where $\CH=(\CD Q)(\CD Q)^T$. (This is equivalent to (\ref{BQQrelintro}), recalling that $\CH=\CD+\CB$.) We normalize the kets $|\CH\rangle$ such that the decomposition of unity in $\CCH$ is $1 = \int [d\CH] \, |\CH \rangle \langle \CH|$, where $\CH>0$ and $[d\CH]$ is the natural measure invariant under field redefinitions $\CH \to R^T \CH R$, as considered already in section \ref{sec:introSpN}. Then we have the {\it exact} relation $\psi_{\rm HH}(\CH) = \langle \CH|\psi_0\rangle = \int dQ \, \langle \CH|Q\rangle \langle Q|\psi_0\rangle$, with $\langle Q|\psi_0\rangle$ as defined in (\ref{psi0defintro}) and  $\psi_{\rm HH}(\CH)$ as defined in (\ref{psiHHintro}).
The factor $(\det \CH)^{\frac{N}{2}}$ in (\ref{psiHHintro}) arises essentially as a Jacobian for the coordinate transformation $Q \to \CH$. {This relation can be thought of as a quantum version of the well-known relation between Gaussian and Wishart matrix ensembles.}  The change of variables becomes singular when $K>2N$, and the $\CH$-space description becomes singular in this case, because the matrix $\CH$ has reduced rank $2N$. However the $Q$-space description remains well-defined. 
\end{enumerate}

\subsection{Probabilities and correlation functions}

{To illustrate how to use this framework in practice, we provide a number of sample computations in section \ref{sec:probandcorr}.} 

{We begin by computing the probability $P(b_0)$ of constant scalar mode excursions in global de Sitter with all other field modes integrated out. One of the most striking apparent pathologies of the Sp(N) wave function $\tilde\psi_{\rm HH}$ defined in (\ref{psiHHintro}), interpreted (incorrectly) as a wave function on a Hilbert space with a flat inner product measure $d\CB$, is its exponential divergence for large negative $b_0$ \cite{Anninos:2012ft}. This pathology is completely eliminated in our framework. We explicitly compute the probability density $P(b_0)$, for any value of $N$, and find it is normalizable. In the large $N$ limit it satisfies, up to ${\cal O}(1)$ constants in the exponential, $P(b_0) \sim e^{-N b_0^2}$ for small $b_0$, $P(b_0) \sim e^{-N b_0}$ for large positive $b_0$, and $P(b_0) \sim e^{-N |b_0|^3}$ for large negative $b_0$.}

We proceed by explicitly computing the scalar-scalar-graviton three-point function and the scalar three- and four-point functions in the Hartle-Hawking vacuum. {The scalar four-point function reduces to the computation of a three-dimensional four-mass box integral previously considered in the scattering amplitude literature \cite{Lipstein:2012kd}. The final result presented in \cite{Lipstein:2012kd} spans half a dozen of pages of complicated Mathematica output. We solve this integral in a different way, using the methods of \cite{Bzowski:2013sza}, and obtain a remarkably simple result (\ref{crazysimple}) that fits in a single line.}

\subsection{Perturbative bulk QFT reconstruction}

{We discuss the reconstruction of perturbative bulk quantum fields from the fundamental operator algebra of $\CCH$ in section \ref{breakdown}. A prerequisite for this is the reconstruction of the bulk QFT Heisenberg algebra. On conformal boundary fields $\alpha_I$, $\beta^J$ this is realized as $[\beta^I,\alpha_J] \propto i \delta^I_J$, $[\beta^I,\beta^J] = 0 = [\alpha_I,\alpha_J]$. The $\beta^I$ were identified with the $QQ$ bilinears $b^I = G^{IK} B_K$ in (\ref{bIBJidintro}). For several reasons, it is however manifestly impossible to find self-adjoint operators $a_J$ in the operator algebra of $\CCH$ exactly realizing these commutation relations. One reason is that if there were such operators, we could exponentiate them to unitary operators acting as translations $b^I \to b^I + c^I$, for arbitrary constants $c^I$, inconsistent with the positivity of $\CH = \CD + b^I \CD_I = \CD Q (\CD Q)^T$. Another reason is that in discretized models with $K > 2N$ (so in particular in the continuum limit $K=\infty$), the $b_I$ are not independent.} 

When $K,N \to \infty$ with $\kappa \equiv \frac{K}{2N}$ fixed and $\kappa \lesssim 0.17$, it is nevertheless possible to define self-adjoint operators $a_I$ such that the Heisenberg algebra is satisfied up to exponentially small corrections $e^{-g(\kappa) N}$, where $g(\kappa)$ is some ${\cal O}(1)$ function derived from the asymptotics of the Tracy-Widom distribution. We relate this exponential error directly to the probability of occurrence of nonperturbatively large fluctuations, defined in a precise sense.  

When $K>2N$, including in the continuum limit $K=\infty$, the naive construction fails, but it is still possible to define ``coarse grained'' self-adjoint operators $\bar b^I$, $\bar a_I$, reducing the number of effectively accessed spatial ``pixels'' to a finite number $K_{\rm eff}$, allowing again to reconstruct the perturbative bulk Heisenberg algebra up to exponentially small errors. This indicates however that  perturbative quantum field theory in higher spin de Sitter space breaks down beyond a $K_{\rm eff}$ of more than ${\cal O}(N)$ pixels, a number of the order of the de Sitter entropy $S_{\rm dS} \sim \ell_{\rm dS}^2/G_{\rm Newton}$.

\subsection{Physical Hilbert space} \label{sec:Hphys}

The physical Hilbert $\CCH_{\rm phys}$ space is the subspace of $\CCH$ invariant under higher spin gauge transformations. As we explain in section \ref{sec:GIaPHS}, the choice of the operator $\CD$ can be thought of as a partial gauge fixing of these gauge transformations. For example, instead of taking $\CD$ to be minus the Laplacian on $\IR^3$, we could alternatively have chosen $\CD$ to be minus the conformal Laplacian on the round 3-sphere. These two choices  are related by a linear field redefinition $\chi \to R \chi$ in the Sp(N) model or $Q \to R Q$ in the $Q$-model, where $R$ is some combination of a Weyl rescaling and a spatial diffeomorphism, effectively transforming $\CD \to R^T \CD R$. From the bulk point of view, this corresponds to an asymptotic spacetime diffeomorphism mapping planar coordinates to global coordinates. Picking a particular $\CD$ almost completely fixes this gauge symmetry, but not quite, since there exists combinations $R$ of Weyl rescalings and spatial diffeomorphisms leaving $\CD$ unchanged, i.e.\ satisfying $R^T \CD R = \CD$. These residual gauge transformations  form a finite-dimensional group, namely the conformal or dS isometry group SO(1,4). In quantum gravity on global de Sitter space (unlike AdS space), these residual symmetries must be viewed as gauged, that is to say, physical states must be SO(1,4)-invariant \cite{DeWitt:1967yk,Higuchi:1991tk,Higuchi:1991tm}. Analogous considerations hold for the higher spin generalization of these symmetries: picking a particular $\CD$ fixes the higher spin generalizations of bulk diffeomorphisms up to global higher spin transformations, i.e.\ the group $\CG$ of linear field redefinitions $R$ satisfying $R^T \CD R = \CD$ \cite{Eastwood:2002su,Mikhailov:2002bp,Joung:2014qya,Segal:2002gd}. Just like its conformal subgroup, this residual higher spin symmetry group must be viewed as gauged in higher spin de Sitter space. Thus, $\CCH_{\rm phys}$ is the $\CG$-invariant subspace of $\CCH$. 

{We approach the problem of explicitly constructing $\CCH_{\rm phys}$ by first considering the discretized finite $K$ models, where this construction is straightforward because the residual gauge symmetry group $\CG$ is isomorphic to O(K), hence compact, and then taking the limit $K \to \infty$. This allows us to explicitly construct a basis of gauge invariant $n$-particle states, and to interpret them as ``group-averages'' of ordinary $n$-particle states on $\CCH$, generalizing the analogous SO(1,4) group-averaging procedure of perturbative gravity in dS \cite{Higuchi:1991tm,Giddings:2007nu,Marolf:2008hg}. We find that for each $n$, there is a {\it finite} number of such states. Consistent with this, we show that all gauge invariant quantities can be computed by a $2N \times 2N$ O(2N)-invariant matrix integral, effectively reducing the number of degrees of freedom to the $2N$ eigenvalues of this matrix. This eliminates all UV divergences of perturbative QFT, and makes it plausible there exists a microscopic identification  of the (finite) de Sitter entropy in this framework.}

\section{Preliminaries and review} \label{sec:prelandrev}

{With the goal of making this work accessible to readers who aren't higher spin gravity experts, and to introduce some notation that will be useful in the remainder of the paper, we review in this section some basic properties of free massless higher spin fields in a fixed de Sitter background, as well as their  Bunch-Davies/Hartle-Hawking vacuum wave function. We emphasize the use of conformal boundary fields and their canonical Heisenberg algebra in this setting, as such fields will form the natural interface with operators defined in the boundary theory we propose in subsequent sections. We then  review the dS-CFT idea relating the Hartle-Hawking wavefunction of an interacting bulk gravity theory to the partition function of a boundary conformal field theory, and its explicit realization in higher spin gravity as proposed in \cite{Anninos:2012qw}.}

\subsection{Free higher spin fields in de Sitter space} \label{sec:freehsfids}

We work in planar coordinates with $\ell_{\rm dS} \equiv 1$, so the background de Sitter metric in $(d+1)$-dimensions takes the form 
\begin{equation}
 ds^2 = \frac{-d\eta^2+dx^2}{\eta^2} \, ,
\end{equation}
with $\eta<0$ and $x \in \mathbb{R}^d$. Future infinity corresponds to $\eta = 0$, while $\eta=-\infty$ is the past horizon of the planar patch. 

We will frequently work in momentum space. Our conventions for the $d$-dimensional Fourier transform are
\begin{align}\label{fourier1}
 f(x) \equiv \int \frac{d^d k}{(2\pi)^d} \, f(k) \, e^{i k \cdot x} \, .
\end{align} 
To avoid dragging along factors $(2\pi)^d$, we will use the notations
\begin{align}\label{fourier2}
 \int_k \equiv \int \frac{d^d k}{(2\pi)^d}  \, , \qquad \delta_k \equiv (2\pi)^d \delta^d(k) \, .
\end{align}

\subsubsection{$m^2=2$ free scalar in dS$_{3+1}$}

Vasiliev gravity in dS$_{3+1}$ has a scalar field of mass $m^2=2$. The action of a free scalar with this mass is 
\begin{align}
 S = \frac{1}{2\gamma} \, \int \frac{d\eta \, d^3 x}{\eta^4} \left( \eta^2 (\partial_\eta \phi)^2 - \eta^2 (\partial_i \phi)^2 - 2 \phi^2 \right) \, .
\end{align}
Canonical normalization corresponds to $\gamma = 1$, but for some purposes it is useful to consider a different normalization (for example ``gravity'' normalization, in which $\gamma$ is proportional the the Newton constant), so we keep it arbitrary here. The mode expansion is
\begin{align}
 \phi(\eta,x) = \sqrt{\gamma} \int_k \frac{1}{\sqrt{2k}} \bigl(  a_k \, \eta \, e^{-i k \eta+i k \cdot x} +  a_k^\dagger \, \eta \, e^{i k \eta-i k \cdot x} \bigr) \, ,
\end{align}
where the coefficients satisfy the canonical commutation relations
\begin{align} \label{cancomaadag}
 [a_k,a_{k'}^\dagger] = \delta_{k-k'} \, , \qquad [a_k,a_{k'}] = 0 = [a_k^\dagger,a_{k'}^\dagger] \, .
\end{align}
The free Bunch-Davies vacuum $|0\rangle$ is the state annihilated by all of the $ a_{k}$. The vacuum 2-point function in momentum space is
\begin{align}
 \langle 0|\,  \phi(\eta,k) \, \phi(\eta',k') |0\rangle = \frac{\gamma}{2k} \, \eta\eta' {e^{-ik(\eta-\eta')}} \, \delta_{k+k'} \, .
\end{align}
By decomposing $e^{i k \eta}=\cos(k\eta) + i \sin(k\eta)$, we can alternatively write the mode expansion in the form
\begin{align} \label{phialphatildebeta}
 \phi(\eta,x) =  -\eta \int_k \bigl(  \alpha(k)  \cos(k \eta)  -  \tilde\beta(k)   \sin(k \eta) \bigr) e^{i k \cdot x} \, , 
\end{align}
where the signs are picked for convention consistency with (\ref{alphabetamodes}) below and
\begin{align}
 \alpha(k) \equiv \sqrt{\frac{\gamma}{k}} \, \frac{a_k + a_{-k}^\dagger}{\sqrt{2}} \, , \qquad \tilde\beta(k) \equiv i \sqrt{\frac{\gamma}{k}} \, \frac{a_k - a_{-k}^\dagger}{\sqrt{2}} \, .
\end{align}
Asymptotically for $\eta \to 0$, we have, up to $(1+O(\eta^2))$ corrections,
\begin{align}
 \phi(\eta,x) \approx   -\alpha(x) \, \eta + \beta(x) \, \eta^2  \, = \, \int_k  \bigl(  -\alpha(k) \, \eta   +  \beta(k) \, \eta^2  \bigr) e^{i k \cdot x}  \, , \qquad \beta(k) \equiv k \tilde \beta(k) \, .
\end{align}
Note that $\alpha(k)$ and $\beta(k)$ are the Fourier coefficients of {\it local} boundary fields $\alpha(x)$ and $\beta(x)$, i.e.\ boundary fields that can be obtained locally from the bulk field $\phi(\eta,x)$ in the limit $\eta \to 0$. These boundary fields transform under the SO(1,4) de Sitter isometry group as $d=3$ conformal primary fields of dimension $\Delta=1$ and $\Delta=2$, respectively. In contrast, the boundary field $\tilde\beta(x)$ with Fourier coefficients $\tilde\beta(k) = \frac{1}{k} \beta(k)$ is non-locally related to $\beta$ and thus $\phi$:
\begin{align} \label{shadowtransformbetatildebeta}
 \tilde\beta(x) \equiv \int_k \, \tilde\beta(k) \, e^{i k \cdot x}  = \frac{1}{2\pi^2} \int d^3 y \, \frac{1}{|x-y|^2} \, \beta(y) \, ,
\end{align}
where we used the Fourier transform formula (\ref{fouriertransformxk}). 
The relation between $\beta$ and $\tilde\beta$ is known in the context of conformal field theories as the ``shadow'' transform \cite{Ferrara:1972uq}. We review this in appendix \ref{shadowapp}; the general shadow transform for scalar operators is given in equation (\ref{scalarshadowtr}). In the case at hand this maps the dimension $\bar\Delta=2$ conformal primary field $\beta(x)$ to the dimension $\Delta=3-\bar\Delta=1$ conformal primary field $\tilde\beta(x)$. So although $\tilde\beta$ is non-locally related to the bulk field $\phi$, it nevertheless transforms locally under the de Sitter isometry group. 

Expressed in terms of the local boundary operators $\alpha$ and $\beta$, the canonical commutation relations (\ref{cancomaadag}) take the conventional local form for canonically conjugate fields:
\begin{align}
 [\beta(k),\alpha(k')] = i \, \gamma \, \delta_{k+k'}  \, , \qquad 
  [\beta(x),\alpha(x')] = i \, \gamma\, \delta^3(x-x') \, ,
\end{align} 
with all other commutators vanishing.
In terms of the the $\Delta=1$ boundary operators $\alpha$ and $\tilde\beta$ on the other hand, we get
\begin{align} 
 [\tilde\beta(k),\alpha(k')] = i  \, \gamma \, G(k,k')  \, , \qquad 
  [\tilde\beta(x),\alpha(x')] = i \, \gamma \, G(x,x') \, ,\label{canoncomalphatildebeta}
\end{align} 
where
\begin{align} \label{GkkGxxdef}
  G(k,k') = \frac{1}{k} \delta_{k+k'} \, , \qquad G(x,x') =  \frac{1}{2\pi^2} \frac{1}{|x-x'|^2} \, .
\end{align}
Note that these take the form of a $\Delta=1$ 3d CFT 2-point function, consistent with the SO(1,4) symmetry constraints. The vacuum 2-point functions of $\alpha$ and $\tilde\beta$ are similarly
\begin{align}
 \langle 0|\alpha \alpha |0\rangle &= \frac{\gamma}{2} \, G  & \langle 0|\alpha \tilde\beta |0\rangle &= -i \, \frac{\gamma}{2} \, G \nn \\ \label{scalartwoptfunctionsvac}
 \langle 0|\tilde\beta  \tilde\beta |0\rangle &= \frac{\gamma}{2} \, G & \langle 0|\tilde\beta \alpha |0\rangle &= + i \, \frac{\gamma}{2}\, G \, ,
\end{align}
which should be read for example as $\langle 0|\alpha(x) \alpha(x') |0\rangle = \frac{\gamma}{2} G(x,x')$ with $G$ as in (\ref{GkkGxxdef}). 
Again all of these take the form of $\Delta=1$ CFT 2-point functions,  as required by symmetry. However these correlation functions cannot possibly be reproduced by correlation functions of a conventional CFT, i.e.\ by a Euclidean path integral with insertions of $\alpha$ and $\tilde\beta$, since such insertions would necessarily commute. In particular this would imply $\langle \alpha(x) \tilde\beta(x') \rangle = \langle \tilde\beta(x') \alpha(x) \rangle$, inconsistent with the above. Indeed as we will discuss in more detail below, in contrast to AdS-CFT, a complete framework for dS-CFT, capable in particular of reproducing quantum mechanics in the bulk, requires more ingredients than just a boundary CFT.

\subsubsection{Free higher spin fields in dS$_{d+1}$} \label{sec:freehsdS}

A free massless spin $s$ field in dS$_{d+1}$ is a totally symmetric tensor $\phi_{\mu_1 \cdots \mu_s}$ satisfying the double-tracelessness conditions
\begin{equation}
 {\phi^{\nu_1 \nu_2}}_{\nu_1 \nu_2 \mu_3 \cdots \mu_s} = 0 \, ,
\end{equation}
and the equations of motion \cite{Fronsdal:1978rb}
\begin{equation} \label{hseom}
 \nabla_\nu \nabla^\nu \phi_{\mu_1 \cdots \mu_s} 
 - s \nabla_\nu \nabla_{(\mu_1} {\phi^\nu}_{\mu_2 \cdots \mu_s)} + \tfrac{1}{2} s(s-1) \nabla_{(\mu_1} \nabla_{\mu_2} {\phi^\nu}_{\nu \mu_3 \cdots \mu_s)}
 -2(s-1)(s+d-2) \phi_{\mu_1 \cdots \mu_s} = 0 \, , \nn
\end{equation}
where $\nabla_\mu$ is the covariant derivative and the symmetrization is over the $(\mu_1 \cdots \mu_s)$ indices only. The equations of motion are invariant under the gauge transformations 
\begin{equation} \label{gaugetransformations}
 \phi_{\mu_1 \cdots \mu_s} \to \phi_{\mu_1 \cdots \mu_s} + \nabla_{(\mu_1} \Lambda_{\mu_2 \cdots \mu_s)} \, , \qquad {\Lambda^\nu}_{\nu \mu_2 \cdots \mu_s} = 0 \, .
\end{equation}
In what follows we will assume $d$ is odd (and we will mostly have $d=3$ in mind). In a suitable gauge, a basis of solutions is provided by the following set of canonically normalized positive frequency mode functions:
\begin{align}\label{psimodes} 
 \psi^{k \sigma}_{i_1 \cdots i_s}(\eta,x)  &= e^\sigma_{i_1 \cdots i_s}(k) \, (-\eta)^{\frac{d}{2}-s} \, \sqrt{\tfrac{\pi}{4}} \, H^{(1)}_{\frac{d}{2}+s-2}\bigl(-k \eta\bigr) \, e^{i k \cdot x} \, , 
\end{align}
with all field components involving one or more time-indices equal to zero.
The functions $H_\nu^{(1)}(z)$ are Hankel $H$ functions of the first kind, $k \in {\mathbb R}^d$ labels the momentum of the mode, and $\sigma$ labels an orthonormal basis of polarization tensors $e_{i_1 \cdots i_s}(k)$ satisfying the tracelessness and transversality conditions ${e^j}_{j i_3 \cdots i_s} = 0$, $ k_j \, {e^j}_{i_2 \cdots i_s} = 0$, where indices are raised with the flat Euclidean metric $\delta^{ij}$, and orthonormality means $(e^\sigma,e^\tau) \equiv  e^{\sigma*}_{i_1 \cdots i_s} e^{\tau \, i_1 \cdots i_s} = \delta^{\sigma \tau}$.
In the $d=3$ case of interest, $\sigma$ takes on two values, corresponding to the two helicity states of a massless spinning particle.
In the far past $\eta \to -\infty$ we have, up to an overall phase
\begin{equation} \label{farpastpsi}
 \psi^{k\sigma}_{i_1 \cdots i_s}(\eta,x) \approx e^\sigma(k)_{i_1 \cdots i_s} \, \eta^{-s+\frac{d-1}{2}} \, \frac{1}{\sqrt{2k}} \, e^{-i k \eta + i k \cdot x} \, ,
\end{equation}
which we recognize as the canonically normalized positive frequency mode associated with the standard Bunch-Davies free vacuum.

The mode expansion of the free spin $s$ field takes the form
\begin{equation} \label{hsqf}
 \phi_{i_1 \cdots i_s}(\eta,x) =  \sqrt{\gamma} \sum_\sigma  \int_k \,  a_{k \sigma}  \, \psi^{k\sigma}_{i_1 \cdots i_s}(\eta,x) +  a_{k \sigma}^\dagger  \, \psi^{k\sigma*}_{i_1 \cdots i_s}(\eta,x)  \, , 
\end{equation}
satisfying canonical commutation relations 
\begin{equation}
 [ a_k^\sigma, a_{k'}^{\sigma'\dagger}] = \delta^{\sigma \sigma'} \, \delta_{k-k'} \, .
\end{equation} 
The Bunch-Davies vacuum $|0\rangle$ is the state annihilated by all of the $ a_{k \sigma}$. The vacuum 2-point function in momentum space is 
\begin{align}
 &\langle 0|\, {\phi}_{i_1 \cdots i_s}(\eta,k) \, {\phi}_{i'_1 \cdots i'_s}(\eta',k') \,  |0\rangle \nonumber \\
 & \qquad    = \gamma \, (\eta \eta')^{\frac{d}{2}-s}   \, \frac{\pi}{4} \, 
 H^{(1)}_{\frac{d}{2}+s-2}\bigl(-k \eta\bigr) \, H^{(2)}_{\frac{d}{2}+s-2}\bigl(-k \eta'\bigr) \, \Pi_{i_1 \cdots i_s,i'_1 \cdots i'_s}(k) \, \delta_{k+k'} \, ,
\end{align}
where 
$\Pi_{i_1 \cdots i_s,i'_1 \cdots i'_s}$ is the projector onto spin $s$ transverse traceless polarizations,
\begin{equation} \label{projector}
 \Pi_{i_1 \cdots i_s,i'_1 \cdots i_s'}(k) = \sum_\sigma e^\sigma_{i_1 \cdots i_s}(k) \,  e^{\sigma*}_{i'_1 \cdots i'_s}(k) \, . 
\end{equation}
For example for spin $s=1$ and $s=2$, we have, respectively, 
\begin{align}
  \Pi_{i,i'}(k) = \delta_{ii'} - \frac{k_i k_{i'}}{k^2} \, , \qquad
  \Pi_{ij,i'j'}= \frac{1}{2} \bigl( \Pi_{i,i'} \Pi_{j,j'} + \Pi_{i,j'} \Pi_{j,i'} \bigr)  - \frac{1}{d\!-\!1} \, \Pi_{i,j} \Pi_{i',j'} \, .
\end{align}
Decomposing $H^{(1)}_\nu(z) = J_\nu(z) + i Y_\nu(z)$, we can alternatively write the mode expansion in terms of boundary fields analogous to (\ref{phialphatildebeta}):
\begin{align} \label{alphabetamodes}
 \phi_{i_1 \cdots i_s}(\eta,x) =  (-\eta)^{d-2} \int_k \,  \left( \alpha_{i_1 \cdots i_s}(k)  \, \bar J_{\frac{d}{2}+s-2}(-k\eta)   +  \tilde\beta_{i_1 \cdots i_s}(k) \, \bar Y_{\frac{d}{2}+s-2}(-k\eta) \right) e^{i k \cdot x} \, ,
\end{align}
where
\begin{align}
 \bar J_\nu(z) \equiv \sqrt{\tfrac{\pi}{2}} \, z^{-\nu} J_{\nu}(z) \, , \qquad 
 \bar Y_\nu(z) \equiv \sqrt{\tfrac{\pi}{2}} \, z^{-\nu} \,Y_{\nu}(z) \, ,
\end{align}
and
\begin{align} \label{alphabetadefinition}
 \alpha_{i_1 \cdots i_s}(k) &= \sqrt{\gamma} \, k^{\frac{d}{2}+s-2}\,\, \bigl(a_{k \sigma} \, e^\sigma_{i_1 \cdots i_s} + a_{k \sigma}^\dagger \, e^{\sigma*}_{i_1 \cdots i_s} \bigr)/\sqrt{2} \nn \\
 \tilde\beta_{i_1 \cdots i_s}(k) &= \sqrt{\gamma} \, k^{\frac{d}{2}+s-2} \, i \, \bigl(a_{k \sigma} \, e^\sigma_{i_1 \cdots i_s} - a_{k \sigma}^\dagger \, e^{\sigma*}_{i_1 \cdots i_s} \bigr)/\sqrt{2} \, .
\end{align}
The late time $\eta \to 0$ behavior is inferred from  
\begin{align}
 \bar J_\nu(z) \approx \sqrt{\tfrac{\pi}{2}} \, \tfrac{1}{2^{\nu} \Gamma(\nu+1)}  \, , \qquad \bar Y_\nu(z) \approx - \sqrt{\tfrac{\pi}{2}} \, \tfrac{2^\nu \Gamma(\nu)}{\pi} \, z^{-2\nu}  \qquad (z \to 0, \nu>0) \, ,
\end{align}
yielding
\begin{equation} \label{futurephi}
 \phi_{i_1 \cdots i_s}(\eta,x) \approx   \int_k \,  \left( c_1 \, \alpha_{i_1 \cdots i_s}(k) \,  \eta^{d-2}   +  c_2 \,  k^{-d-2s+4} \, \tilde\beta_{i_1 \cdots i_s}(k) \, \eta^{2-2s}  \right) e^{i k \cdot x}
\end{equation}
where $c_1=(-1)^{d-2} \sqrt{\tfrac{\pi}{2}} \, \tfrac{1}{2^{\nu} \Gamma(\nu+1)}$,  $c_2=-\sqrt{\tfrac{\pi}{2}} \, \tfrac{2^\nu \Gamma(\nu)}{\pi}$, $\nu=\frac{d}{2}+s-2$. Note that for $s>2-\frac{d}{2}$, the mode with coefficient $\tilde\beta$ is the dominant one at late times, so we are departing here from the  customary notation in the AdS-CFT literature, where $\alpha$ usually refers to the dominant mode.

In terms of the boundary fields $\alpha$, $\tilde\beta$, the nonzero canonical commutators become
\begin{align} \label{betaalphacanconj}
 [\tilde\beta_{i_1 \cdots i_s}(k),\alpha_{i_1' \cdots i_s'}(k')] \, = \, i \, \gamma \, G_{i_1 \cdots i_s,i_1' \cdots i_s'}(k,k')  \, , 
\end{align}
where
\begin{align} \label{Gi1isip1sips}
 G_{i_1 \cdots i_s,i_1' \cdots i_s'}(k,k') \, \equiv  \, k^{d+2s-4} \, \Pi_{i_1 \cdots i_s,i_1' \cdots i_s'}(k) \, \delta_{k+k'}  \, .
\end{align}
The vacuum correlation functions are proportional to this same $G$; suppressing indices,
\begin{align} \label{vacvactpf}
 \langle 0|\alpha \alpha |0\rangle &= \frac{\gamma}{2} \, G  & \langle 0|\alpha\tilde\beta |0\rangle &= -\frac{i \gamma}{2} \, G \nn \\
 \langle 0|\tilde\beta \tilde\beta |0\rangle &= \frac{\gamma}{2} \, G \,  & \langle 0|\tilde\beta\alpha |0\rangle &= +\frac{i \gamma}{2} \, G \, .
\end{align}
Consistent with the transformation properties of $\alpha$ and $\tilde\beta$ under the SO(1,d+1) de Sitter isometry group, these correlators take the form of CFT 2-point functions of spin-$s$, $\Delta=d-2+s$ primary fields, that is to say higher spin conserved currents (for $s>0$). However, just like for the scalar case discussed earlier, they cannot possibly all be reproduced from a single conventional Euclidean CFT, because the operators insertions do not commute in the above. 

\begin{figure} 
 \begin{center}
   \includegraphics[width=\textwidth]{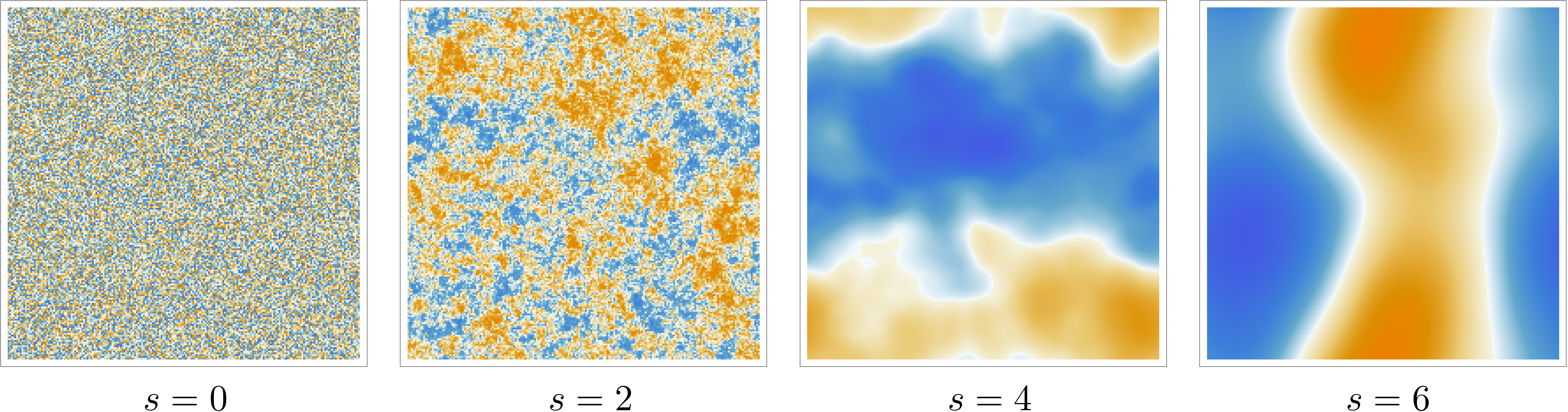}
 \end{center}
 \caption{Random cosmological samples of dS$_{3+1}$ higher spin field modes $\beta$ of spin $s=0,2,4,6$ in position space. Drawn from Gaussian distribution with power spectrum $\CD^{-s+\frac{1}{2}} \sim k^{-2s+1}$ where $\CD$ is the discrete Laplacian on a 3d torus of linear size $L$, with lattice spacing $\frac{L}{200}$ (so $K = 200^3 = 8 \times 10^6$ lattice points), with the zero mode IR-regulated by adding a mass term of order $\frac{1}{L}$ to $\CD$. \label{fig:randomfields}}
\end{figure}

The Fourier transform $\alpha_{i_1 \cdots i_s}(x)$ of $\alpha_{i_1 \cdots i_s}(k)$ is locally related to the $\eta \to 0$ asymptotic bulk field $\phi_{i_1 \cdots i_s}(\eta,x)$, but the Fourier transform $\tilde\beta_{i_1 \cdots i_s}(x)$ of $\tilde\beta_{i_1 \cdots i_s}(k)$ is not. As can be seen from (\ref{futurephi}), the boundary field $\beta$ locally obtained from  $\phi$ is related to $\tilde\beta$ by the following transformation in momentum space:
\begin{align} \label{hsshadowrel}
 \tilde\beta_{i_1 \cdots i_s}(k) = k^{d+2s-4} \, \beta_{i_1 \cdots i_s}(k) = \int_{k'} G_{i_1 \cdots i_s,i_1' \cdots i_s'}(k,k') \, \beta_{i_1' \cdots i_s'}(k')  \, .
\end{align} 
For the second equality we used the fact that $\beta_{i_1 \cdots i_s}(k)$ is transverse and traceless. Since $G$ is the 2-point function of spin-$s$ conserved currents, this can again be interpreted as the CFT shadow transform discussed in appendix \ref{shadowapp}. Thus $\beta$ transforms as a spin-$s$ primary field of conformal dimension $\Delta=d-(d-2+s)=2-s$. In particular, although $\tilde\beta(x)$ is non-locally related to $\beta(x)$, both transform locally under the de Sitter isometry group SO(1,d+1). Note that since the 2-point function of a conserved current has null directions, the the relation $\tilde\beta = G \cdot \beta$ by itself does not uniquely specify $\beta$ given $\tilde\beta$ --- one has to specify a particular gauge to invert it (transverse traceless gauge in the case at hand).  
This can be viewed as a manifestation of the gauge redundancy (\ref{gaugetransformations}). Such a gauge ambiguity in the value of $\beta_{i_1 \cdots i_s}$ for $s>0$ is to be expected: for example the spin-1 boundary field $\beta_i(x)$ parametrizes fluctuations of the late time asymptotic value of a bulk gauge field $A_i(x)$, the spin-2 boundary field $\beta_{ij}(x)$ parametrizes linearized fluctuations of the asymptotic metric $g_{ij}(x)$ on spatial slices, and so on, and these are physically defined only up to to U(1) gauge transformations, linearized diffeomorphisms, and so on. 

The commutator of $\beta$ with $\alpha$ takes the canonical $\delta$-function form,
\begin{align}
 [\beta_{i_1 \cdots i_s}(k),\alpha_{i_1' \cdots i_s'}(k')] \, =  \, i \, \gamma \, \Pi_{i_1 \cdots i_s,i_1' \cdots i_s'}(k) \, \delta_{k+k'} \, ,
\end{align}
and the $\beta$-$\beta$ 2-point function is 
\begin{align} \label{betabetatptf}
 \langle 0|\beta \beta|0\rangle = \frac{\gamma}{2} \, \tilde G \, , \qquad \tilde G_{i_1 \cdots i_s,i_1' \cdots i_s'}(k,k') = k^{-(d+2s-4)} \, \Pi_{i_1 \cdots i_s,i_1' \cdots i_s'}(k) \, \delta_{k+k'} \, .
\end{align}
Note that due to the power of $k$ growing large and negative in the momentum space 2-point function, in position space, $\tilde G(x,x')$ acquires IR divergences for $s \geq 2$ (logarithmic for the graviton, positive powers for the higher spin fields). Physically this can be thought of as being due to ``drift'' of the higher spin gauge fields due to accumulation of  frozen-out, quasi-pure-gauge modes during the effectively stochastic time evolution in de Sitter space. This manifests itself in a treelike organization of late time configurations sampled from the Bunch-Davies vacuum \cite{Anninos:2011kh,Denef:2011ee,Shaghoulian:2013qia,Roberts:2012jw}.   Correlation functions of local gauge invariant quantities such as curvatures on the other hand can be expected to remain IR finite, due to the presence of $s$ derivatives in their definition, canceling off the $k^{-2s}$. 

A visualization of real space Gaussian random fields in $d=3$ with a discretized version of the $\beta$-$\beta$ power spectrum (\ref{betabetatptf}) is given in fig.\ \ref{fig:randomfields}. The $s=2$ power spectrum is scale invariant, whereas for $s>2$ the low frequency modes increasingly dominate, manifest in the fact that the samples look increasingly smooth. For the shadows $\tilde\beta$, this behavior gets inverted, since the power spectrum gets inverted.

\subsection{Vacuum wave function and dS-CFT}

\subsubsection{Free theory}

The wave function of the free Bunch-Davies vacuum state $|0\rangle$ in an eigenbasis $|\beta\rangle$ of the boundary field operators $\beta_{i_1 \cdots i_s}(x)$ appearing in (\ref{hsshadowrel}) is 
\begin{align} \label{freevacpsi}
 \psi_{0,\rm free}[\beta] \equiv \langle \beta|0\rangle \, \propto \, e^{-\frac{1}{2\gamma} \int \beta G \beta} = e^{-\frac{1}{2\gamma} \int_k  \, k^{2s+d-4}  \, \Pi_{i_1 \cdots i_s,i_1' \cdots i_s'}(k)  \, \beta_{i_1 \cdots i_s}(k) \beta_{i_1' \cdots i_s'}(-k)} \, .
\end{align}
Indeed, with the canonical conjugate $\alpha$ represented on wave functions $\psi(\beta)$ in the usual way as a derivative, we get from (\ref{alphabetadefinition}) combined with (\ref{hsshadowrel}) that the annihilation operators can be represented as
\begin{align}
 a_\sigma e^\sigma \propto \tilde\beta + i \alpha = G \cdot \beta + \gamma \frac{\partial}{\partial \beta} \, ,
\end{align}
making (\ref{freevacpsi}) the unique solution to the defining equations $a|0\rangle = 0$. Alternatively, one easily checks that this wave function reproduces the vacuum 2-point functions listed in the previous section. 

Minimal Vasiliev higher spin gravity in dS \cite{Vasiliev:1990en,Vasiliev:2003ev,Fradkin:1987ks,Vasiliev:1999ba,Iazeolla:2007wt} has one massless higher spin field for each even spin, 
\begin{align}
 s = 0,2,4,\ldots \, .
\end{align} 
Odd spin fields appear in non-minimal higher spin gravity, but for simplicity we will only consider the minimal case here. In the free (Gaussian) approximation, the wave function of this theory is simply the product of the free wave functions given in (\ref{freevacpsi}) for all even spins:
\begin{align} \label{psi0freehs}
 \psi_{0,\rm free}[\beta] \, \propto \, e^{-\frac{1}{2\gamma} \sum_{s \in 2\IZ} \beta^{(s)} G^{(s)} \beta^{(s)} } \, . 
\end{align}
Of course, Vasiliev gravity is an interacting theory, so the full, interacting vacuum wave function will not be Gaussian. We turn to this next.

\subsubsection{Interacting theory and dS-CFT}

In interacting gravitational theories, a natural definition of a preferred cosmological vacuum state is given by the Hartle-Hawking wave function $\psi_{\rm HH}[\beta]$, semiclassically obtained as a path integral with asymptotic boundary conditions specified by $\beta$ in the future, and by ``no boundary'' Euclidean boundary conditions in the past \cite{Hartle:1983ai,Hartle:2007gi}. This generalizes the representation of the ground state wave function in time translation invariant quantum mechanical systems as a Euclidean path integral. Perturbatively, the Hartle-Hawking wave function takes the general form
\begin{align} \label{psiHHexpeprt}
 \psi_{\rm HH}[\beta] \, \propto \, e^{-\frac{1}{2\gamma} \left( g_{IJ} \beta^I \beta^J +  g_{IJK} \beta^I \beta^J \beta^K + g_{IJKL} \beta^I \beta^J \beta^K \beta^L + \cdots\right)} \, ,
\end{align}
where we collectively denoted all asymptotic field degrees of freedom by $\beta^I$, with $I$ labeling both spatial points and field species. If the perturbation theory is around a de Sitter background and the $\beta^I$ are  asymptotic boundary fields defined along the lines of our definitions above, the $\beta^I$ transform as conformal primary fields under the background de Sitter isometry group SO(1,d+1). Invariance of the wave function under this group then requires the coefficients $g_{I_1 \cdots I_n}$ to satisfy the same kind of conformal invariance constraints as $d$-dimensional CFT $n$-point functions. For example the coefficients $g_{IJ}$ in the Gaussian wave function (\ref{psi0freehs}) are CFT 2-point functions of spin-$s$ traceless conserved currents. 

A simple (but by no means unique) way of realizing these constraints more generally is to identify $\psi_{\rm HH}[\beta]$ with the partition function of an actual CFT, where the fields $\beta$ appear as sources:
\begin{align} \label{psiHHZCFT}
 \psi_{\rm HH}[\beta] \, \propto \, Z_{\rm CFT}[\beta] = \int \CD\chi \, e^{-S_{\rm CFT}[\chi] + \beta^I \CO_I}  \, ,
\end{align}
where the CFT operators $\CO_I$ are conformal primary fields of dimension $\Delta_{\CO_I} = d-\Delta_{\beta^I}$. The coefficients $g_{I_1 \cdots I_n}$ are then proportional to the connected correlation functions of the $\CO_I$.
This is the basic idea of the dS-CFT correspondence \cite{Strominger:2001pn,Witten:2001kn,Maldacena:2002vr}. In this form, it is roughly speaking an analytic continuation of the AdS-CFT correspondence: in Euclidean AdS, $\psi_{\rm HH}[\beta]$ becomes a bulk path integral with spatial boundary conditions parametrized by $\beta$, and the above equality becomes just the standard GKPW prescription \cite{Gubser:1998bc,Witten:1998qj}.

Unfortunately, this is only morally speaking an analytic continuation, because the unitarity constraints in AdS and those in dS are not analytic continuations of each other: to obtain a unitary theory in Lorentzian AdS$_{d+1}$ perturbation theory, the CFT must furnish unitary representations of SO(2,d), whereas to obtain a unitary theory in dS$_{d+1}$ perturbation theory, the CFT must furnish unitary representations of SO(1,d+1) \cite{dixmier,Dobrev:1977qv,Joung:2006gj,Joung:2007je}. {In the context of dS-CFT, this point was emphasized in \cite{Guijosa:2003ze,Chatterjee:2016ifv}.} In particular this gives rise to rather different allowed spectra of primary conformal dimensions $\Delta$. This is manifest for example in the relation between bulk field mass $m$ and boundary field conformal dimension $\Delta$; for say scalar fields in AdS$_{d+1}$ this is $m^2 =\Delta(\Delta-d)$, whereas for scalar fields in dS$_{d+1}$ this is $m^2 =-\Delta(\Delta-d)$ (in units with curvature radius equal to 1). In most CFTs familiar as duals of AdS spacetimes, the spectrum of single-trace scalar primary operators includes arbitrarily large $\Delta$, which evidently renders them unacceptable as duals to dS, since it would imply the existence of bulk scalar fields of arbitrarily large negative $m^2$. This forms a significant obstacle to constructing examples of CFTs potentially dual to dS, and as a result examples are rare.  

A striking exception to these obstacles in relating AdS duals to dS duals, pointed out in \cite{Anninos:2011ui}, is the free bosonic O(N) vector model dual to parity-even minimal AdS$_4$ Vasiliev gravity \cite{Klebanov:2002ja,Giombi:2009wh}, and cousins thereof \cite{Sezgin:2003pt,Chang:2012kt}. This theory has $N$ real bosonic fundamental scalar fields $\chi^a(x)$, $a=1,\ldots,N$, with Euclidean action on $\IR^3$ given by
\begin{align} \label{ONaction}
 S = \int d^3 x \, (\partial_i \chi^a)^2 \, .
\end{align}
The space of physical operators consists of O(N)-invariant combinations of the $\chi^a$. The building blocks of those are primary single-trace (i.e.\ bilinear) operators. These consist of a $\Delta=1$ scalar $\CO(x)$ and spin $s$, $\Delta=1+s$, traceless conserved currents for all even $s>0$:  
\begin{align} \label{primsingtrace}
 \CO = \,c_0  :\! \chi^2 \! : \, , \qquad \CO_{ij} = \, c_2  :\! \chi \partial_i \partial_j \chi - 3 \partial_i \chi \partial_j\chi + \delta_{ij} (\partial \chi)^2 \! : \, , \qquad \ldots
\end{align}
where the $c_s$ are at this point arbitrary normalization constants. 
The operator $\CO_{ij}$ is the conserved traceless energy-momentum tensor, satisfying $\partial_j \CO_{ji} = 0$, $\CO_{jj} = 0$. Similarly constructed  spin $s$ conserved currents $\CO_{i_1 \cdots i_s}$ satisfy $\partial_{j} \CO_{ji_2 \cdots i_s} = 0$, $\CO_{jj i_3 \cdots i_s}=0$. This operator spectrum is consistent with perturbative unitarity both in AdS and in dS. For example in AdS, a $\Delta=1$ scalar primary corresponds to a bulk scalar of mass $m^2=-2$, whereas in dS it corresponds to a bulk scalar of mass $m^2=+2$. A spin-2, $\Delta=3$ primary  (i.e.\ the energy-momentum tensor) corresponds to a massless spin-2 bulk field in both cases (i.e.\ the graviton), and the higher spin currents correspond to massless higher spin fields reviewed in the previous sections. Indeed the  operator spectrum exactly coincides with the spectrum of higher spin fields in minimal Vasiliev gravity in both AdS$_4$ and dS$_4$.  

Nevertheless, using $Z_{\rm CFT} = Z_{O(N)}$ in the identification (\ref{psiHHZCFT}) does {\it not} lead to a sensible wave function $\psi_{\rm HH}[\beta]$: direct comparison with perturbative bulk computations of $g_{IJ}$ and $g_{IJK}$ in (\ref{psiHHexpeprt}) reveals that one gets the wrong sign for these coefficients in this way \cite{Anninos:2011ui}. One does get the correct sign if, as in \cite{Hertog:2011ky}, instead one identifies 
\begin{align} \label{psiHHinvZON}
  \psi_{\rm HH}[\beta]=\frac{1}{Z_{\rm O(N)}[\beta]} \, .
\end{align}
In \cite{Anninos:2011ui} it was furthermore shown that if one grants the O(N) - Vasiliev duality in AdS, then the identification (\ref{psiHHinvZON}) gives the correct Hartle-Hawking wave function for Vasiliev in dS, at least in perturbation theory.

An alternative and more readily generalizable way of expressing this identification is to consider {\it anti}-commuting Grassmann scalars instead of the bosonic scalars of the O(N) model. Since the theory is Gaussian, the fermionic partition function will simply be the inverse of the bosonic partition, effectively reproducing (\ref{psiHHinvZON}). Moreover, the correspondence then continues to work for the interacting $\chi^4$ model (dual to Vasiliev with boundary conditions on $\alpha$ rather than $\beta$ for the scalar), in which case the bosonic and fermionic partition functions are no longer each others inverse \cite{Anninos:2011ui}. We will review the free Grassmann model in more detail next.

\subsection{The Sp(N) model} \label{sec:SpNmodel}

To establish notation which will be useful further on, we spell out some details here about the free fermionic Grassmann model. Consider Grassmann-valued scalar fields $\chi^a_x$, $x$ parametrizing spatial points, $a=1,\ldots,\N$,  with action
\begin{align}
 S_0 = \frac{1}{2} \int  \chi \CD \chi \,  ,
\end{align}
where $\CD$ equals minus the Laplacian on flat $\IR^3$, minus the conformal Laplacian on the round $S^3$, or any other similar operator defining a free CFT. In what follows we will assume for simplicity we are considering flat $\IR^3$. Note that the index contractions of the $\chi^a$ can no longer be performed with the O(N)-invariant metric $\delta_{ab}$, as this gives zero action for anticommuting fields. Instead of the symmetric $\delta_{ab}$, we must contract by a constant antisymmetric $\epsilon_{ab}$, which we can take to be of the following standard symplectic form:
\begin{align} \label{symplecticform1}
 \chi \chi \equiv \epsilon_{ab} \chi^a \chi^b \, , \qquad \epsilon_{ab} = \begin{pmatrix} 0_{\frac{\N}{2} \times \frac{\N}{2}} & 1_{\frac{\N}{2} \times \frac{\N}{2}} \\ -1_{\frac{\N}{2} \times \frac{\N}{2}} & 0_{\frac{\N}{2} \times \frac{\N}{2}} \end{pmatrix} \, .
\end{align} 
This is invariant under the group Sp(N) of linear symplectic transformations preserving $\epsilon_{ab}$. We are assuming $\N$ is even here. Thus the group Sp(N) takes over the role of O(N) in this model. In particular, we view this group as gauged, in the sense that physical operators are restricted to be Sp(N)-invariant combinations of the $\chi^a$. 

The two-point function is
\begin{align}
 \llang \chi^a_x \chi^b_y \rrang = \epsilon^{ab} (\CD^{-1})_{xy} \, ,
\end{align}
where $\epsilon^{ab}=-\epsilon_{ab}$, and (on $\IR^3$) 
\begin{align} \label{inverseCD}
  (\CD^{-1})_{xy} = \frac{1}{4 \pi \, |x-y|} \, .
\end{align}
Normal ordering is defined such that $\llang \, :\! \chi^a_x \chi^b_y \! : \, \rrang \equiv 0$,
that is 
\begin{align} \label{NOsubtr}
:\! \chi^a_x \chi^b_y \! : \,\, \equiv \chi^a_x \chi^b_y - \llang \chi^a_x \chi^b_y \rrang =  \chi^a_x \chi^b_y - \epsilon^{ab} (\CD^{-1})_{xy} \, .
\end{align}
The single-trace primary fields $\CO_{i_1 \cdots i_s}(x)$ consist of even spin $s$ traceless conserved current bilinears, as in (\ref{primsingtrace}), i.e.\ $\CO =  c_0\! :\! \chi^2 \! :$, $\CO_{ij} = c_2\! :\! \chi \partial_i \partial_j \chi - 3 \partial_i \chi \partial_j\chi + \delta_{ij} (\partial \chi)^2 \!:$, and so on. 
The partition function $Z_{\rm Sp(N)}[\beta]$ is the generating function for correlation functions of these operators:
\begin{align} \label{ZSPNbbb}
 Z_{\rm Sp(N)}(b,b^{ij},b^{ijkl},...) = \frac{1}{Z_0} \int d \chi \, e^{-\frac{1}{2}  \int \chi \CD \chi + b  \CO + b^{ij} \CO_{ij} + b^{ijkl} \CO_{ijkl} + ...  } \, ,
\end{align} 
where $b_x$, $b^{ij}_x$, $b^{ijkl}_x$, ... are source fields and $Z_0 = \int d \chi \, e^{-\frac{1}{2} \int \chi\CD\chi}$. Since all of the $\CO_{i_1 \cdots i_s}$ are normal ordered bilinears, the deformed theory remains Gaussian, and we can naturally rewrite the partition function in the form
\begin{align} \label{ZSpNdet}
 Z_{\rm Sp(N)}(\CB) = \frac{1}{Z_0} \int d \chi \, e^{-\frac{1}{2} \left( \chi_x \CD^{xy} \chi_y + \CB^{xy} : \chi_x \chi_y :  \right) } = \det\bigl(1+\CD^{-1}\CB\bigr)^{\frac{N}{2}} e^{-\frac{N}{2} \Tr(\CD^{-1}\CB)} \, .
\end{align} 
Here contracted $x,y$ indices are integrated over, $\CD^{xy} = -\partial^2 \delta^3(x-y)$, and the trace term arises due to the normal ordering subtraction (\ref{NOsubtr}).
In (\ref{ZSPNbbb}), the source deformation $\CB$ takes the form
\begin{align}
 \CB^{xy} = \int d^3 z \Bigl( b_z \, \CD_{z}^{xy} + b^{ij}_z \, \CD^{xy}_{ij,z} + \cdots   \Bigr) \, ,
\end{align}
where, denoting $\delta_z^x = \delta^3(z-x)$ and $\partial_i = \partial_{z^i}$,
 \begin{align} \label{CDIdefinitions}
  \CD_{z}^{xy} \equiv c_0\, \delta_z^x \delta_z^y \, , \qquad
  \CD_{ij,z}^{xy} \equiv c_2 \bigl(\partial_i \partial_j \delta_z^{(x}  \, \delta_z^{y)} - 3 \, \partial_i \delta_z^{(x}  \, \partial_j \delta_z^{y)} + \delta_{ij} \, \partial_k \delta_z^{(x} \, \partial_k \delta_z^{y)}\bigr)  \, , \qquad \cdots .
\end{align}
The round brackets on the indices $x,y$ denote symmetrization: $A^{(x} B^{y)} \equiv \frac{1}{2}(A^x B^y + A^y B^x)$.
Collectively labeling the spin indices and spatial points by $I = (i_1 \cdots i_s,z)$, we can succinctly rewrite the above expressions as
\begin{align} \label{COIdef}
 \CB = \sum_I b^I \CD_I \, ,  \qquad Z_{\rm Sp(N)}(\CB) = \frac{1}{Z_0} \int d\chi \, e^{-\frac{1}{2}(\chi \CD \chi + b^I \CO_I)}  \, , \qquad \CO_I = :\! \chi \CD_I \chi \! : \, .
\end{align} 
In this notation, the primary 2-point functions are
\begin{align} \label{tptfcoicoj}
 \llang \CO_I \CO_J \rrang = - 2 N G_{IJ} \, , \qquad G_{IJ} \equiv \Tr(\CD^{-1} \CD_I \CD^{-1} \CD_J)  \, .
\end{align}
For example for $s=0$ we have $G^{(0)}_{xy} = \frac{c_0^2}{(4\pi)^2} \frac{1}{|x-y|^2}$, or in momentum space $G^{(0)}_{kk'} = \frac{c_0^2}{8} \frac{1}{k} \delta_{k+k'}$. 

For $s>0$, the bare 2-point function $G_{IJ}$ is actually UV divergent in momentum space. One way to see this is that the position space 2-point functions for spin $s$ primaries diverge at small separation $r=|x-y|$ as $r^{-2\Delta_s} = r^{-2(1+s)}$, whose Fourier transform diverges. Alternatively, working directly in momentum space in the Sp(N) model, these divergences arise from the one-loop  momentum integral computing the 2-point function. We  renormalize $G_{IJ}$ by analytic continuation, by first considering the position space 2-point function and then using the analytically continued Fourier transform (\ref{fouriertransformxk}). Equivalently, working directly in momentum space, this can be implemented for example by replacing the propagator $\CD^{-1} = p^{-2}$ by $\CD^{-\kappa} = p^{-2\kappa}$  for some value of $\kappa$ for which the integral converges, and the analytically continuing $\kappa \to 1$. This leads to finite momentum space two-point functions with the expected conformal invariance properties. (In general one should be more careful \cite{Bzowski:2013sza} when logarithmic divergences occur, but this is not the case in the $d=3$ case of interest to us.)  In what follows this renormalization of $G_{IJ}$ and higher-point generalizations will always be tacitly assumed. 

Due to the conformal symmetry selection rules, $G_{IJ}=0$ for $I,J$ labeling different primaries (i.e.\ for different values of $s$), so $G_{IJ}$ becomes fully diagonal in momentum space. Upon suitable choice of the operator normalization factors $c_s$, the nonzero components of $G_{IJ}$ exactly equal the 2-point functions $G_{i_1 \cdots i_s,j_1 \cdots j_s}(k,k')$ defined in (\ref{Gi1isip1sips}). For example for the scalar this means picking $c_0 = \sqrt{8}$. 

It will be very useful for our purposes further on to have an explicit formula extracting the sources $b^I$ in $\CB=\sum_I b^I \CD_I$ from a given $\CB$. Using the above considerations, we have
\begin{align} \label{bIfromCB}
 \tilde{b}_I \equiv G_{IJ} b^J = \Tr(\CD_I \CD^{-1} \CB \CD^{-1}) \, .
\end{align}
Notice that since $G_{IJ}$ equals the 2-point function (\ref{Gi1isip1sips}), $\tilde b_I$ is precisely the shadow transform of $b^I$ as defined in (\ref{hsshadowrel}). Upon picking a gauge for the $s>0$ source gauge fields $b^I$ (most conveniently, transverse traceless gauge), $G_{IJ}$ can be inverted to $b^I = G^{IJ} \tilde b_J$, allowing explicit extraction of individual source fields $b^I$ from any given $\CB$.

Expressions similar to (\ref{tptfcoicoj}) can be written for all connected $n$-point functions, as can be seen most easily by rewriting (\ref{ZSpNdet}) as 
\begin{align} \label{ZSpNexpansion}
 Z_{\rm Sp(N)}(\CB) &= \det\bigl(1+\CD^{-1}\CB\bigr)^{\frac{N}{2}} e^{-\frac{N}{2} \Tr(\CD^{-1}\CB)} \nn \\
 &= e^{\frac{N}{2} \Tr [\log(1+ \CD^{-1}\CB) - \CD^{-1} \CB]} \nn \\
 &= 
 e^{-\frac{N}{2} \Tr [\frac{1}{2}(\CD^{-1}\CB)^2 - \frac{1}{3}(\CD^{-1}\CB)^3 + \cdots ]} \, \nn \\
 &= e^{-\frac{N}{2}\Tr[\frac{1}{2} G_{IJ} b^I b^J - \frac{1}{3} G_{IJK} b^I b^J b^K + \cdots]} \, ,
\end{align}
where 
\begin{align} \label{GI1Indef}
 G_{I_1 \cdots I_n} \equiv \Tr\bigl(\CD^{-1} \CD_{(I_1} \cdots \CD^{-1} \CD_{I_n)} \bigr) \, .
\end{align}
Again this object is to be understood in a renormalized sense.
Since $\log Z$ is the generating function of connected correlation functions, the coefficients $G_{I_1 \cdots I_n}$ are proportional to the connected $n$-point functions of the $\CO_I$ in the Sp(N) model. The 1-point function vanishes by construction due to the normal ordering subtraction, whence the wave function has a local maximum at $\CB=0$. Since we have picked the normalizations $c_s$ of the operators $O_I$ to be such that $G_{IJ}$ coincides with (\ref{Gi1isip1sips}), we see that the Gaussian part of $Z_{\rm Sp(N)}(\beta)$ coincides with the free higher spin wave function (\ref{psi0freehs}), $\psi_{0,\rm free}(\beta)$, provided we identify
\begin{align} \label{gammaistwooverN}
 \gamma = \frac{2}{N} \, .
\end{align}
This is essentially the expected relation between CFT central charge and the Newton constant,  $N \propto \ell_{\rm dS}^2/G_{\rm Newton}$.

The conjecture of \cite{Anninos:2011ui} can now be phrased more precisely as 
\begin{align} \label{psiHHSpN}
  \psi_{\rm HH}(\beta) \propto Z_{\rm Sp(N)}(\beta)\, .
\end{align} 
Here $\psi_{\rm HH}$ is the perturbative Hartle-Hawking wave function of parity-even minimal Vasiliev higher spin gravity in dS$_4$, expressed in terms of the boundary fields $\beta$, and $Z_{\rm Sp(N)}$ is the perturbative Sp(N) partition function (\ref{ZSpNexpansion}).
This is consistent with (\ref{psiHHinvZON}), since the analogously constructed bosonic O(N) model partition function  satisfies $Z_{\rm O(N)}=1/Z_{\rm Sp(N)}$. This relation holds because the model is free. Although (\ref{psiHHinvZON}) and (\ref{psiHHSpN}) are equivalent for the free model, they do become different for various interacting generalizations. The form (\ref{psiHHSpN}) has been argued to  agree with perturbative bulk computations for such interacting CFTs as well \cite{Anninos:2014hia,Chang:2013afa}. Roughly speaking this works because analytic continuation from AdS to dS entails a continuation of the curvature radius $\ell \to i\ell$, so $\frac{\ell^2}{G_N} \to - \frac{\ell^2}{G_N}$, i.e.\ $N \to -N$, which is effectively realized by replacing commuting scalars by anticommuting scalars. In this work, we will only consider the free case.

\section{Problems}\label{sec:missingHilbertspace}

\begin{figure} 
 \begin{center}
   \includegraphics[width=0.6\textwidth]{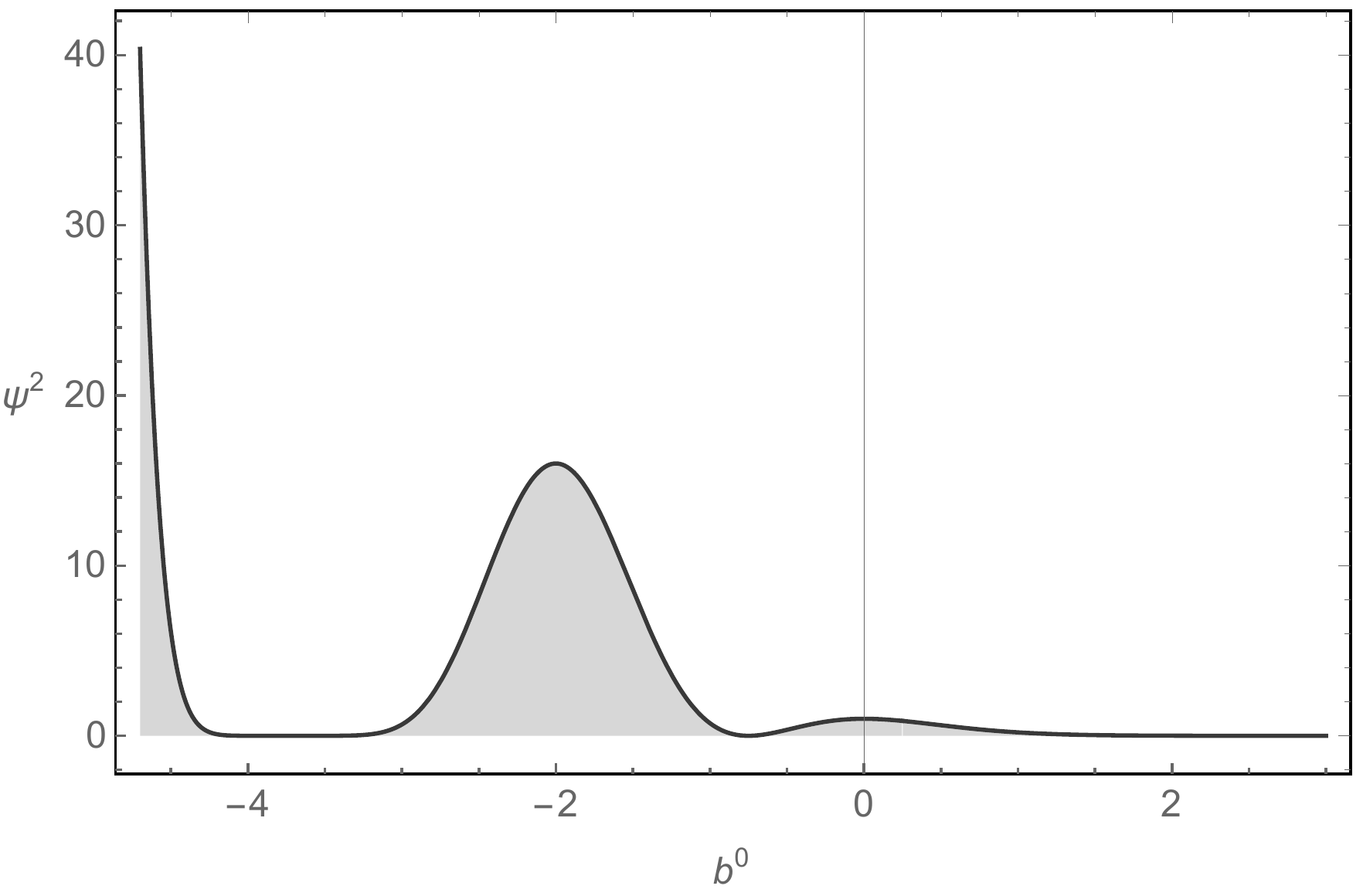}
 \end{center}
 \caption{Wave function squared $|\tilde\psi_{\rm HH}(b^0)|^2$ for $N=2$ according to naive interpretation of the Sp(N) model. Peaks of exponentially increasing heights appear on the negative $b^0$-axis, rendering the wave function naively non-normalizable. \label{fig:spnwave}}
\end{figure}

Pleasing as it is, the above concrete realization of the idea (\ref{psiHHZCFT}) is by no means a complete answer of what higher spin quantum gravity is in de Sitter space. It is at most half the answer:  
\begin{enumerate}
 \item The main problem is that just specifying a wave function without specifying the Hilbert space to which it belongs is physically meaningless. This problem is a practical one: without a Hilbert space inner product, it is impossible to compute probabilities, expectation values or cosmological observables such as vacuum correlation functions of field fluctuations. In AdS-CFT, the Hilbert space of AdS is identified with the Hilbert space of the Lorentzian version of the CFT. No such identification can be made for the dS Hilbert space to which $|\psi_{\rm HH}\rangle$ belongs. The Sp(N) model does not even have a sensible unitary Lorentzian version, as it violates the spin-statistics theorem, and even if it did, its states would live on two-dimensional spatial slices, whereas the wave function $\psi_{\rm HH}(\beta)$ lives on a three-dimensional slice. Related to this, time evolution in AdS corresponds to translations along the boundary, hence coincides with time evolution in the CFT, whereas is dS time evolution must emerge holographically. Another related issue is that in dS, the CFT sources $\beta$ are themselves dynamical, again in contrast to AdS. 
 \item To specify a Hilbert space, one needs to define an inner product. In principle this could be as straightforward as specifying a domain for $\CB$ and an integration measure $[d\CB]$ in $\langle \psi_1|\psi_2\rangle \equiv \int [d\CB] \, \psi_1(\CB)^* \, \psi_2(\CB)$. Usually when we quantize classical systems, the integration measure is determined by the symplectic structure on the classical phase space: picking canonical Darboux coordinates, the measure is flat. However in the case at hand, we aren't directly quantizing classical Vasiliev gravity, and moreover the structure of the phase space of Vasiliev gravity is unknown. 
 \item One could proceed naively and assume the sources $\beta^I$ to be half of a set of Darboux coordinates $(\beta^I,\alpha_I)$, as is the case in the free higher spin field theory discussed in section \ref{sec:freehsdS}. In the quantum theory these are then promoted to self-adjoint operators satisfying $[\beta^I,\alpha_J]=i\delta^I_J$. The appropriate measure $[d\CB]$ in this case is flat and the domain of the $\beta^I$ unrestricted. Unfortunately this immediately leads to a host of fatal non-normalizability problems when going beyond perturbatively small values of $\beta$ \cite{Anninos:2012ft,Banerjee:2013mca}. For example, $Z_{\rm Sp(N)}(b^0)$ for a constant scalar deformation on the 3-sphere diverges roughly as $e^{N|b^0|}$ for large negative $b^0$ \cite{Anninos:2012ft}; see fig.\ \ref{fig:spnwave}. One could entertain the possibility of restricting the domain of $\beta$, but this would be inconsistent with the assumed existence of a self-adjoint conjugate operator $\alpha$, since if $\alpha$ is self-adjoint, $U_b \equiv e^{i b^I \alpha_I}$ is a unitary operator generating arbitrary translations $\beta^I \to \beta^I + b^I$, including translations violating the assumed restrictions on the domain of $\beta$.
 \item An alternative approach to determining the appropriate integration measure $[d\CB]$ is to require it to be invariant under transformations of $\CB$ we wish to implement as unitary operators in the quantum theory. A natural set of transformations of $\CB$ are those induced by general linear field redefinitions $\chi_x \to {R_x}^y \chi_y$ of the fundamental fields $\chi_x$ in the partition function. Such reparametrizations effectively act on the total differential operator $\CH \equiv \CD + \CB$ in $Z(\CH) = \int d\chi \, e^{-\frac{1}{2} \chi \CH \chi}$ as $\CH^{xy} \to {R_{x'}}^x \CH^{x'y'} {R_{y'}}^y$, i.e.\ $\CH \to R^T \CH R$. For suitable $R$ these can be interpreted as bulk (higher spin) gauge transformations  acting on the boundary fields $\beta^I$. For example a diffeomorphism $\chi \to R \chi$ acts as a spatial diffeomorphism on late time slices in the bulk, while a Weyl rescaling acts as a change of time slicing. It is therefore more natural to take the measure $[d\CB]$ (or equivalently the measure $[d\CH]$ where $\CH=\CD+\CB$) to be invariant under such transformations rather than under translations as in naive canonical quantization. As we will discuss in more detail further on, for $K$-dimensional real symmetric matrices $\CH^{xy}$, the unique measure invariant under GL(K) transformations $\CH \to R^T \CH R$ is given by $[d\CH] = d\CH / (\det \CH)^{\frac{K+1}{2}}$, where $d\CH$ is the flat measure. In this case, the integration domain is naturally restricted to positive definite matrices $\CH$, since this domain is GL(K) invariant, and since we do not require translations of $\CH$ to be a symmetry of the theory. Evidently though, the proper $K=\infty$ continuum counterpart of this is tricky to define, and its existence  is far from clear. 
 \item There are other technical issues with the definition and interpretation of $\psi_{\rm HH}(\beta)=Z_{\rm Sp(N)}(\beta)$ in the continuum, once we go beyond perturbation theory, and want to make sense of it as a finite functional for general continuous sources $\beta^I$ of all spins. Indeed this requires adding infinitely many local counterterms to cancel off UV divergent contact terms. Even at the quadratic level in the sources, determining those is already a formidable task \cite{Bekaert:2010ky}, and simple dimensional analysis makes it clear that the problem becomes in particular intractable for the higher spin sources $\beta_{i_1 \cdots i_s}$, since these have negative dimension $\Delta_s = 2-s$, allowing in principle infinitely many local counterterms of arbitrarily high order. 
\end{enumerate}
We will take a different point of view here, and start instead with a description of a Hilbert space and a vacuum state $|\psi_0\rangle$, to be identified with the Hartle-Hawking state, which is manifestly well-defined, and then show that at least in a formal sense it implements the invariant measure suggested above, that it satisfies the required symmetry properties, and that it reproduces the predictions of the bulk $\psi_{\rm HH}(\beta)$ in perturbation theory in the large $N$ limit.

\section{Proposal} \label{sec:Proposal}

{In this section, guided by an explicit construction of the generating function of cosmological correlation functions to leading order in the large $N$ limit, we formulate our proposal for the fundamental degrees of freedom and Hilbert space of minimal Vasiliev de Sitter gravity, as well as  the exact Hartle-Hawking state $|\psi_0\rangle$ within this Hilbert space. We give some simple concrete examples to illustrate the idea. We proceed by showing that the Hartle-Hawking wave function as computed by the Sp(N) model is exactly recovered from our proposal upon a suitable change of variables, together with the appropriate natural measure in this description.}

\subsection{Large $N$ argument} \label{sec:largeNsaddle}

To get a crucial hint as to what the proper, well-defined description of the Hilbert space might be, consider again the wave function (\ref{psiHHSpN}):
\begin{align} \label{HHpsi}
 \psi_{\rm HH}(\CB) \, \propto \, \det(1+\CD^{-1} \CB)^{\frac{N}{2}} e^{-\frac{N}{2} \Tr(\CD^{-1} \CB)} \, , \qquad \CB=b^I \CD_I \, .
\end{align}
The generating function for correlation functions of the (shadow) boundary fields $\tilde b_I=G_{IJ} b^J$ in this state is formally
\begin{align}
 \langle e^{N a^I \tilde b_I} \rangle \equiv \langle \psi_{\rm HH}|e^{N a^I \tilde b_I}|\psi_{\rm HH}\rangle = \int [d\CB] \, \bigl|\psi_{\rm HH}(\CB) \bigr|^2 \, e^{N \Tr(\CD^{-1} \CA \CD^{-1} \CB)} \, , \qquad \CA \equiv a^I \CD_I \, ,
\end{align}
where we used (\ref{bIfromCB}), and we leave the measure unspecified at this point. Substituting (\ref{HHpsi}), this can be written as
\begin{align}
 \langle e^{N a^I \tilde b_I} \rangle \, \propto \, \int [d\CB] \, e^{N \Tr\left[ 
 \log(1+\CD^{-1}\CB) - (1-\CD^{-1}\CA) \CD^{-1} \CB \right]} \, .
\end{align}
To leading order in the $N \to \infty$ limit, we can evaluate this in saddle point approximation, without knowledge of the measure, just by extremizing the exponent with respect to variations of $\CB$. The saddle point equations are
\begin{align}
 (1+\CD^{-1}\CB)^{-1} = 1 - \CD^{-1} \CA \, .
\end{align}
Thus to leading order in the large $N$ limit, we find
\begin{align} \label{leadingorderlargeNapprox}
 \langle e^{N a^I \tilde b_I} \rangle \approx \det(1-\CD^{-1}\CA)^{-N} e^{-N\Tr(\CD^{-1}\CA)}\, .
\end{align}
We now recognize the right hand side of (\ref{leadingorderlargeNapprox}) as the generating function for correlation functions of bilinears in a {\it bosonic} O(2N) vector model:
\begin{align}
 \det(1-\CD^{-1}\CA)^{-N} e^{-N\Tr(\CD^{-1}\CA)} = \frac{1}{Z_0} \int dQ \, e^{-\int Q (\CD - \CA) Q} \, e^{-N\Tr(\CD^{-1} \CA)} \, , 
\end{align}
where the integration variables are $2N$ bosonic fields $Q_x^\alpha$, $\alpha=1,\ldots,2N$. This can now be interpreted as the generating function for correlation functions of normal-ordered $QQ$ bilinears in a very simple Gaussian vacuum state in a well-defined Hilbert space, as follows: 
\begin{enumerate}
 \item Define a Hilbert space $\CCH_0$ with standard (flat measure) inner product
\begin{align} \label{psiQinnerproduct}
 \langle \psi_1|\psi_2\rangle \equiv \int dQ \, \psi_1(Q)^* \, \psi_2(Q) \, .
\end{align}
 \item Define a vacuum wave function 
\begin{align} \label{psi0Qdefin}
 \psi_0(Q) \, \equiv c \, e^{-\frac{1}{2} \int  Q \CD Q} \, ,
\end{align}
with $c$ a normalization constant such that $\langle \psi_0|\psi_0\rangle = 1$.
\item  Define normal ordering of $QQ$ bilinears in the usual way by subtracting their vacuum expectation value:
\begin{align} \label{normalorderingQQ}
 :\! Q^\alpha_x Q^\beta_y \! : \,\, \equiv  \, Q^\alpha_x Q^\beta_y - \langle \psi_0|Q^\alpha_x Q^\beta_y |\psi_0 \rangle \, = \, Q^\alpha_x Q^\beta_y - \frac{1}{2} \delta^{\alpha\beta} (\CD^{-1})_{xy} \, .
\end{align}
\item Define single trace primary operators (analogous to (\ref{COIdef}))
\begin{align} \label{BIdefinnnn}
 B_I \, \equiv \, \frac{1}{N}\! :\! \Tr(Q \CD_I Q) \!: \, = \, \frac{1}{N} :\! Q_x^\alpha \CD_I^{xy} Q_y^\alpha \! : \, .
\end{align}
\item Then we may write (\ref{leadingorderlargeNapprox}) as
\begin{align} \label{treelevelcorrespondence}
 \langle \psi_{\rm HH}| e^{N a^I \tilde b_I} |\psi_{\rm HH} \rangle \, \approx \, \langle \psi_0|e^{N a^I  B_I} |\psi_0\rangle  \, \qquad (N \to \infty) \, ,
\end{align}
\end{enumerate}
Note that the large-$N$ limit considered above does not correspond to the free limit in the bulk, but to the (interacting) tree level approximation. Thus, in view of the results of \cite{Anninos:2011ui}, the left hand side of (\ref{treelevelcorrespondence}) should correspond to the generating function for vacuum correlation functions computed at tree level in the dS$_4$ higher spin Vasiliev theory. We can't really do any better than that on the left hand side, because we have not defined the measure, and because Vasiliev gravity at this point is only defined as a low-energy classical effective field theory, not as a complete perturbative quantum field theory. (There is no known action, no known phase space to quantize, and the theory breaks down as an effective theory at de Sitter curvature scale.)

The right hand side of (\ref{treelevelcorrespondence}) on the other hand is perfectly well-defined. The Hilbert space $\CCH_0$ has a positive definite inner product, manifest higher spin symmetry, and a spectrum of single-trace primary operators in one-to-one correspondence with the field content of the bulk theory. This suggests we should take the O(2N)-singlet sector $\CCH$ of the  Hilbert space $\CCH_0$ constructed above, modulo residual spacetime gauge symmetries, to be the Hilbert space of quantum gravity in Vasiliev de Sitter space, with the simple Gaussian $\psi_0(Q)$ identified as the Hartle-Hawking wave function of the universe. {The gauging of the spacetime symmetries and the resulting physical Hilber space $\CCH_{\rm phys}$ will be discussed in section \ref{sec:conclandHphys} and in more detail in section \ref{sec:GIaPHS}. For now we consider $\CCH$.}  Note that the single-trace primaries $B_I = \frac{1}{N} \!:\!Q_x \CD_I^{xy} Q_y\!:$ take the same higher spin current form as those in the usual O(N) model dual to Vasiliev-AdS, but their interpretation is quite different: in AdS, they correspond to the boundary fields $\alpha_I$ corresponding to normalizable modes in AdS, whereas here, they correspond to the shadows $\tilde \beta_I = G_{IJ} \beta^J$ of the boundary fields $\beta^I$, as defined more explicitly in  (\ref{hsshadowrel}). In AdS, these correspond to non-dynamical sources, but in dS, they these modes are dynamical, dominating in fact the late-time structure of the universe.

\subsection{Some simple illustrations} \label{sec:somesimpleexamples}

Before continuing to further refine this proposal and to provide more arguments in favor of it, let us pause here and give some simple examples to more concretely illustrate the perhaps somewhat abstract considerations given above. 

According to our prescription, the exact generating function for vacuum correlation functions of the boundary fields is
\begin{align} \label{gefufovacofu}
 \bigl\langle e^{\N a^I \tilde \beta_I} \bigr\rangle 
 \, = \, \det(1-\CD^{-1}\CA)^{-\N} e^{-\N \Tr(\CD^{-1}\CA)} \, = \, e^{+\N \left[ \frac{1}{2} G_{IJ} a^I a^J + \frac{1}{3} G_{IJK} a^I a^J a^K  \, + \,  \cdots \right]} \, ,
\end{align}
where $G_{I_1 \cdots I_n} \equiv \Tr\bigl(\CD^{-1} \CD_{(I_1} \cdots \CD^{-1} \CD_{I_n)} \bigr)$ as in (\ref{GI1Indef}). In particular the vacuum 2-point functions are
\begin{align}
 \langle \tilde\beta_I \tilde\beta_J \rangle = \frac{1}{N} \, G_{IJ} \, ,
\end{align}
which, with the appropriate choice of normalization of the $\tilde\beta_I = \frac{1}{N}\! :\! Q \CD_I Q \! :$ bilinears, exactly reproduce the  $\tilde\beta\tilde\beta$ 2-point functions (\ref{vacvactpf}) under the $\gamma=\frac{2}{N}$ identification (\ref{gammaistwooverN}):
\begin{align}
 \bigl\langle \tilde \beta_{i_1 \cdots i_s}(k) \,\tilde\beta_{i_1' \cdots i_s'}(k') \bigr\rangle = \frac{1}{N} \, k^{2s-1} \, \Pi_{i_1 \cdots i_s,i_1' \cdots i_s'}(k) \, \delta_{k+k'} \, ,
\end{align}
with the transverse traceless projectors defined in (\ref{projector}). Here we implemented again the renormalization prescription of defining correlation functions of spin $s>0$ bilinear operators in momentum space by analytic continuation, as explained under below (\ref{tptfcoicoj}).  Shadow transforming back to the local boundary fields $\beta^I = G^{IJ} \beta_J$ in transverse traceless gauge, this becomes (\ref{betabetatptf}):
\begin{align} 
 \bigl\langle \beta_{i_1 \cdots i_s}(k) \,\beta_{i_1' \cdots i_s'}(k') \bigr\rangle = \frac{1}{N} \, k^{-(2s-1)} \, \Pi_{i_1 \cdots i_s,i_1' \cdots i_s'}(k) \, \delta_{k+k'} \, .
\end{align}
A visualization of real space Gaussian random fields with a discretized version of these $\beta$-$\beta$ power spectra was given in fig.\ \ref{fig:randomfields}. The physical harmlessness of the growing IR fluctuations of higher spin fields $\beta$ is reflected in our framework by the fact that their backreaction on the other modes is small when $N$ is large, in the sense that higher point functions are suppressed by powers of $1/N$ and (renormalized) perturbation theory remains well-posed in momentum space.

Going one order beyond the Gaussian approximation, we can read off from (\ref{gefufovacofu}) that the vacuum 3-point functions are given by
\begin{align}
 \langle \tilde\beta_I \tilde\beta_J \tilde\beta_K \rangle = \frac{2}{N^2} \, G_{IJK} \, , 
\end{align}
where $G_{IJK} = \Tr(\CD^{-1} \CD_{(I} \CD^{-1} \CD_J \CD^{-1} \CD_{K)})$. 
Recalling that $\CD=-\partial^2$, so in momentum space $(\CD^{-1})^{pq} = \frac{1}{p^2} \delta_{p+q}$, and that in this normalization, the scalar $\tilde\beta(x) = \sqrt{8} :\!Q_x Q_x\!:$, so in momentum space $\tilde\beta(k)=\sqrt{8} \int_{p,q} \delta_{p+q-k} :\! Q_p Q_q \! :$, i.e.\ $\CD_k^{pq} = \sqrt{8} \, \delta_{p+q-k}$, we thus get
\begin{align}
 \bigl\langle \tilde\beta(k_1) \tilde\beta(k_2) \tilde\beta(k_3) \bigr\rangle &= \frac{2}{N^2} \, \sqrt{8}^3 \int_{pqr} \frac{1}{p^2} \delta_{p-q-k_1} \frac{1}{q^2} \delta_{q-r-k_2} \frac{1}{r^2} \delta_{r-p-k_3} \, \nn \\
 &= \frac{2 \sqrt{8}^3}{N^2} \int_{p} \frac{1}{p^2} \frac{1}{(p-k_1)^2} \frac{1}{(p+k_3)^2} \, \delta_{k_1+k_2+k_3} \nn \\
 &= \frac{2\sqrt{8}}{N^2} \frac{1}{k_1 k_2 k_3} \, \delta_{k_1+k_2+k_3} \, .
\end{align}
Shadow transforming $\tilde\beta(k)$ back to {${\beta}(k)=k \,\tilde{\beta}(k)$}, we obtain
\begin{align}
 \bigl\langle \beta(k_1) \beta(k_2) \beta(k_3) \bigr\rangle = \frac{2 \sqrt{8}}{N^2} \, \delta_{k_1+k_2+k_3} \, .
\end{align}
In position space this is a pure contact term $\propto \delta^3(x_1-x_2) \delta^3(x_2-x_3)$. This behavior is non-generic, and related to the well-known subtleties of scalar 3-point functions in 3+1-dimensional AdS-Vasiliev theories \cite{Sezgin:2003pt,Petkou:2003zz}. For other spins, the 3-point function is nonvanishing for separated points. We compute the graviton-scalar-scalar 3-point function and scalar 4-point function in section \ref{sec:cosmocor}. 

\subsection{Equivalence of descriptions for $K$ degrees of freedom}

In this section we consider toy models in which the boundary field $\CB^{xy}$ in $\psi_{\rm HH}(\CB)$ is replaced by a finite-dimensional $K \times K$ matrix, i.e.\ the continuous spatial indices $x,y$ of the original model are replaced by discrete indices $x,y = 1,\ldots,K$. This can be thought of as a lattice regularization of space with $K$ lattice points. We will show that in this case, the proposed description in terms of $\psi_0(Q)$ is {\it exactly} equivalent to the original description in terms of $\psi_{\rm HH}(\CB)$, for a well-defined and natural choice of measure $[d\CB]$, {\it provided} $K \leq 2N$. In section \ref{breakdown} we will see that this bound has significance in the continuum model as well: it is roughly speaking the maximal number of degrees of freedom that can be adequately described by a  local field theory in the bulk.

\subsubsection{$K=1$} \label{sec:toymodel}

We begin by considering the simplest case, $K=1$. In this case we can drop the $x,y$ indices altogether, and $\psi_0$ in (\ref{psi0Qdefin}) reduces to the ground state wave function of a $2\N$-dimensional isotropic harmonic oscillator with coordinates $q^\alpha$, $\alpha=1,\ldots,2\N$:
\begin{align} \label{psi0}
 \psi_0(q) = \langle q|\psi_0\rangle = 
 \sqrt{\tfrac{1}{\pi^N}} \,\, e^{- q^2/2} \, , \qquad q^2 = q^\alpha q^\alpha \, , 
\end{align}
where we have put $\CD \equiv 1$ for simplicity.
The inner product on the Hilbert space $\CCH_0$ is the usual $\langle \psi_1|\psi_2\rangle = \int d^{2\N}  q \, \psi_1(q)^* \psi_2(q)$, and $\langle \psi_0|\psi_0\rangle = 1$. 
The state $|\psi_0\rangle$ is O(2\N)-invariant. We take this O(2\N) symmetry to be gauged, which means we take the  Hilbert space $\CCH$ to be the subspace of O(2\N)-invariant wave functions $\psi(q)=\phi(q^2)$. A basis of the invariant Hilbert space is provided by the states
 $|n\rangle \, \propto \, (a^\dagger_\alpha a^\dagger_\alpha)^n |\psi_0\rangle$, where $a_\alpha^\dagger = \frac{1}{\sqrt{2}}(q^\alpha-\partial_{q^\alpha})$.

Alternatively, we may construct $\CCH$ by quantizing the  O(2\N)-invariant classical phase space. This phase space is two-dimensional. A natural pair of canonical coordinates is given by $u \equiv  \log \frac{q^2}{\N}$, $v \equiv \frac{1}{2} q \cdot p$.
The $\frac{1}{\N}$ normalization factor is chosen to guarantee a well-defined large-$\N$ limit, as $\langle \psi_0| \frac{q^2}{\N} |\psi_0\rangle=1$. The phase space coordinates $(u,v)$ are canonical (Darboux) because they have canonical Poisson brackets: $[u,v]_{\rm PB} = \partial_{q^\alpha} u  \, \partial_{p_\alpha} v - \partial_{p_\alpha} u  \,\partial_{q^\alpha} v = 1$. The corresponding canonically conjugate quantum operators take the same form, with the appropriate symmetrization of $v$ to ensure hermiticity:
\begin{align}
 \hat u = \log \frac{\hat q^2}{\N} \, , \qquad \hat v = \frac{1}{4}(\hat q \cdot \hat p + \hat p \cdot \hat q) \, , \qquad [\hat q^\alpha,\hat p_\beta] = i \delta^\alpha_\beta  \qquad \Rightarrow \qquad [\hat u,\hat v]=i \, .
\end{align}
Denote the basis of delta-function orthonormal eigenstates of $\hat u$ by $|u\rangle$, i.e.\
\begin{align} \label{uupdeltanorm}
 \hat u|u\rangle = u |u\rangle \, , \qquad \langle u|u'\rangle = \delta(u-u') \, , \qquad 1 = \int du \, |u\rangle \langle u| \, .
\end{align}
We can alternatively parametrize these canonical kets by 
\begin{align}
 h \equiv e^{u} = \frac{q^2}{\N} \, , \qquad |h\rangle \, \equiv \, \bigl|u\!=\!\log h \bigr\rangle \, .
\end{align}
In terms of this coordinate, 
(\ref{uupdeltanorm}) becomes
\begin{align} \label{hbasis}
 \hat h|h\rangle = h|h\rangle \, , \qquad \langle h|h'\rangle = \delta(\log h - \log h') = h \, \delta(h-h') \, , \qquad 1= \int \frac{dh}{h} \, |h\rangle \langle h| \, .
\end{align}
Note that $h>0$, so, unlike $\hat u=\log \hat h$, the operator $\hat h$ itself cannot have a well-defined hermitian canonical conjugate operator on the Hilbert space. 
Equivalently, unlike translations $u \to u + a$, translations $h \to h + a$ are not a symmetry of the Hilbert space. Instead, translations $u \to u+a$ map to scale transformations $h \to e^a h$, which do indeed preserve the half-line $h>0$. This explains the appearance of the scale-invariant measure $\frac{dh}{h}$ in the decomposition of unity in (\ref{hbasis}).

To express the ground state $|\psi_0\rangle$ in terms of the $|h\rangle$ basis, we cannot simply substitute the change of variables $q^2=\N h$ into (\ref{psi0}). Rather we have to compute
\begin{align} \label{psi0h}
 \tilde\psi_0(h) \equiv \langle h|\psi_0\rangle = \int d^{2N} q \, \langle h|q\rangle \langle q|\psi_0\rangle \, . 
\end{align}
The matrix element $\langle q|h\rangle$ is the wave function of $|h\rangle$ interpreted as a state on the original Hilbert space. Since $\hat q^2 |h\rangle = \N h|h\rangle$, it must take the form $\langle q|h\rangle = f(h) \, \delta(q^2 - \N h)$, where $f(h)$ is fixed by the normalization specified in (\ref{hbasis}). A short computation yields
\begin{align}
 \langle q|h\rangle = \sqrt{\tfrac{\Gamma(N+1)}{\pi^\N \N^{\N-1}}} \,\, h^{1-\frac{1}{2} \N} \, \delta(q^2-\N h) \,  , 
\end{align}
and 
\begin{align} \label{psi0inhbasis}
  \tilde\psi_0(h) = \langle h|\psi_0\rangle = \sqrt{\tfrac{\N^{\N}}{\Gamma(\N)}} \,\, h^{\N/2} \,  e^{-\N h/2} \,  .
\end{align}
For more general O(2\N)-invariant wave functions $\psi(q) = \phi(q^2/\N)$,
we get similarly
\begin{align} \label{wavetransf}
 \tilde\psi(h) = \sqrt{\tfrac{\N^{\N}}{\Gamma(\N)}} \,\, h^{\N/2} \,  \phi(h) \, . 
\end{align}
Expanding $h=1+b$ around its expectation value $\langle \psi_0|h|\psi_0\rangle = 1$, we may write 
\begin{align}
 \tilde\psi_0(b) \, \propto \, (1+b)^{\frac{N}{2}} \, e^{-\frac{N}{2} b} \, ,
\end{align}
which we recognize as the $K=1$ analog of the wave function (\ref{HHpsi}). It can be represented as an Sp(N)-invariant Grassmann integral in the obvious way. What the present analysis tells us is that this form of the wave function will give exactly the same results as the $q$-space form, provided we use the natural measure $[dh] = \frac{dh}{h}$, integrated over $h>0$.

As a check, let us compare the generating function $\langle e^{\N a\hat b} \rangle \equiv \langle \psi_0|e^{\N a\hat b}|\psi_0\rangle$ for moments of the fluctuation variable $\hat b$, 
computed using $\psi_0(q)$ versus using $\tilde\psi_0(h)$. First we use 
 $\psi_0(q)$:
\begin{align} \label{genfunctpsiq}
 \langle e^{\N a\hat b}\rangle = \int d^{2\N}q \, |\psi_0(q)|^2 e^{a(\hat q^2-\N)} =\pi^{-\N} \int d^{2\N} q \, e^{-(1-a) q^2} e^{-a \N} = (1-a)^{-\N} e^{-\N a} \, .
\end{align}
Next we use $\tilde\psi_0(h)$ to compute the same:
\begin{align} \label{genfunctpsih}
 \langle e^{\N a\hat b}\rangle = \int \frac{dh}{h} \, |\tilde\psi_0(h)|^2 \, e^{\N a(h-1)} = \tfrac{\N^{\N}}{\Gamma(\N)}  \int \frac{dh}{h} \, h^\N e^{-\N (1-a) h} \, e^{-\N a} = (1-a)^{-\N} e^{-\N a} \, ,
\end{align} 
in agreement with (\ref{genfunctpsiq}).

It should be kept in mind that O(2\N)-invariant operators $o$, for example the self-adjoint operators
\begin{align}
 h \equiv \frac{q^2}{N}, \qquad  v \equiv \frac{1}{2}(q \cdot p + p \cdot q) \, , \qquad w \equiv p^2 \, ,
\end{align}
acting on O(2\N)-invariant wave functions $\psi(q)=\langle q|\psi\rangle = \phi(q^2/N)$, 
map nontrivially to operators $\tilde o$ acting on wave functions $\tilde\psi(h) = \langle h|\psi\rangle \propto h^{N/2} \phi(h)$, in the following way: 
\begin{align}
 \tilde o = h^{N/2} o \, h^{-N/2} \, .
\end{align}
For the above examples this yields
\begin{align} \label{tildeVexpr}
 \tilde h = h \, , \qquad \tilde v = -2 i h \partial_h \, , \qquad 
  \tilde w = -\frac{4}{\N} \, h \, \partial_h^2  \, \, + \frac{\N-2}{h} \, .
\end{align}
Note that these operator are self-adjoint with respect to the measure $\frac{dh}{h}$. In particular $\tilde v$ generates scale transformations $h \to e^{2\lambda} h$.  
As a check one can verify that the operators $\tilde h$, $\tilde v$ and $\tilde w$ close under commutation, satisfying the same algebra as $h$, $v$, $w$.

\subsubsection{$K \leq 2\N$} \label{sec:equivforKlessN}

We will now do the same exercise starting from a Gaussian wave function depending on $2\N \times K$ coordinates $Q^\alpha_x$, $\alpha=1,\ldots,2\N$, $x=1,\ldots,K$,
\begin{align} \label{psi0Qbasis}
 \psi_0(Q) \,\propto\, e^{-\frac{1}{2} {\rm Tr}(Q^T \CD Q)} = e^{-\frac{1}{2} Q^\alpha_x \CD^{xy} Q^\alpha_y} \, ,
\end{align}
where $\CD$ is an arbitrary positive definite $K\times K$ matrix. (With a view on making contact with the continuum version (\ref{psi0Qdefin}), this can be thought of as a discretized version of minus the Laplacian, defined on a lattice with $K$ points.) The wave function lives in a Hilbert space $\CCH_0$ with standard inner product
$\langle \psi_1|\psi_2\rangle = \int dQ \, \psi_1(Q)^* \psi_2(Q)$.
Both the state $|\psi_0\rangle$ and the inner product are invariant under O(2\N) rotations acting on the $\alpha$-indices and under O(K) rotations acting on the $x$-indices. More specifically the latter are linear transformations $Q_x^\alpha \to {R_x}^y Q_y^\alpha$ leaving the ``Laplacian'' $\CD^{xy}$ invariant, i.e.\
\begin{align}
 R^T \CD R = \CD \, .
\end{align}

To construct a basis for the O(2N)-invariant Hilbert space $\CCH$, we consider
O(2\N)-invariant operators analogous to $h$ in section \ref{sec:toymodel}:
\begin{align} \label{Hxydefi}
 H_{xy} \equiv \frac{1}{\N} \, Q_x^\alpha Q_y^\alpha = \frac{1}{\N} \, (Q Q^T)_{xy} \, .
\end{align}
The matrix $H_{xy}$ is symmetric and non-negative. The normalization is chosen such that the expectation value of $H_{xy}$ in the state $|\psi_0\rangle$ is
\begin{align} \label{Hxyvev}
 \langle \hat H_{xy}\rangle = (\CD^{-1})_{xy} \, .
\end{align}
Examples of random samples in a $d=1$ model with $Q$ drawn from the probability distribution $P(Q)=|\psi(Q)|^2$ are shown in fig.\ \ref{fig:HBbdim1}. It is clear from the top row images  that a random $H_{xy}$ closely approximates the inverse Laplacian at large $N$, while at smaller $N$ there are larger fluctuations deviating from it.  

\begin{figure} 
 \begin{center}
   \includegraphics[width=\textwidth]{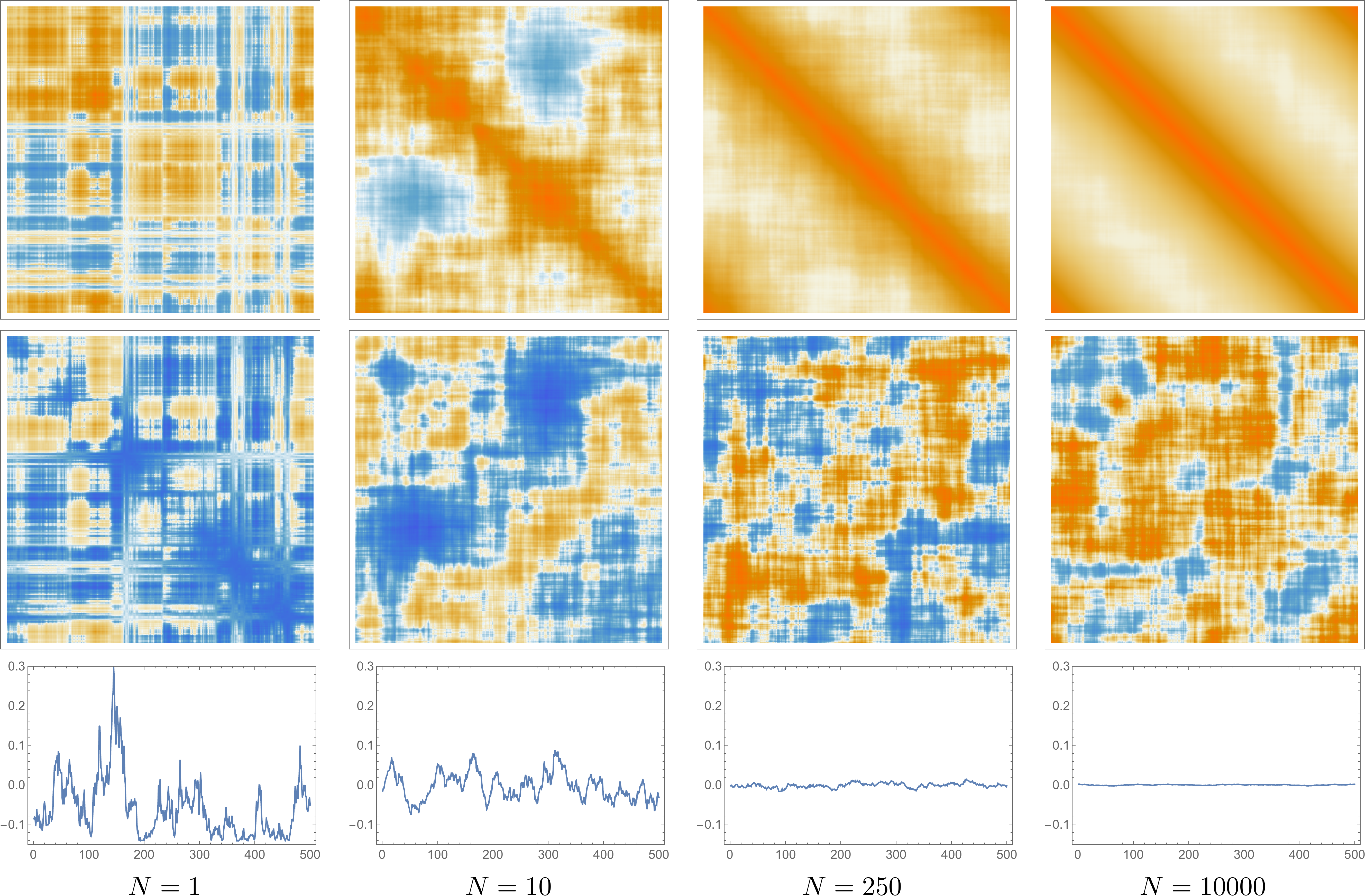}
 \end{center}
 \caption{Samples of, from top to bottom, $H_{xy}$, $B_{xy}=H_{xy}-\langle H_{xy}\rangle$, $\tilde\beta(x) = B_{xx}$ on 1d circle of size $L$ and lattice spacing $\frac{L}{K}$, $K=500$, with $\CD$ = discrete Laplacian plus small mass term $\sim \frac{1}{L}$. From left to right, $N=1,10,250,10000$. $H$ for $N=1,10$ has reduced rank $2N$. \label{fig:HBbdim1}}
\end{figure}

There are two qualitatively distinct cases to consider:
\begin{enumerate}
 \item $K \leq 2\N$: In this case, $H_{xy}$ generically has no zero eigenvalues and is hence generically strictly positive definite.
 \item $K>2\N$: In this case the $K \times K$ matrix $H_{xy}$ has reduced rank, equal to $2\N$ (or less), and (at least) $K-2\N$ zero eigenvalues.  
\end{enumerate}
In what follows we will assume we are in the first case, $K \leq 2\N$. A discussion of the case $K>2N$ and the continuum limit is postponed to section \ref{sec:Kgt2M}.

In analogy with (\ref{hbasis}) we define O(2\N)-invariant eigenkets of the hermitian operator $\hat H_{xy}$, satisfying 
\begin{align} \label{HHpnormaliz}
 \hat H_{xy} |H\rangle = H_{xy} |H\rangle \, , \qquad \langle H|H'\rangle = (\det H)^{\frac{K+1}{2}} \delta(H-H') \, .
\end{align}
The normalization of the eigenkets is chosen such that the corresponding decomposition of unity on the physical Hilbert space $\CCH$ has a scale invariant measure, as in (\ref{hbasis}):
\begin{align} \label{decompunityH}
 1 = \int [dH] \, |H\rangle \langle H| \, , \qquad [dH] \equiv \frac{dH}{(\det H)^{\frac{K+1}{2}}} \, .
\end{align}
Here the integral is over real positive symmetric matrices $H_{xy}$. To see that this is scale invariant, observe that real symmetric matrices have $\frac{K(K+1)}{2}$ independent components, so under a scale transformation $H \to \lambda H$, we have $dH \to \lambda^{\frac{K(K+1)}{2}} dH$, while at the same time $(\det H)^{\frac{K+1}{2}} \to \lambda^{\frac{K(K+1)}{2}} (\det H)^{\frac{K+1}{2}}$ as well, leaving the measure invariant. In fact this measure is invariant under a much larger symmetry group, namely arbitrary GL(K) transformations $R$ acting as
\begin{align}
 H \to R H R^T \, .
\end{align}
Note that this is a symmetry of the domain of $H$, i.e.\ the space of positive definite symmetric matrices, and (unlike, say, translations of $H$) a well-defined symmetry of the Hilbert space: it acts on $Q$ as $Q \to R Q$. In fact, the above measure is the unique one invariant under  this GL(K) symmetry.

More explicitly, the wave functions of these kets in the original Hilbert space $\CCH_0$ are of the form $\langle Q|H\rangle = f(H) \, \delta(Q Q^T-\N H)$, where $f(H)$ is fixed by the normalization condition (\ref{HHpnormaliz}), yielding 
\begin{align} \label{QHmatrixel}
 \langle Q|H\rangle \, \propto \, (\det H)^{\frac{K+1-\N}{2}} \, \delta(Q Q^T-\N H) \, .
\end{align} 
Thus the wave function of the state $|\psi_0\rangle$ expressed in the basis $|H\rangle$ is
\begin{align} \label{psi0Hdown}
 \tilde\psi_0(H) \equiv \langle H|\psi_0\rangle = \int dQ \, \langle H|Q\rangle \langle Q|\psi_0\rangle   = c \, (\det \CD H)^\frac{\N}{2} \, e^{-\frac{\N}{2} \, {\rm Tr} (\CD H) } \, ,
\end{align}
normalized such that $1 = \langle \psi_0|\psi_0\rangle = \int [dH] \, \langle \psi_0|H\rangle \langle H|\psi_0\rangle$,
that is to say
\begin{align}\label{ceq}
 \frac{1}{c^2}  
 = \int [dH] \, (\det H)^\N \, e^{-\N \, {\rm Tr} \, H} \, ,
\end{align}
where we used the GL(K)-invariance of the measure. Although we won't need it in this section, the normalization constant can be computed explicitly. We do this in appendix \ref{app:normalization}, the final result being (\ref{final}):
\begin{equation} \label{normalizationconstantKM}
 \frac{1}{c^2} =\frac{\pi^{\frac{K(K-1)}{4}}}{\N^{\N K}}\prod_{j=1}^K\Gamma \left( \N-\tfrac{K-j}{2} \right) \, .
\end{equation}
Note that the restriction $K \leq 2\N$ we made is necessary for this to be finite. 

Instead of $H_{xy}$, we can also consider the closely related matrices $\CH^{xy}$ obtained from $H_{xy}$ by raising indices using the ``metric'' $\CD^{xy}$:
\begin{align} \label{HCHrel}
 \CH^{xy} \equiv \CD^{xx'} H_{x'y'} \CD^{y'y} \, , \qquad {\rm i.e.} \quad \CH = \CD H \CD \, .
\end{align}
The ket $|\CH\rangle$ equals the ket $|H\rangle$ with $H=\CD^{-1} \CH \CD^{-1}$. 
The measure in the decomposition of unity (\ref{decompunityH}) remains unchanged by this transformation, i.e.\ $[d\CH] = [dH]$,
due to its GL(K)-invariance. In terms of $\CH$, the wave function (\ref{psi0Hdown}) becomes
\begin{align} \label{psi0Hup}
 \tilde\psi_0(\CH) =\langle \CH|\psi_0\rangle = c \, \det (\CD^{-1} \CH)^\frac{\N}{2} \, e^{-\frac{\N}{2}  {\rm Tr} (\CD^{-1} \CH) } \, ,
\end{align}
with the same normalization constant $c$ as in (\ref{psi0Hdown}). Expanding this in the fluctuation 
\begin{align} \label{fluctuationdef}
 \CB^{xy} = \CH^{xy} - \langle \CH^{xy} \rangle = \CH^{xy} - \CD^{xy} \, ,
\end{align}
we get
\begin{align}
 \tilde\psi_0(\CB) \, \propto \, \det \bigl(1+\CD^{-1} \CB \bigr)^\frac{\N}{2} \, e^{-\frac{\N}{2} {\rm Tr} (\CD^{-1} \CB) } \, ,
\end{align}
which we recognize as the finite-dimensional analog of the wave function (\ref{HHpsi}). Again this can be written as an Sp(N)-invariant partition function in the obvious way. What the present analysis tells us is that this form of the wave function will give exactly the same results as the $Q$-space form, provided we use the natural GL(K)-invariant measure $[d\CH]$ defined in (\ref{decompunityH}), integrated over $\CH>0$.

As a check, we can compare the generating functions for vacuum correlation functions of normal ordered $QQ$ bilinears $B_I$ defined as in (\ref{BIdefinnnn}) and (\ref{normalorderingQQ}), but now with the $\CD_I^{xy}$ a set of $K \times K$ matrices. This reproduces all the results of the saddle point analysis of section \ref{sec:largeNsaddle}, except that now the equality of generating functions is {\it exact}. In both descriptions it is given by (\ref{gefufovacofu}), for any $N \geq K/2$, not just in the $N \to \infty$ limit, provided we use the natural GL(K)-invariant measure $[d\CH]$ defined in (\ref{decompunityH}), integrated over $\CH>0$.

\subsubsection{$K>2\N$ and continuum limit} \label{sec:Kgt2M}

The case of actual interest has $K>2\N$ --- in the continuum limit, $K=\infty$. When $K > 2\N$, the $Q$-space description of the Hilbert space $\CCH$ remains nondegenerate, but the $H$-space description as we defined it becomes singular. This can be seen explicitly for example from (\ref{normalizationconstantKM}), which diverges for $K>2\N$. The reason for the breakdown at $K>2N$ is that the matrix $H_{xy} = \frac{1}{\N} \, Q_x^\alpha Q_y^\alpha$ has (at most) rank $2\N$, and (at least) $K-2\N$ zero eigenvalues. In particular this implies that the factors $\det H$ in (\ref{QHmatrixel}) vanish, while at the same time the delta-function overconstrains $Q$, rendering the kets $|H\rangle$ ill-defined.   

One can still proceed formally and make sense of e.g.\ vacuum correlation functions in the $H$-space description by defining various quantities at intermediate steps through analytic continuation. For example the normalization constant (\ref{normalizationconstantKM}) becomes finite when we analytically continue $\N \to \N + \epsilon$. Since the normalization constant drops out of the generating function of vacuum correlation functions (\ref{gefufovacofu}), we can at the end take $\epsilon \to 0$. Then we arrive again at the conclusion that the (analytically continued) generating function of correlation functions in the $H$-space description coincides with the one computed in the $Q$-space description, which remains given by (\ref{gefufovacofu}) irrespective of whether $K$ is smaller or larger than $2\N$.  

For non-perturbative questions the $H$-space description will likely be inadequate. The point of view we take here is that the $Q$-space description is the more fundamental one. Happily, it is also the simpler one.

\subsection{Conclusion and comments on $\CCH$ vs $\CCH_{\rm phys}$} \label{sec:conclandHphys}

{We have defined a Hilbert space $\CCH$ consisting of O(2N)-invariant wave functions $\psi(Q)$, with inner product $\langle \psi_1|\psi_2\rangle = \int dQ \, \psi_1(Q)^* \psi_2(Q)$ and vacuum state $\psi_0(Q) \, \propto \, e^{-\frac{1}{2} \Tr( Q \CD Q)}$. We argued that the vacuum correlation functions of $B_I = \frac{1}{N} \!:\! \Tr(Q\CD_I Q)\!:$ in this model coincide with the vacuum correlation functions of $\tilde b_I \equiv G_{IJ} b^J$ computed in the large-$N$ limit starting from the Sp(N)-model wave function $\tilde\psi_{\rm HH}(b) \, \propto \, \int d\chi \, e^{-\frac{1}{2} \Tr( \chi \CD \chi + b^I :\chi \CD_I \chi:)}$. Here $G_{IJ} = \Tr(\CD^{-1} \CD_I \CD^{-1} \CD_J)$ is proportional to the 2-point function of the bilinears in either model, so the relation between $B_I$ and $b^I$ can be thought of in CFT language as a shadow transform. We showed that in discretized models with a finite number $K \leq 2N$ of spatial points, there is an exact equivalence between the two descriptions, provided we choose the natural GL(K)-invariant measure on the Hilbert space to which $\tilde\psi_{\rm HH}(b)$ belongs. For $K>2N$ the $Q$-space description remains well-defined, while the description in terms of $\tilde\psi_{\rm HH}(b)$ becomes singular. However the equivalence persists in perturbation theory defined by analytic continuation. Thus we will use the $Q$-space description as the fundamental definition of the Hilbert space $\CCH$.}

{The Hilbert space $\CCH$ is not yet the {\it physical} Hilbert space $\CCH_{\rm phys}$ of higher spin de Sitter quantum gravity. To define  $\CCH_{\rm phys}$, we need to take into account additional  constraints related to bulk higher spin gauge invariance. We will discuss these gauge symmetries and the construction of $\CCH_{\rm phys}$ in section \ref{sec:GIaPHS}. We will also explain there that the choice of $\CD$ appearing in the definition of $\psi_0(Q)$ and the Sp(N) model can be thought of as a partial gauge fixing choice: for example, taking $\CD$ to be the flat Laplacian on $\IR^3$ corresponds to picking planar coordinates in de Sitter, while taking it to be the conformal Laplacian on $S^3$ corresponds to picking global coordinates. As we will see, the choice of $\CD$ does not fully fix the gauge: it leaves a residual gauge group $\CG$ of linear transformations $R:Q \to RQ$ satisfying $R^T \CD R = \CD$, which can be thought of as the global higher spin symmetry group. The physical Hilbert space $\CCH_{\rm phys}$ is then the $\CG$-invariant subspace of $\CCH$.}

{In the dual perturbative bulk QFT, there is in principle a parallel 2-step construction of the physical Hilbert space. One first constructs a Fock space $\CCH_{\rm Fock}$ of free higher spin fields in a de Sitter background along the lines of section \ref{sec:freehsfids}. The higher spin symmetry group $\CG$ (of which the dS isometry group SO(1,4) is a subgroup) is represented unitarily on $\CCH_{\rm Fock}$. Generalizing the arguments of \cite{Higuchi:1991tk,Higuchi:1991tm} from SO(1,4) to $\CG$, the physical Hilbert space $\CCH_{\rm Fock,\, phys}$ of the perturbative bulk QFT is then the $\CG$-invariant subspace of $\CCH_{\rm Fock}$. 
} 

{Thus our Hilbert space $\CCH$ should be viewed as a nonperturbative completion of $\CCH_{\rm Fock}$, while $\CCH_{\rm phys}$ should be viewed as a nonperturbative completion of $\CCH_{\rm Fock,\,phys}$. The following two sections will exclusively pertain to $\CCH$ rather than $\CCH_{\rm phys}$, and the results we obtain should therefore be compared to analogous computations on $\CCH_{\rm Fock}$ rather than $\CCH_{\rm Fock,\,phys}$.}

\section{Probabilities and correlation functions} \label{sec:probandcorr}

{In this section we give some examples of concrete computations of probabilities and correlation functions on the Hilbert space $\CCH$. The first one is the probability distribution of the constant scalar mode. The second one includes cosmological vacuum 3-point functions of scalars and gravitons, as well as the scalar 4-point function.}

\subsection{Probability distribution of constant scalar on $S^3$} 

{As mentioned in section \ref{sec:missingHilbertspace}, one of the most striking pathologies of the Sp(N) wave function $\tilde\psi_0(\CB)$ reviewed in section \ref{sec:SpNmodel}, interpreted naively as a wave function on a Hilbert space with a flat inner product measure $[d\CB]$, is its apparent non-normalizability, as pointed out in \cite{Anninos:2012ft}. This pathology is completely eliminated in our setup. In what follows, we will compute more specifically the probability distribution for the constant scalar mode on $S^3$, and find that in contrast to the results of \cite{Anninos:2012ft}, it is normalizable.} 

We will switch from planar to the global de Sitter gauge here, for which the operator $\CD$ appearing in the wave function $\psi_0(Q) \, \propto \, e^{-\frac{1}{2} \int Q \CD Q}$ becomes minus the conformal Laplacian on $S^3$. The flat Laplacian on $\IR^3$ is mapped to the conformal Laplacian on the round sphere of radius $L$ in stereographic coordinates by a Weyl transformation $Q_x \to \frac{\sqrt{2L}}{\sqrt{1+x^2}} Q_x$. A further spatial diffeomorphism $x(u)$ maps this to the conformal Laplacian in any desired coordinate system $u$ with metric $ds^2=h_{ij} du^i du^j$ on the round sphere of radius $L$. In more detail, under these field redefinitions, the planar dS wave function $\psi_0(Q) \, \propto \, e^{\frac{1}{2} \int d^3 x \, Q \partial^2 Q} = e^{-\frac{1}{2} \int d^3 x \, \partial_i Q \partial_i Q}$ gets mapped to the global dS wave function $\psi_0(Q) \, \propto \, e^{-\frac{1}{2} \int d^3 u \, \sqrt{h}  \left( h^{ij} \partial_i Q \partial_j Q + \frac{1}{8} {\cal R}(h) Q^2 \right)} = e^{-\frac{1}{2} \int d^3 u \, Q \, \CD \, Q}$ where $\CD=\partial_i \sqrt{h} h^{ij} \partial_j + \frac{1}{8} \sqrt{h} {\cal R}(h)$, and ${\cal R}(h) =  \frac{6}{L^2}$ is the Ricci scalar.

\subsubsection{Sp(N) model} \label{sec:constmodeHspace}

In \cite{Anninos:2012ft} it was pointed out that the Sp(N) wave function $\tilde\psi_0(\CB)$ on $S^3$ diverges exponentially at large negative constant scalar deformation $\CB^{xy} = b^0 \delta^{xy}$. This is easy to see. The spectrum of the minus the conformal Laplacian $\CD$ on the round sphere with unit radius is $\lambda_\ell = \ell(\ell+2) + \frac{3}{4}$ with $\ell=0,1,2,\dots$ and degeneracy $d_\ell = (\ell+1)^2$. Equivalently, putting $\ell=k-1$, we have $\lambda_k = k^2 - \frac{1}{4}$ with $k=1,2,\ldots$ and degeneracy $d_k = k^2$. Deforming the conformal Laplacian by a constant $b^0$ just shifts this spectrum to $\lambda_k = k^2 - \frac{1}{4} + b^0$, so the corresponding wave function becomes
\begin{align} \label{tildepsi0b}
 \tilde\psi_0(b^0) \, \propto \, \det(1+\CD^{-1} b^0)^\frac{\N}{2} e^{-\frac{\N}{2}\Tr(\CD^{-1} b^0)} = e^{-\frac{\N}{2} F(b^0)} \, , 
\end{align}
where
\begin{align}
 F(b^0) = -\sum_k k^2 \bigl[ \log\bigl(1+\tfrac{b^0}{k^2-\frac{1}{4}} \bigr) - \tfrac{b^0}{k^2-\frac{1}{4}} \bigr] \, .
\end{align}
This can be evaluated by first computing $\partial_b F(b)$, which sums to an elementary function, and then integrating this back up to $F(b^0)$ with integration constant fixed by $F(0)=0$. The result is
\begin{align} \label{Fbsol}
 F(b^0) = \frac{\pi}{2} \int_0^{b^0} db \sqrt{b-1/4} \, \coth\bigl( \pi \sqrt{b-1/4} \bigr) \, .
\end{align}
The resulting $\tilde\psi_0(b^0) \, \propto \, e^{-\frac{\N}{2} F(b^0)}$ falls off as $\psi_0(b^0) \sim e^{-\frac{\N}{2} \frac{\pi}{3} b_0^{3/2}}$ at large positive $b^0$ but diverges exponentially (in an oscillatory way) at large negative $b^0$, rendering the wave function apparently non-normalizable {(see fig.\ \ref{fig:spnwave})}.

However in our current setup, this problem is completely eliminated because the domain of $\CH = \CD + \CB$ consists of positive definite operators $\CH$ only. This means $b^0$ is effectively restricted to 
\begin{align} \label{bbound}
 b^0 > -\frac{3}{4} \, .
\end{align} 
The value $b^0=-\frac{3}{4}$ corresponds to the first zero of $\tilde\psi_0(b^0)$. The resulting wave function on this proper domain looks like fig.\ \ref{fig:psi0b}. 

\begin{figure} 
 \begin{center}
   \includegraphics[width=\textwidth]{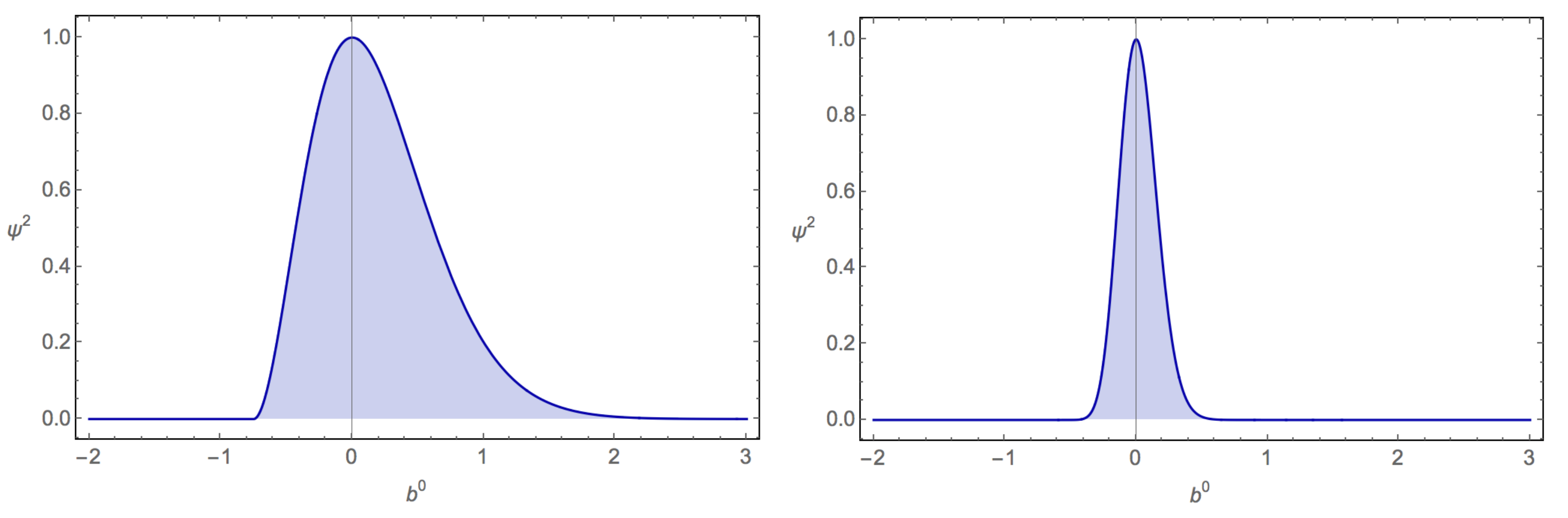}
 \end{center}
 \caption{Wave function squared $|\tilde\psi_0(b^0)|^2$ for $\N=2$ (left) and $\N=20$ (right). The positivity constraint on $\CH=\CD+\CB$ restricts $b^0>-\frac{3}{4}$, unlike in the naive version shown in fig.\ \ref{fig:spnwave}.  \label{fig:psi0b}}
\end{figure}

It should be kept in mind that this is just the wave function for a single deformation, with all other deformations kept zero. As such, $|\tilde\psi_0(b^0)|^2$ does {\it not} have an immediate interpretation as the probability distribution for measuring $b^0$, tracing over everything else. To obtain this probability distribution, we would have to integrate over all possible other deformations as well. At first sight this seems like an impossible task, since this involves integrating over an infinite number of higher spin degrees of freedom, all coupled to each other. Remarkably, in the $Q$-model description, this actually becomes entirely straightforward. We turn to this next. 

\subsubsection{Q-model} \label{sec:constscalarQspace}

Let us work in coordinates $\omega=(\omega^1,\omega^2,\omega^3,\omega^4)$, $\omega \cdot \omega = 1$ on the unit 3-sphere, for which the metric is $ds^2 = d\omega^2$.
The operator $\CD$ is the conformal Laplacian and its inverse is the free $\Delta=\frac{1}{2}$ scalar 2-point function on the sphere:
\begin{align}
 (\CD^{-1})_{\omega\omega'} = \frac{1}{4\pi} \frac{1}{\sqrt{2}(1-\omega \cdot \omega')^{1/2}} \, .
\end{align}
(The normalization can be checked by taking the small separation limit, i.e.\ considering $\omega \cdot \omega' = \cos\theta$ in the limit $\theta \to 0$, and comparing to the flat space 2-point function (\ref{inverseCD}).) In spherical harmonic space, i.e.\ expanding the $Q$-fields in orthonormal $S^3$ spherical harmonics $Y_k^{mm'}(\omega)$, $\CD$ becomes diagonal with eigenvalues $k^2-1/4$ on the diagonal, so the 2-point function $\CD^{-1}$ is likewise diagonal, with $\frac{1}{k^2-1/4}$ on the diagonal. The constant mode corresponds to $k=1$.  


Recall that the relation between Sp(N)-model sources $b^I$ in the expansion $\CB = b^I \CD_I$ on the one hand and $QQ$-bilinears $B_I=Q_\omega \CD_I^{\omega \omega'} Q_{\omega'}$ on the other hand is given in general by the shadow transform $B_I = G_{IJ} b^J$, where $G_{IJ} = \Tr(\CD^{-1} \CD_I \CD^{-1} \CD_J)$ is the 2-point function. Turning on a constant scalar source $b^0$ on $S^3$ means taking $\CB = b^0 \CD_0$, where $\CD_0^{\omega \omega'} =  \delta^{\omega\omega'}$. Since $G_{IJ}$ does not mix different spins and is diagonal in angular momentum space, we get the simple relation $B_0 = G_{00} b^0$, where  
\begin{align}
 B_0 = \frac{1}{\N} :\! Q_\omega \CD_0^{\omega\omega'} Q_{\omega'} \! : = \frac{1}{\N} \int d\omega \, :\! Q_\omega Q_{\omega} \! : \, ,
\end{align}
and $G_{00}$ is the $\Delta=1$ scalar 2-point function in the zero angular momentum sector on $S^3$.  Explicitly we can compute $G_{00}$ directly from its definition either in position space,
$G_{00} = \Tr(\CD^{-1} \CD_0 \CD^{-1} \CD_0) = \frac{1}{32 \, \pi^2} \int d\omega \, d\omega' \, \frac{1}{1-\omega \cdot \omega'} =  \frac{\pi^2}{4}$, or in angular momentum space, $G_{00} = \sum_k k^2 \, \frac{1}{(k^2-1/4)^2} = \frac{\pi^2}{4}$.
We conclude that
\begin{align} \label{B0CB0rel}
 B_0 = \frac{\pi^2}{4} \, b^0 \, .
\end{align}
The probability distribution for the bilinear $B_0$ in the $Q$-model is computed by
\begin{align} \label{Probb0}
 P(B_0) = \langle \psi_0|\delta(\hat B_0 - B_0)|\psi_0 \rangle = \N \int \frac{d\lambda}{2\pi} \, \bigl\langle e^{i \, \N \lambda \left(B_0 - \frac{1}{\N} \int d\omega \, :Q_\omega Q_\omega: \right) } \bigr\rangle \, .
\end{align}
We may evaluate
\begin{align}
 \bigl\langle e^{-i \lambda \int d\omega \, :Q_\omega Q_\omega: } \bigr\rangle = \det\bigl(1+\CD^{-1} i \lambda\bigr)^{-\N} e^{\N \lambda \Tr(\CD^{-1} i \lambda)} \, .
\end{align}
Notice that this is the same expression as (\ref{tildepsi0b}), except for the replacements $ \N \to -2 \N$, $b \to i \lambda$.
Therefore we can just copy (\ref{Fbsol}) with the appropriate substitutions to obtain
\begin{align}
 \bigl\langle e^{-i \lambda \N \hat B_0} \bigr\rangle =  \bigl\langle e^{-i \lambda \int d\omega \, :Q_\omega Q_\omega: } \bigr\rangle = 
 e^{\N F(i\lambda)} \, ,
\end{align} 
where
\begin{align}
 F(i\lambda) = \frac{\pi}{2} \int_0^{i\lambda} dz \, \sqrt{z-1/4} \, \coth\bigl(\pi \sqrt{z-1/4}\bigr) \, ,
\end{align}
where the $z$ contour runs along the imaginary axis.
\begin{figure} 
 \begin{center}
   \includegraphics[width=\textwidth]{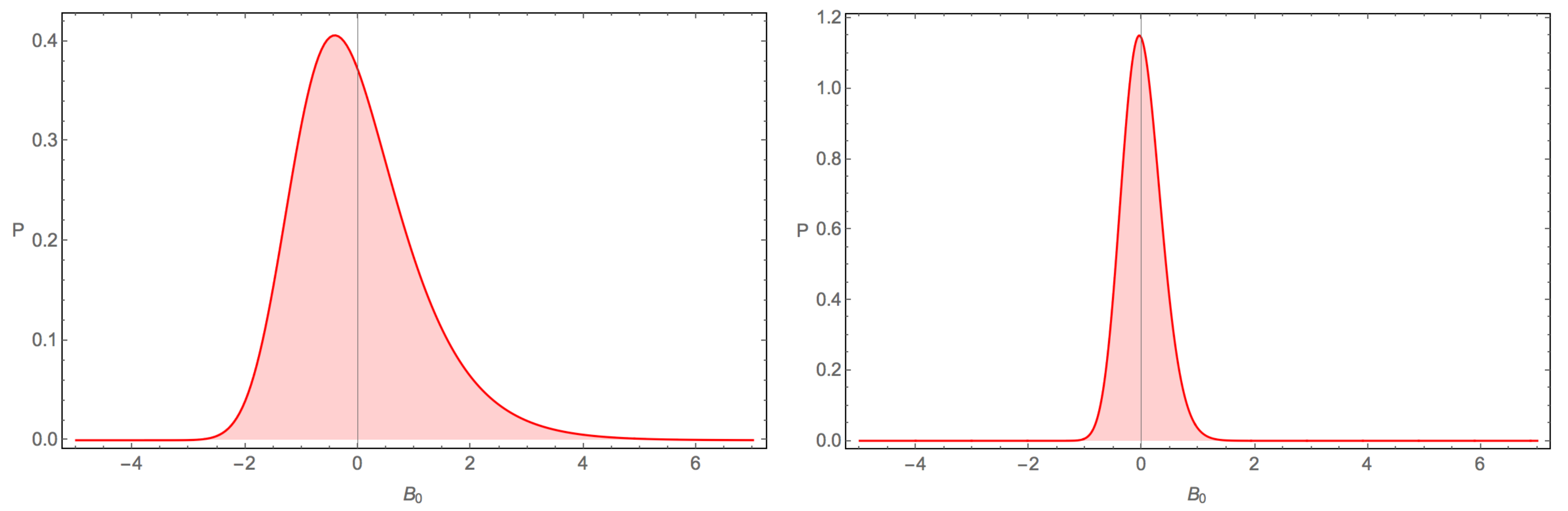}
 \end{center}
 \caption{Probability distribution $P(B_0)$ for $\N=2$ (left) and $\N=20$ (right). (For comparison to fig.\ \ref{fig:psi0b}, recall that $B_0 = \frac{\pi^2}{4} b^0 \approx 2.5 \, b^0$.) \label{fig:P0b}}
\end{figure}
Returning to (\ref{Probb0}), we thus get
\begin{align} \label{Pb0integral}
 P(B_0) = \frac{\N}{2\pi} \int d\lambda \, e^{\N\left(i \lambda B_0 + F(i\lambda)\right)} \, .
\end{align}
Asymptotically for $\lambda \to \pm \infty$, we have $F(i\lambda) \approx -\frac{\pi}{3 \sqrt{2}}(1 \mp i) |\lambda|^{3/2}$ so this integral converges. 
We can integrate (\ref{Pb0integral}) numerically for any value of $\N$. The results for $\N=2$ and $\N=20$ are shown in figure (\ref{fig:P0b}). We checked the numerical result by verifying the distribution integrates to 1, and that $\langle B_0 \rangle = 0$. 

At large $\N$, we can evaluate the integral in saddle point approximation. The saddle point equation is
\begin{align}
 B_0 = - F'(u) = -\frac{\pi}{2} \sqrt{u-1/4} \, \coth\bigl(\pi \sqrt{u-1/4}\bigr) \, , \qquad u \equiv i \lambda \, .
\end{align}
This is useful in particular to get the large-$\N$ probability distribution in various limits:
\begin{align}
 B_0 \to -\infty: P(B_0) &\sim e^{-\N  \frac{20}{3 \pi^2} \, |B_0|^3} \nn \\
 B_0 \to 0:   P(B_0) &\sim e^{-\N  \frac{2}{\pi^2} \, B_0^2} \nn \\
 B_0 \to +\infty:    P(B_0) &\sim e^{-\N  \frac{3}{4} \, B_0} \, .
\end{align}
After making the change of variables (\ref{B0CB0rel}), this coincides for {\it small} values of $B_0$ with the wave function squared $|\tilde\psi_0(b)|^2$ discussed in section \ref{sec:constmodeHspace}. For large and small values of $B_0$ on the other hand the behavior is quite different. This is to be expected since at small $B_0$ and large $\N$, the leading Gaussian term dominates in the wave function, whereas at large $|B_0|$, the nonlinear couplings to other fields become important. 

Note in particular that the probability $P(B_0)$ is nonzero for arbitrarily negative values of $B_0$. At first this might seem to contradict the positivity bound (\ref{bbound}). However this bound applies only to the case in which all other modes of the scalar and higher spin fields are set to zero. When the other fields are allowed to fluctuate, as is the case in the current computation since we are tracing out all other modes, the constant scalar mode may go arbitrarily far below this, as long as other fields ensure the total $\CH=\CD+b^0$ remains positive. As a result, although the probability becomes highly suppressed for large negative $B_0$, it does not become zero. 

Finally note that the plots of $B_{xx}$ in figure \ref{fig:HBbdim1} for the $d=1$ case are qualitatively consistent with the above results: the scalar is more often negative than positive, but its positive fluctuations tend to be larger, with this being most pronounced for small $N$. 

\subsection{Probabilities of general field profiles} \label{sec:PBgenprofile}

The above computation is readily generalized to general scalar field profiles. For example the probability distribution for a general shadow scalar profile $\tilde\beta(x)$ on $\IR^3$ is given by
\begin{align}
 P(\tilde\beta) &= \bigl\langle  \delta( B - \tilde\beta ) \bigr\rangle \, \propto \, \int d\lambda \,  \bigl\langle e^{i \N \int d^3 x  \, \lambda(x) \left( \tilde\beta(x) - \frac{c_0}{\N}  :Q_x Q_x:  \right) } \bigr\rangle \nn \\
 &= \int d\lambda  \, e^{i \N \int  \lambda \, \tilde\beta} \, \det\bigl(1 + i c_0 \CD^{-1}   \lambda \bigr)^{-\N} e^{i c_0 \N  \Tr(\CD^{-1} \lambda)} \, ,
\end{align}
where the integral over $\lambda$ is now a functional integral, and we recall $c_0 = \sqrt{8}$ if we want $G_{IJ}$ to coincide with (\ref{Gi1isip1sips}) and $\tilde\beta$ to be normalized like the scalar boundary field introduced in (\ref{phialphatildebeta}) or (\ref{alphabetamodes}). (Recall from (\ref{gammaistwooverN}) that the bulk coupling constant $\gamma=2/N$ in these conventions.) The functional determinant can in principle be evaluated by a variety of  techniques, including numerical methods, as in \cite{Anninos:2013rza}.
The large $N$ saddle point equation for $u(x) \equiv i \lambda(x)$ is
\begin{align}
 \bigl((\CD + c_0 u)^{-1} - \CD^{-1}\bigr)_{xx}  = \frac{1}{c_0} \tilde\beta(x) \, .
\end{align}
This can be solved numerically or analytically in suitable limits by standard methods, e.g.\ in long wavelength, large field regimes, or perturbatively at small $B$. As a check, note that to lowest order in small $B$ perturbation theory, this equation becomes $\int d^3 y \, G(x,y) \, u(y) = \frac{1}{c_0^2} \tilde\beta(x) = \frac{1}{8} \tilde\beta(x)$, where $G(x,y) = [(\CD^{-1})_{xy}]^2 = \frac{1}{(4\pi)^2}\frac{1}{|x-y|^2}$. In momentum space this becomes $u(k) = k \tilde\beta(k) = \beta(k)$, that is to say the saddle point value of $u(x)$ equals the local boundary field $\beta(x)$ of the bulk theory to this order, and the probability density for $\beta$ becomes $P(\beta) \propto e^{-N \int_k \frac{1}{k} \, \beta(k) \, \beta(-k)}$ to this order, as expected from the free bulk theory. Non-Gaussian corrections to this are obtained by going to higher orders in perturbation theory. In principle we can evaluate non-perturbatively as well, by evaluating the functional determinant non-perturbatively, as we did for the constant scalar mode on $S^3$ in the previous section. 

Similar considerations can be made for spin $s>0$, provided we interpret the resulting functional determinant in a renormalized sense by analytic continuation (e.g.\ zeta-function renormalization), similar to how we defined the renormalized $n$-point functions $G_{I_1 \cdots I_k}$ in momentum space, as explained below (\ref{tptfcoicoj}). Because of this, the precise physical interpretation of the ``renormalized probabilities'' thus computed is less clear. The safest interpretation of $P(\tilde\beta)$ in this case is as objects reproducing the correct renormalized $n$-point functions in perturbation theory for the fields under consideration (with the density $P(\tilde\beta)$ defined with respect to a flat measure $d\tilde\beta$).  

Of course it is more efficient to compute such $n$-point functions directly in the $Q$ model rather than to first compute the full $P(\beta)$. We turn to this next.

\subsection{Cosmological correlation functions} \label{sec:cosmocor}


{In this section we compute the vacuum scalar-scalar-graviton 3-point function as well as the scalar 4-point function. Although we won't consider spins higher than two, let us begin by briefly outlining how in principle these computations can be systematically generalized to arbitrary spins. To do so, we need an efficient way of generating higher spin currents. Here we will consider the generating function used in \cite{Sleight:2016dba}.} Given $2N$ real fields $Q^\alpha_x$, one employs the equivalence between traceless symmetric tensors and functions of a complex null vector $z$ \cite{Costa:2011mg} to construct
\begin{equation}
B^s(x|z) \equiv B^s_{i_1 \ldots i_s}(x) z^{i_1} \ldots z^{i_s} \propto \sum_{k=0}^s a^{(s)}_k(z\!\cdot\!\partial)^k Q^\alpha_x(z \cdot \partial)^{s-k}Q^\alpha_x,
\end{equation}
where the coefficients $a_k$ are given in terms of Gegenbauer polynomials $C^{(a)}_b(u)$ as
\begin{equation}
\sum_{k=0}^sa_k^{(s)}x^ky^{s-k}=(x+y)^sC^{(d-3)/2}_s\left(\frac{x-y}{x+y}\right)~.
\end{equation}
The definition is somewhat degenerate at $d=3$, and we should rescale the currents by some function that is singular at $d=3$. In order to get the correct normalization for the two-point functions, we choose
\begin{equation}
f_s(x,y)=\lim_{d\to 3} \frac{2^{(5-s)/2}s!}{N (d-3)_s}(x+y)^sC^{(d-3)/2}_s\left(\frac{x-y}{x+y}\right)= \frac{2^{(5-s)/2}}{N} (x+y)^sT_s\left(\frac{x-y}{x+y}\right)
\end{equation}
where $T_s(u)$ is Chebyshev polynomial of order $s$, and $(M)_n$ denotes the Pochhammer symbol.
With this choice, the spin-two $B_I$ operator is
\begin{equation}
B_2(x|z)=\frac{2\sqrt{2}}{N}:\!\left[(z\!\cdot\!\partial)^2Q^{\alpha}_xQ^{\alpha}_x+Q^{\alpha}_x(z\!\cdot\!\partial)^2Q^{\alpha}_x-6(z\cdot\partial)Q^{\alpha}_x(z\!\cdot\!\partial)Q^{\alpha}_x\right]\!:
\end{equation}
We separately choose the normalization of the scalar operator $B_0$ as follows
\begin{align}
B_0(x)=\frac{2\sqrt{2}}{N}:\!Q^\alpha_x Q^\alpha_x\!:
\end{align}
The two point function of higher spin fields $B_I$ with these normalizations are then \cite{Sleight:2016dba}
\begin{equation}
    \langle B_s(x_1|z_1)B_{s'}(x_2|z_2)\rangle=\frac{(2s)!}{\pi^2N}\frac{(z_1\!\cdot\!H(x_{12})\cdot\! z_2)^s}{(x_{12}^2)^{1+s}} \delta^{ss'},\quad s\ge 1
\end{equation}
where $H_{ij}(x) = \delta_{ij} - 2 x_i x_j / x^2$. For scalar and spin two operators, this gives two-point functions consistent with the normalization of section \ref{sec:freehsdS},
\begin{equation}
    \langle B_0(x_1)B_0(x_2)\rangle=\frac{1}{N}\frac{1}{2\pi^2 x_{12}^2},\quad   \langle B_2(x_1|z_1)B_{2}(x_2|z_2)\rangle=\frac{1}{N}\frac{4!(z_1\!\cdot\! H(x_{12})\!\cdot\! z_2)^2}{\pi^2x_{12}^6}~.
\end{equation}


\subsubsection{Three-point functions in momentum space}

Three-point functions of higher spin conserved currents in three dimensions are extensively discussed in \cite{Giombi:2011rz}. Conformal symmetries fix much, but not all, of their structure. In appendix \ref{shadowapp} it is shown how the shadow transform simplifies in momentum space. Given the relevance of shadow transforms to our discussion, we are particularly interested in momentum space expressions for cosmological correlators. Conformally invariant correlations in momentum space have been considered very generally in \cite{Bzowski:2013sza}. In dS-CFT they are related to non-Gaussian features of the wavefunction (see for instance \cite{Maldacena:2002vr,Falk:1992sf,Maldacena:2011nz,McFadden:2010vh,McFadden:2011kk,Mata:2012bx,Anninos:2014lwa,Arkani-Hamed:2015bza,Arkani-Hamed:2017fdk} for  calculations of non-Gaussianities in momentum space). 

As a simple example, we consider the three point function of two scalars and a graviton. The relevant Fourier transforms are
\begin{eqnarray}
B_0(p) &=& \frac{2\sqrt{2}}{N}\int_q\,:\!Q^\alpha_{q}Q^{\alpha}_{p-q}\!: \\
B_2(p) &=& -\frac{2\sqrt{2}}{N}\int_q\, :\!Q^\alpha_{q}Q^{\alpha}_{p-q}\!:[(z\!\cdot\!q)^2+(z\!\cdot\! (p-q))^2+6 (z\!\cdot\! q) z\!\cdot\!(q-p)]~.
\end{eqnarray}
The three point function is
\begin{align}
    \langle B_{0}(p_1)B_{0}(p_2)B_{2}(p_3|z)\rangle&=-\frac{16\sqrt{2}}{N^3}\int_{k_1}\int_{k_2}\int_{k_3}\,\bigl\langle \, :\!Q_{k_1}Q_{p_1-k_1}\!: \, :\!Q_{k_2}Q_{p_2-k_2}\!: \, :\!Q_{k_3}Q_{p_3-k_3}\!: \, \bigr\rangle\nn \\
    &\times [(z\cdot k_3)^2+(z\cdot(p_3-k_3))^2+6(z\cdot k_3)z\cdot(k_3-p_3)]
\end{align}
There are eight types of contractions among the three groups of normal ordered $Q$ bilinears. They all have the same contribution, which is
\begin{align}
    \langle B_{0}(p_1)B_{0}(p_2)B_{2}(p_3|z)\rangle&=-\frac{{32}\sqrt{2}}{N^2}\int_{k_1}\int_{k_2}\int_{k_3}\, k^{-2}_1 k^{-2}_2 k^{-2}_3 \, \delta_{p_1-k_1+k_2}\delta_{p_2-k_2+k_3}\delta_{p_3-k_3+k_1} \nn \\
    &\times [(z\!\cdot\!k_3)^2+(z\!\cdot\!(p_3-k_3))^2+6(z\!\cdot\! k_3)z\!\cdot\!(k_3-p_3)]
\end{align}    
The integrals can be performed following the methods of \cite{Bzowski:2013sza}. One finds
\begin{align}
I_1(p_1,p_2,p_3;z)&=\int_k\!\frac{(z\!\cdot\!k)^2}{k^2(k-p_2)^2(k+p_3)^2}+p_2\leftrightarrow p_1 \nn \\
&=\frac{(p_1\!\cdot\! z)^2(p_1+p_3)+(p_2\cdot z)^2(p_2+p_3)}{2^3(p_1+p_2+p_3)^2p_1p_2}+\frac{(p_3\!\cdot\! z)^2(p_1p_2+p_1p_3+p_3p_2+p^2_1+p^2_2)}{2^3(p_1+p_2+p_3)^2p_1p_2p_3} \nn
\end{align}
and
\begin{align}
I_2(p_1,p_2,p_3;z)&=\int_k \, \frac{6(z\!\cdot \!k)(z\!\cdot\! k+z\!\cdot \!p_3)}{k^2(k-p_2)^2(k+p_3)^2}\nn\\
&=3 \, \frac{(p_1\!\cdot\! z)^2(p_1+p_3)+(p_2\!\cdot \!z)^2(p_2+p_3)}{2^3(p_1+p_2+p_3)^2p_1p_2} 
-3 \, \frac{(p_3\!\cdot\! z)^2(p_1p_2+p_1p_3+p_3p_2+p^2_3)}{2^3(p_1+p_2+p_3)^2p_1p_2p_3}. \nn
\end{align}
The three point function can now be expressed more compactly as
\begin{equation}\label{3B}
 \langle B_{0}(p_1)B_{0}(p_2)B_{2}(p_3|z)\rangle=-\frac{{32}\sqrt{2}}{N^2}\delta_{p_1+p_2+p_3}(I_1(p_1,p_2,p_3;z)+I_2(p_1,p_2,p_3;z))
\end{equation}
{Finally, we rewrite the three-point function in terms of the standard local boundary fields $\CB_0$ and $\CB_2^{ij}$ (which appear as local scalar and spin-2 sources in the Sp(N) model). By our general prescription, they are related to $B_0$ and $B_{ij}$ by the shadow transform $B_I = G_{IJ} \CB^J$:}
\begin{equation}\label{BCB}
    B_0(p)=\frac{1}{p}\CB_0(p) \, , \quad B_{ij}(p)=p^3 \, \Pi_{ij,i'j'}({p}) \,\CB_2^{i'j'}(p) \, ,
\end{equation}
{where $\Pi_{ij,i'j'}({p})$ is the projector (\ref{projector}). Picking a transverse traceless gauge for $\CB_2$, this simplifies to $B_{ij}(p) = p^3 \CB_2^{ij}(p)$, with inverse $\CB_2^{ij}(p)= \frac{1}{p^3} B_{ij}(p)$.
Putting everything together we end up with the following simple expression for the scalar-scalar-graviton 3-point function:}
\begin{equation}
 \bigl\langle \CB_{0}(p_1) \, \CB_{0}(p_2) \, \CB_{2}^{ij}(p_3) \bigr\rangle=\frac{{16}\sqrt{2}}{N^2}\frac{p_1+p_2+2p_3}{(p_1+p_2+p_3)^2p^3_3} \, \Pi_{ij,i'j'}({p}_3) \, p_1^{i'} p_2^{j'} \, \delta_{p_1+p_2+p_3} \, .
\end{equation}

\subsubsection{Scalar four-point function}

\def\m{{p_{21}}}
\def\n{{p_{23}}}

It is also straightforward to calculate higher point functions in our setup. These are far less constrained by conformal symmetries. As a final example, we give here the scalar four-point function. For inflationary theories these were considered, for example, in \cite{Seery:2008ax,Ghosh:2014kba}. In our case we have
\begin{align}
\mathcal{G}_4(p_i) &\equiv \langle \CB_{0}(p_1)\CB_{0}(p_2)\CB_0(p_3)\CB_0(p_4)\rangle  \nonumber \\
&= \frac{2^6}{N^4} \, {p_1 p_2 p_3 p_4} \, \int_{q_1, q_2, q_3, q_4}{\bigl\langle : \! Q_{q_1}^\alpha Q_{p_1 - q_1}^\alpha\! : \, : \! Q_{q_2}^\beta Q_{p_2 - q_2}^\beta \! : \, : \! Q_{q_3}^\gamma Q_{p_3 - q_3}^\gamma \! : \, : \! Q_{q_4}^\delta Q_{p_4 - q_4}^\delta \! : \, \bigr\rangle} \ .
\end{align}
There are 60 possible Wick contractions, 12 of which are the product of two-point functions and the other 48 are equivalent up to permutations. The later can be expressed in terms of an integral over a single momentum $q$,
\begin{equation}
\mathcal{G}_4(p_i)  = \left[ \frac1{N^2} \, p_1 p_3 \delta_{p_1 + p_2} \delta_{p_3 + p_4} + \frac{64}{N^3} \, p_1 p_2 p_3 p_3 \delta_{p_1 + p_2 + p_3 + p_4} I \right] + (p_2 \leftrightarrow p_3) + (p_2 \leftrightarrow p_4)~, 
\end{equation}
with
\begin{equation}\label{boxintegral}
I \equiv \int_q \, \frac{1}{q^{2} (q + p_1)^{2}(q + p_1 + p_2)^{2} (q + p_1 + p_2 + p_3)^{2}}~.
\end{equation}
This integral in can be calculated\footnote{
The approach we take to solve (\ref{boxintegral}) is due to explanations from Adam Bzowski, who we would like to gratefully acknowledge.} by considering the inverse momenta as follows \cite{Bzowski:2011ab}. Let us define $P_i \equiv \sum_{j=1}^i p_i$ and denote the inverse of a vector with a tilde, for example $\tilde{q}_i = q_i / q^2$. We then have that
\begin{equation}
	\frac1{(q + P_2)^2} = \frac{\tilde{q}^2 \tilde{P}_2^2}{(\tilde{q} + \tilde{P}_2)^2}~, \quad\quad  \int_q = \int_{\tilde{q}} \frac1{\tilde{q}^6} \ .
\end{equation}
Using this, we can rewrite the integral appearing in the four-point function in terms of three-point function integrals,
\begin{align}
	I &= \tilde{P}_1^2 \tilde{P}_2^2 \tilde{P}_3^2 \int_{\tilde{q}} \frac{\tilde{q}^2}{(\tilde{q} + \tilde{P}_1)^2 (\tilde{q} + \tilde{P}_2)^2 (\tilde{q} + \tilde{P}_3)^2}  \nonumber \\
	&= \tilde{P}_1^2 \tilde{P}_2^2 \tilde{P}_3^2 \int_{\tilde{p}} \frac{\tilde{p}^2 - 2 \tilde{p} \cdot \tilde{P_1} + \tilde{P}_1^2}{\tilde{p}^2 (p + \tilde{P}_2 - \tilde{P}_1)^2 (\tilde{p} + \tilde{P}_3  - \tilde{P}_1)^2} \ .
\end{align}
{We introduce the quantities 
\begin{align}
 \m \equiv |p_2+p_1| \, , \qquad \n \equiv |p_2 + p_3|
\end{align}}
and calculate the three-point function integrals using the methods of \cite{Bzowski:2013sza},
\begin{align}
	\int_{\tilde{p}}{\frac1{(\tilde{p} + \tilde{P}_2 - \tilde{P}_1)^2 (\tilde{p} + \tilde{P}_3  - \tilde{P}_1)^2}} &= \frac1{8 |\tilde{P}_3 - \tilde{P}_2|} = \frac{\m p_4}{8 p_3} \ ,  \nonumber \displaybreak[0] \\
	\int_{\tilde{p}}{\frac{2 \tilde{p} \cdot \tilde{P}_1}{\tilde{p}^2 (\tilde{p} + \tilde{P}_2 - \tilde{P}_1)^2 (\tilde{p} + \tilde{P}_3  - \tilde{P}_1)^2}} 
	&= \m p_4 \frac{\n p_4 (\m^2 - p_1^2 + p_2^2) + \m p_2 (\n^2 - p_1^2 + p_4^2)}{8 p_2 p_3 \n(p_1 p_3 + p_2 p_4 + \m \n)} \ ,  \nonumber \displaybreak[0] \\
	\int_{\tilde{p}}{\frac{\tilde{P}_1^2}{\tilde{p}^2 (\tilde{p} + \tilde{P}_2 - \tilde{P}_1)^2 (\tilde{p} + \tilde{P}_3  - \tilde{P}_1)^2}} 
	&= \frac{\m^2 p_4^2}{8 p_2 \n p_3} \ .
\end{align}
This leads to the scalar four-point function
\begin{multline}\label{crazysimple}
 \mathcal{G}_4(p_i)  = \left[ \frac1{N^2} \, p_1 p_3 \delta_{p_1 + p_2} \delta_{p_3 + p_4} + \frac{8}{N^3} \,\frac{(p_1 p_2 + p_3 p_4) \m + (p_1 p_4 + p_2 p_3) \n}{\m \n (p_1 p_3 + p_2 p_4 + \m \n)}\,  \delta_{p_1 + p_2 + p_3 + p_4}  \right]   
	\\ + (p_2 \leftrightarrow p_3) + (p_2 \leftrightarrow p_4)~.
\end{multline}
As a simple check, the above result is permutation invariant under exchange of the momenta and has the correct scaling properties. {We have also verified the result numerically.} 


{In the scattering amplitudes literature,  the integral (\ref{boxintegral}) is referred to as a four-mass box integral. Explicit analytic results in the $d=3$ case of interest to us were previously obtained in \cite{Lipstein:2012kd}. Perhaps the approach we have taken here, based on \cite{Bzowski:2013sza,Bzowski:2011ab}, may be useful more broadly in that context as well, given that our expression (\ref{crazysimple}) is dramatically simpler than the one obtained in \cite{Lipstein:2012kd}.}

\section{Perturbative bulk QFT reconstruction}\label{breakdown}


\subsection{Bulk reconstruction and the Heisenberg algebra}

{So far we have shown how to compute probability distributions and vacuum correlation functions of higher spin de Sitter boundary field modes $\beta^I$, identified in our framework with bilinear operators $B_I = \frac{1}{N} :\! Q_x \CD_I^{xy} Q_y$ acting on the Hilbert space $\CCH$, related more precisely by the shadow transform $B_I = \tilde\beta_I = G_{IJ} \beta^J$.  We found results consistent with general expectations from perturbative bulk QFT, and illustrated for the constant scalar mode $\beta^0$ on $S^3$ how to go beyond the perturbative bulk QFT regime.} 

{However, to go beyond observables that can be described entirely in terms of the boundary fields $\tilde \beta_I$, and in particular to reconstruct local bulk quantum fields $\phi_{i_1 \cdots i_s}(\eta,x)$, we also need to identify the operators canonically conjugate to $\tilde\beta_I$, that is to say the operators $\alpha_I$ appearing in the perturbative bulk QFT Heisenberg algebra (\ref{betaalphacanconj}), which in our condensed notation reads 
\begin{align} \label{pertbulkQFTHA}
  [\tilde\beta_I,\alpha_J] \, = \, i \, \gamma \, G_{IJ}, \qquad [\tilde\beta_I,\tilde\beta_J]=0=[\alpha_I,\alpha_J] \, ,
\end{align}
where in our conventions, as in (\ref{gammaistwooverN}), $\gamma= \frac{2}{N}$. 
Once these conjugate operators have been identified, we can define free local quantum fields $\phi_{i_1 \cdots i_s}(\eta,x)$ in the bulk simply by copying the free field expression (\ref{alphabetamodes}), that is, in $d=3$,
\begin{align} \label{alphabetamodes2}
 \phi_{i_1 \cdots i_s}(\eta,x) \equiv  (-\eta) \int_k \,  \left( \alpha_{i_1 \cdots i_s}(k)  \, \bar J_{s-\frac{1}{2}}(-k\eta)   +  \tilde\beta_{i_1 \cdots i_s}(k) \, \bar Y_{s-\frac{1}{2}}(-k\eta) \right) e^{i k \cdot x} \, ,
\end{align}
where $\bar J_\nu(z) \equiv \sqrt{\tfrac{\pi}{2}} \, z^{-\nu} J_{\nu}(z)$ and  
$\bar Y_\nu(z) \equiv \sqrt{\tfrac{\pi}{2}} \, z^{-\nu} \,Y_{\nu}(z)$. 
Transformed to position space, these definitions express $\phi_{i_1 \cdots i_s}(\eta,x)$ as  convolutions of $\alpha_{i_1 \cdots i_s}(x')$ and $\tilde\beta_{i_1 \cdots i_s}(x')$ with certain boundary-to-bulk kernels $K_\alpha(\eta,x-x')$ and $K_\beta(\eta,x-x')$, providing a dS analog to the HKLL construction \cite{Hamilton:2006az} in AdS (except that in dS, we have two dynamical modes rather than the single normalizable mode in AdS; {see also \cite{Xiao:2014uea,Sarkar:2014dma}}). The boundary fields $\alpha$, $\beta$ must satisfy the Heisenberg algebra in order for the free bulk fields thus constructed to be local and causal, i.e.\ in order for the fields commute at spacelike separations. Bulk interactions can in principle be reconstructed order by order by comparing vacuum correlation functions computed in the $Q$-model to vacuum correlation functions computed perturbatively in the bulk interaction picture using these free fields.}

{
However, as we will see, {\it the exact bulk perturbative Heisenberg algebra (\ref{pertbulkQFTHA}) cannot be realized on $\CCH$.} Indeed this is obvious for a number of reasons:
\begin{enumerate}
 \item The algebra (\ref{pertbulkQFTHA}) implies an infinite number of independent degrees of freedom per spatial point: one for the scalar, and two for each of the infinite tower of higher spin fields. However, the $Q$-model has only $2N$ degrees of freedom per spatial point. Hence $\CCH$ cannot possibly accommodate (\ref{pertbulkQFTHA}).
 \item Even in discretized models with a finite number $K$ of spatial points, including the case $K=1$ of section \ref{sec:toymodel}, for which we have just a single operator $\beta=\frac{1}{N} \!:\! q^2 \! :$, there cannot possibly exist a self-adjoint operator $\alpha$ satisfying $[\beta,\alpha]=i$. For indeed if such a self-adjoint operator did exist, we could exponentiate it to a unitary operator $U_\lambda = e^{i\lambda \alpha}$ acting on $\beta$ as an arbitrary translation $\beta \to \beta + \lambda$. But the existence of such a unitary  operator on $\CCH$ is inconsistent with the fact that $\beta=\frac{1}{N} \! :\! q^2 \! :$ is manifestly positive.  
\end{enumerate}
Nevertheless, we will show below that at large $N$, in discretized models with $K \lesssim O(N)$, it is possible to define self-adjoint operator $\alpha_I$ such that (\ref{pertbulkQFTHA}) is satisfied up to non-perturbatively small corrections, in the sense that on states not too far from the vacuum, the algebra is satisfied up to an operator-valued error term of order 
\begin{align}
  \delta_{\rm Heis} \sim e^{-g(\kappa) N} \, ,
\end{align}
where $\kappa = \frac{K}{2N}$ and $g(\kappa)$ is some order 1 function which is positive for $\kappa \lesssim 0.17$. Moreover, for $K > 2N$, including in the continuum limit $K=\infty$, the Heisenberg algebra can be realized with similar exponential accuracy provided we consider ``coarse grained'' operators $\bar\beta_I$, $\bar \alpha_I$ and restrict to a finite volume of space, such that the effective number of resolvable spatial ``pixels'' $K_{\rm eff}$ is less than some order $N$ number.}  

{This suggests we can reconstruct perturbative bulk quantum field theory from the fundamental boundary Hilbert space $\CCH$, but only up to a resolution in which at most $\mathcal {O}(N)$ spatial pixels are resolved. This may seem peculiar, but keeping in mind that the de Sitter horizon entropy $S_{\rm dS} \sim \ell^2_{\rm dS}/G_{\rm Newton} \sim N$ in this model, it is in line with general expectations on limitations of bulk effective field theory in de Sitter, based on the holographic principle and related ideas (see e.g.\ \cite{Goheer:2002vf,Albrecht:2002xs,Banks:2003pt,ArkaniHamed:2007ky}), although realized here in an unusual and perhaps surprising way. 
} 

\subsection{More general comments}

{Independent of the construction of a Heisenberg algebra, there are other reasons to expect a breakdown of conventional perturbative bulk QFT in out setup. One is that perturbative single-particle kets such as 
\begin{align}
 |f_s\rangle \equiv \int d^3 x \, f_s^{i_1 \cdots i_s}(x) \, B_{i_1 \cdots i_s}(x) \, |0\rangle \, ,
\end{align}
which for smooth functions $f_s$ with compact support represent normalizable states on the bulk perturbative QFT Fock space $\CCH_{\rm Fock}$, actually do not represent normalizable states when interpreted as kets on the $Q$-model Hilbert space $\CCH$ when $s > 0$. The norm squared of such a state $|f\rangle = f^I B_I |0\rangle$ is $\langle f|f\rangle = f^I f^J \langle 0|B_I B_J |0\rangle$. This is finite for $s=0$ states, but diverges for $s>0$, due to the non-integrability of the position space 2-point function in this case. In the perturbative bulk QFT this is easily resolved by adding the appropriate contact terms to the 2-point function, or equivalently by defining the 2-point function in momentum space, as is usually done in Fock space constructions. This is allowed because the 2-point function is part of the data defining the QFT. However in the $Q$-model, the 2-point function is a derived quantity, fixed by the definition of $B_I$, and it is not possible to modify this definition in such way that $|f_s\rangle$ becomes normalizable for $s>0$.} 

{Of course this does not mean we cannot define normalizable ``single-particle'' states more general than single-scalar states in $\CCH$.
For example the states $|F\rangle \equiv \int_{xy} F^{xy} B_{xy} |0\rangle$, for smooth functions $F^{xy}$ with compact support, are normalizable, and can formally be thought of as infinite superpositions of higher spin particle states by Taylor expanding in $r=(x-y)$. But it does mean that within in the $Q$-model, we cannot truly define an orthonormal basis of single-particle states of definite $s$, and thus cannot truly reproduce the multi-particle Fock space $\CCH_{\rm Fock}$ of perturbative bulk QFT. Keeping in mind that the Fock space is constructed starting from the Heisenberg algebra, this observation is consistent with our earlier claim, to be demonstrated below, that $\CCH$ cannot support the full bulk QFT Heisenberg algebra.}

{We encountered related subtleties earlier in defining the 2-point function $\langle 0|B_I B_J|0\rangle \propto G_{IJ}$ in momentum space. Strictly speaking this is UV divergent for $s>0$. To make sense of this, we defined a renormalized $G_{IJ}$ by standard analytic continuation, as discussed under (\ref{tptfcoicoj}).  What this really means is that we are actually computing $\langle 0|(B_I B_J - \mbox{c.t.})|0\rangle$ instead, where ``c.t.'' are local counterterms canceling off the UV divergences. This makes sense when interpreted as the vacuum expectation value of a renormalized operator product $(B_I B_J - \mbox{c.t.})$, which was adequate for our purposes of giving a prescription for computing finite correlation functions. However it is not adequate for computing the actual inner product of two kets $B_I|0\rangle$, $B_J|0\rangle$ on $\CCH$, because the counterterm subtractions cannot be realized at the level of the individual kets.  For the same reasons, we were not able to assign a straightforward physical interpretation to the (renormalized) probability densities $P(B_I)$ of spin $s>0$ field profiles computed in our framework along the lines of section \ref{sec:PBgenprofile}.}

{In a similar spirit, we noted in section \ref{sec:Kgt2M} that in discretized models with $K$ lattice points, although it was possible to formulate perturbation theory purely in terms of the Sp(N) model wave function $\tilde\psi_0(\CH)$, this required renormalization through analytic continuation as soon as $K>2N$. More generally, we noted various qualitative differences in behavior of such models depending on whether $K \leq 2N$ or $K>2N$. The underlying reason for this was that for $K \leq 2N$, the O(2N)-invariant bilinears $H_{xy} = \frac{1}{N} Q_x^\alpha Q_y^\alpha$ are independent variables, with $H$ generically of rank $K$, while for $K>2N$, the $H_{xy}$ are not independent, as $H$ generically has reduced rank $2N$ (as illustrated in fig.\ \ref{fig:HBbdim1}). This distinction between $K \leq 2N$ and $K>2N$ will also play a crucial role in our construction of an approximate Heisenberg algebra below.}

{Finally, an important thing to keep in mind is that all of these considerations pertain to the Hilbert space $\CCH$ (which is to be compared to $\CCH_{\rm Fock}$, as we do in this section), not to the {\it physical} Hilbert space $\CCH_{\rm phys}$ (which is to be compared to $\CCH_{\rm Fock,\rm phys}$). We discuss the construction of the physical Hilbert space in detail in section \ref{sec:GIaPHS}. As we will see, the difference between $\CCH$ and $\CCH_{\rm phys}$ is rather enormous in higher spin de Sitter, because the residual gauge group $\CG$ on $\CCH$, i.e.\ the higher spin group, is infinite dimensional. This reduces the number of gauge-inequivalent degrees of freedom to such extent that the physical Hilbert space of gauge-inequivalent $n$-particle states becomes {\it finite} dimensional for any given $n$. In particular this effectively removes all UV divergences of norms of single-particle states mentioned above. Moreover we will show that all gauge invariant quantities can be computed by a $2N \times 2N$ matrix integral. Thus, from the point of view of $\CCH_{\rm phys}$, the inability to reconstruct local bulk QFT beyond a certain resolution set by $N$ is not all that dramatic, and is in fact quite natural.}

\subsection{Reconstructing the Heisenberg algebra} 

The goal of this section is to investigate to what extent we can reproduce the perturbative higher spin bulk QFT Heisenberg algebra (\ref{betaalphacanconj}) in our framework. Of course, the  Hilbert space $\CCH_0$ of wave functions $\psi(Q)$ with inner product (\ref{psiQinnerproduct}) has a standard Heisenberg algebra, given by
\begin{align} \label{QPHeis}
 [Q_x^\alpha,P^y_\beta] = i \delta_x^y \delta^\alpha_\beta \, , \qquad 
 [Q_x^\alpha,Q_y^\beta] = 0 = [P^x_\alpha,P^y_\beta] \, ,
\end{align}
where $Q_x^\alpha$ and $P^x_\alpha \equiv - i \frac{\partial}{\partial Q_x^\alpha}$ are manifestly self-adjoint. However, the perturbative bulk QFT Heisenberg algebra (\ref{betaalphacanconj}) we seek to reconstruct here is rather different from this, as it must be realized on the O(2N)-invariant Hilbert space $\CCH$, and therefore must be realized by self-adjoint O(2N)-invariant composite operators $B_I$, $A_J$, where $B_I=\frac{1}{N}\! :\! Q_x^\alpha \CD_I^{xy} Q_y^\alpha\!:$ and $A_J$ is to be determined.  
In conventions such that $G_{IJ}$ as defined in (\ref{tptfcoicoj}) coincides with $G_{IJ}$ as defined in (\ref{Gi1isip1sips}), the bulk QFT Heisenberg algebra (\ref{betaalphacanconj}) is to be realized as  
\begin{align} \label{Heisa}
 [B_I,A_J] = \frac{2}{N} \, i \, G_{IJ} \, , \qquad [A_I,A_J] = 0 = [B_I,B_J] \, ,
\end{align}
The commutator $[B_I,B_J]=0$ is trivially realized, but finding a set of operators $A_I$ realizing the other commutation relations is a nontrivial task. In fact, it is impossible. On general grounds, there can be no well-defined self-adjoint operators $A_I$ on the Hilbert space $\CCH$ {\it exactly} realizing the Heisenberg algebra, for if there were such operators, we could exponentiate them to well-defined unitary operators $U_\lambda = e^{i \lambda^I A_I}$ acting on the Hilbert space $\CCH$, but these would act as arbitrary translations $B_I \to B_I + G_{IJ} \lambda^J$, implying an eigenvalue spectrum of $B_I$ spanning the entire real line, which is inconsistent with the positive definiteness of $H=\CD^{-1} + B = Q Q^T$. In suitable circumstances it is nevertheless possible to find well-defined, self-adjoint operators $A_I$ realize an {\it approximate} Heisenberg algebra. Finding such operators will be our goal in what follows. 

\subsubsection{Free field theory approximation}

It is easy to find operators $A_I^{(0)}$ realizing the Heisenberg algebra at the level of vacuum expectation values, i.e.\ $\langle \psi_0|[B_I,A^{(0)}_J]|\psi_0\rangle = \frac{2}{N} \, i \, G_{IJ}$, $\langle \psi_0|[A^{(0)}_I,A^{(0)}_J]|\psi_0\rangle = 0$, by defining 
\begin{align} \label{freeapproxA0}
 A^{(0)}_I \equiv \frac{1}{N} \! :\! Q_x^\alpha \CD_I^{xy} P_y^\alpha \! : \, , \qquad P_y^\alpha \equiv (\CD^{-1})_{yz} \, P^z_\alpha \, ,
\end{align}
with $P^z_\alpha = - i \partial_{Q_z^\alpha}$ as in (\ref{QPHeis}), so 
 $[Q_x^\alpha,P_y^\beta] = i \, \delta^{\alpha\beta}(\CD^{-1})_{xy} $. Then we have
\begin{align} \label{BIAJ0com}
 [B_I,A_J^{(0)}] &= \frac{2}{N^2} \, i \, \Tr(Q \CD_I \CD^{-1} \CD_J Q) \nn \\
 & = \frac{2}{N} \, i \, G_{IJ} \, + \,  
 \frac{2}{N^2} \, i :\!\Tr(Q \CD_I \CD^{-1} \CD_J Q)\! : \, ,
\end{align}
where we used $\langle \psi_0|Q^\alpha_x Q^\alpha_y|\psi_0\rangle = N (\CD^{-1})_{xy}$ and $G_{IJ}=\Tr(\CD^{-1} \CD_I \CD^{-1} \CD_J)$. 
Thus the vacuum expectation value of the commutator is $\langle \psi_0|[B_I,A^{(0)}_J]|\psi_0\rangle = \frac{2}{N} \, i \, G_{IJ}$, as claimed. Similarly,
\begin{align}
 [A_I^{(0)},A_J^{(0)}] = \frac{i}{N^2} :\!\left( \Tr(Q\CD_J \CD^{-1} \CD_I P) - 
 \Tr(Q\CD_I \CD^{-1} \CD_J P) \right)\!: \, . 
\end{align}
Here we were allowed to add the normal-ordering signs because the bilinear operator inside the brackets has zero vacuum expectation value, as can be seen using $P_x^\alpha |\psi_0\rangle = i Q_x^\alpha |\psi_0\rangle$. Thus $\langle \psi_0|[A^{(0)}_I,A^{(0)}_J]|\psi_0\rangle = 0$, as claimed.

Because these commutation relations realize the exact Heisenberg algebra at the level of 2-point functions, they are good enough if we want to construct {\it free} bulk quantum field theory. To do so, we simply define the free higher spin bulk quantum fields as in (\ref{alphabetamodes}), using $A_I^{(0)}$ and $B_I$ instead of $\alpha_I$ and $\tilde\beta_I$. This already reproduces many interesting characteristics of de Sitter space, including its thermal nature from the point of view of a static observer. However the above commutation relations are not good enough beyond the free limit, not even in lowest order tree level perturbation theory, because the deviations from the Heisenberg algebra are only suppressed by a single power of $\frac{1}{N}$, which is the order at which interactions enter. 

To do better, we can try to correct the $A_I^{(0)}$ order by order in the $\frac{1}{N}$ expansion. We turn to this next, first for the discretized toy models with $K$ spatial points, and then in the continuum limit.

\subsubsection{K=1} \label{sec:pertconj}

We first consider the $K=1$ toy model introduced in section \ref{sec:toymodel}. The analog of (\ref{Heisa}) in this case is
\begin{align}  \label{bacoco}
 [b,a] = \frac{2}{N} \, i \, .
\end{align}
It will be useful to first switch to the $h$-space description, which for $K=1$ is equivalent to the $q$-space description. Recall that in this description $h=\frac{1}{N} q^\alpha q^\alpha$, $b=h-1=\frac{1}{N}\!:\!q^2\!:$, so (\ref{bacoco}) is equivalent to $[h,a]=\frac{2}{N} \, i$. The O(2N)-invariant Hilbert space $\CCH$ consists of wave functions $\tilde\psi(h)$ on the domain $h>0$, normalizable with respect to the inner product with measure $[dh]=\frac{dh}{h}$. 
The $q$-space ground state wave function $\psi_0(q) \, \propto \, e^{-\frac{1}{2} q^2}$ is represented as $\tilde\psi_0(h) \, \propto \, h^{\frac{N}{2}}  e^{-\frac{N}{2} h}$. 

Although the operator $-\frac{2i}{N} \partial_h$ is formally canonically conjugate to $h$, as noted earlier, it cannot possibly be a well-defined self-adjoint operator on the Hilbert space $\CCH$, since if it was, it would generate unitary translations of $h$, violating $h>0$. In fact it is not even hermitian, since
\begin{align}
  \int \frac{dh}{h} \, \tilde\psi^* (-i \partial_h) \tilde\psi = \int \frac{dh}{h} \bigl( (-i\partial_h + \frac{i}{h}) \tilde\psi \bigr)^* \tilde\psi   \qquad \Rightarrow \qquad (-i\partial_h)^\dagger = -i\partial_h + \frac{i}{h} \, .
\end{align}
These manipulations are valid for wave functions $\tilde\psi(h)$ vanishing sufficiently fast at $h=0$, so integrating by parts does not pick up any boundary terms. The non-hermiticity is easily fixed, however, by defining
\begin{align} \label{tildeadefi}
 \tilde a \equiv -\tfrac{2i}{N}  \bigl( \partial_h - \tfrac{1}{2} h^{-1} \bigr) = -\tfrac{2i}{N}\, h^{\frac{1}{2}} \partial_h h^{-\frac{1}{2}}  = \tfrac{1}{2N} \{ h^{-1},\tilde v \} \, , \qquad \tilde v = - 2 i h \partial_h \, ,
\end{align}
where $\{A,B\}=AB+BA$ denotes the anticommutator, and we recall that $\tilde v$ appeared in (\ref{tildeVexpr}) as the $h$-space version of the manifestly self-adjoint $q$-space operator $v=:\! q^\alpha p^\alpha \! :$. The operator $\tilde a$ is still canonically conjugate to $h$, as $[h,\tilde a] = \frac{2i}{N}$.
Although $\tilde a$ is hermitian, in the sense that it maps to itself upon integrating by parts on wave functions vanishing sufficiently fast at the origin, is not self-adjoint, essentially for the same reason as the ordinary momentum operator in quantum mechanics of a particle on a half-line is not self-adjoint. Because of this, it cannot be exponentiated to a unitary operator, and it does not have a spectral decomposition on $\CH$, so it is not a valid canonical conjugate variable.

In the last expression for $\tilde a$ in (\ref{tildeadefi}), the appearance of the singular operator $h^{-1}$ is what causes the failure of $\tilde a$ to be self-adjoint. If instead of $h^{-1}=\frac{1}{q^2}$ we had a self-adjoint operator regular at $q=0$, the resulting operator would have been  self-adjoint as well. This observation immediately suggests a definition for a self-adjoint operator approximating $\tilde a$, approximately realizing the Heisenberg algebra in a perturbative sense. The basic idea is simply to truncate the formal power series $h^{-1} = (1+b)^{-1} = \sum_{k=0}^\infty (-b)^k$ to a finite number of terms, defining
\begin{align} 
 (h^{-1})^{(n)} \equiv \sum_{k=0}^n (-b)^k \, , \qquad \tilde a^{(n)} \equiv \frac{1}{2N} \{ (h^{-1})^{(n)},\tilde v\} \, .
\end{align}
Since $b=\frac{1}{N} :\! q^2 \! :$ is regular at the origin $q=0$, this operator is manifestly self-adjoint and regular at $q=0$. Its counterpart in the $q$-space description is, written out explicitly,
\begin{align}
 a^{(n)} = \frac{1}{2N}  \sum_{k=0}^n \frac{(-1)^k}{N^k} \{(:\!q^2\!:)^n,:\!qp\!:\} \, .
\end{align} 
Notice $a^{(0)} = \frac{1}{N} :\!qp\!:$, which is the $K=1$ analog of the free theory approximation (\ref{freeapproxA0}). The higher order $a^{(n)}$ add corrections to this reducing the error in the Heisenberg algebra: 
\begin{align}
 [b, a^{(n)}] &= \frac{2i}{N} \sum_{k=0}^n  (-b)^k  (1+b)  
 = \frac{2i}{N} \bigl(1 - (-b)^{n+1} \bigr) 
  = \frac{2i}{N} \Bigl(1 + \frac{(-1)^n }{N^{n+1}} \, (:\!q^2\!:)^{n+1} \Bigr) \, ,
\end{align}
where we used $[h,v]=2 i h$, $h=1+b$, and $b = \frac{1}{N}\! :\!q^2\!:$. Note that for $n=0$, this reproduces (\ref{BIAJ0com}). When $n$ increaes, the $\frac{1}{N}$ suppression of the error term increses. A rough estimate based on $\langle b^2 \rangle \sim \frac{1}{N}$ suggests the error term is of order $N^{-(n+1)/2}$, getting smaller indefinitely when $n$ gets larger. However this estimate is not correct for large values of $n$. A more careful examination shows that the vacuum expectation value of the error $\langle b^{n+1} \rangle$ reaches a minimum value at $n \sim \frac{N}{2}$, after which it starts growing again. To see this in more detail, consider
the limit $N \to \infty$ with $\nu \equiv n/N$ fixed. Then we have 
\begin{align} \label{asymptoticbnvec}
 \langle b^n \rangle &= \frac{n!}{N^n} \frac{1}{2\pi i} \oint \frac{dz}{z} \, z^{-n} \, \bigl\langle e^{N z b} \bigr\rangle 
  = \frac{n!}{N^n} \frac{1}{2\pi i} \oint \frac{dz}{z} \, z^{-n} \, (1-z)^{-N} e^{-Nz} \nn \\
  &\sim \oint dz \, e^{-N(\nu \log z + \log(1-z) + z + \nu - \nu\log\nu)} \, ,
\end{align} 
where we used $n! \sim (n/e)^n$ and neglected corrections unimportant at large $N$. Evaluating this in large $N$ saddle point approximation and minimizing the result with respect to variations of $\nu$ is equivalent to minimizing the exponent with respect to both $\nu$ and $z$. This results in $z=\frac{1}{2}$, $\nu = \frac{1}{2}$, with the minimal value being
\begin{align}
 \langle b^n \rangle|_{\rm min} = \langle b^{N/2} \rangle = e^{-N(1-\log 2)} \, .
\end{align}
That is to say, on the vacuum, or on states perturbatively close to it, the self-adjoint operator $ a_{\rm pert} \equiv a^{(N/2)}$ satisfies the Heisenberg algebra up to non-perturbatively small corrections at large $N$: 
\begin{align} \label{bapertccr}
 [b, a_{\rm pert}] = \frac{2i}{N} (1 \pm e^{-\Sigma}) \, , \qquad \Sigma=(1-\log 2)N \approx 0.3 \, N \, .
\end{align}
There is a very simple physical interpretation of this result: the power series $h^{-1} = (1+b)^{-1} = \sum_{k=0}^\infty (-b)^k$ has radius of convergence equal to 1. The probability for $b$ to exit this region of perturbative convergence is $P_{\rm exit} = \frac{N^N}{\Gamma(N)} \int_2^\infty \frac{dh}{h} \, h^N \, e^{-N h} = \frac{\Gamma(N,2N)}{\Gamma(N)}$, where $\Gamma(a,z)$ is the incomplete Gamma-function. Asymptotically at large $N$, this becomes $P_{\rm exit} = e^{-(1-\log 2)N}$, which exactly equals the minimal error term $e^{-\Sigma}$ in the Heisenberg algebra. Thus, rather pleasingly, we conclude that the minimal error term in the perturbative Heisenberg algebra equals the exponentially small probability of a nonperturbatively large fluctuation. Although this is of course a toy model with just one spatial point, this is qualitatively in line with general physical expectations in de Sitter space: nonperturbatively large ``uphill'' fluctuations away from the classical vacuum (e.g.\ of a scalar $\phi$ with some potential $V(\phi)$) are possible in de Sitter, but exponentially suppressed by a factor of $e^{-S}$, where $S$ is the de Sitter entropy. In Vasiliev dS gravity with Sp(N) dual, the dS entropy is proportional to $N$.  


\subsubsection{$K \leq 2N$} \label{sec:pcctH}


\noindent  The above construction for $K=1$ is readily generalized to the case $K \leq 2N$ introduced in section \ref{sec:equivforKlessN}. In this section we won't notationally distinguish operators $\tilde O$ acting on $\tilde\psi(H)$ from operators $O$ acting on $\psi(Q)$ like we did in earlier, trusting the context will make clear which one is meant. As in the $K=1$ case discussed above, it is possible to formally construct hermitian operators $A_I$ satisfying the Heisenberg algebra (\ref{Heisa}). Explicitly,  they are given by
\begin{align} \label{AIdef}
 A_I \equiv \Tr(D_I \Pi) \, ,
\end{align}
where, when acting on wave functions $\tilde \psi(\CH)$,
\begin{align}  \label{Pixydef}
 \Pi_{xy} \equiv - \tfrac{2i}{N} \left( \partial_{\CH^{xy}} - \tfrac{1}{2} \tfrac{K+1}{2} \,  ({\cal H}^{-1})_{xy} \right) =  (\det \CH)^{{\frac{K+1}{4}}} (-\tfrac{2i}{N} \partial_{\CH^{xy}}) (\det \CH)^{-\frac{K+1}{4}} \, ,
\end{align}
and we recall $\CH^{xy} \equiv \CD^{xu} H_{uv} \CD^{vy}$.  It is easy to check that these operators formally obey Heisenberg commutation relations (\ref{Heisa}).
As in the $K=1$ case, the $A_I$ are hermitian with respect to the  inner product with GL(K)-invariant measure $[d\CH]=d\CH/(\det \CH)^{\frac{K+1}{2}}$, when acting on wave functions vanishing sufficiently fast at the boundary of the $\CH>0$ domain. However they are not self-adjoint. Below we will define self-adjoint operators $A_{I}^{(n)}$ approximating $A_I$, which satisfy the Heisenberg  algebra (\ref{Heisa})  up to exponentially small corrections at large $N$.

To write $A_I$ in a form analogous to the last expression for $a$ in (\ref{tildeadefi}), first note that
\begin{align}
 \Pi_{xy} = \frac{1}{2N} \bigl( {R_x}^z V_{zy} + V_{zy} {R_x}^z \bigr) \, , \qquad R \equiv (H\CD)^{-1} = (1+B\CD)^{-1} \, ,
\end{align}
where, when acting on wave functions $\psi(H)$, $ V_{xy} = - 2i {(H\CD)_x}^z \partial_{\CH^{zy}}$, and when acting on O(2N)-invariant wave functions $\psi(Q)$,
$V_{xy} = :\! Q_x \! \cdot \! P_y \! :$. 
Thus the operators $A_I=\Tr(\CD_I \Pi)$ and $B_I=\Tr(\CD_I B)$ are represented on $\psi(Q)$ as
\begin{align} \label{AIQspace}
 A_I = \frac{1}{2N} \bigl( {R_x}^z :\! Q_z D_I^{xy} P_y \! : + :\! Q_z D_I^{xy} P_y \! : {R_x}^z \bigr) \, , \qquad B_I = \frac{1}{N}\! :\! Q_x D_I^{xy} Q_y \! : \, ,
\end{align}
where
\begin{align} \label{Rseries}
 R = \bigl(1+ B\CD \bigr)^{-1} = \sum_{k=0}^\infty (-B\CD)^k  \, , \qquad B_{xy} = \frac{1}{N} \! : \! Q_x Q_y \! : \, .
\end{align}
It can be checked explicitly from these expressions that the Heisenberg algebra  (\ref{Heisa}) is indeed satisfied on O(2N)-invariant wave functions. (On wave functions which are not O(2N)-invariant, $[A_I,A_J] \neq 0$.) Note furthermore that $\langle A_I \rangle = \langle B_I \rangle = 0$.

From the $Q$-space point of view the operators $A_I$ are clearly singular  because $R=(H\CD)^{-1}=(QQ\CD)^{-1}$ is singular at $Q=0$. However as in section \ref{sec:pertconj}, we may truncate the series (\ref{Rseries}) at some finite order $n$,
\begin{align} \label{truncatedRseries}
 R^{(n)} \equiv  \sum_{k=0}^n (-B\CD)^k \, ,
\end{align}
and define perturbative canonical operators $A_{I}^{(n)}$ accordingly by replacing $R$ by $R^{(n)}$ in (\ref{AIQspace}). The case $n=0$ corresponds to the free approximation (\ref{freeapproxA0}). The truncated operators $A_{I}^{(n)}$ are now nonsingular and self-adjoint, but the Heisenberg (\ref{Heisa}) is only approximately satisfied. 
A short computation gives for example for the $[B,A^{(n)}]$ commutators: 
\begin{align} \label{BIAJnn}
  [B_I,A_{J}^{(n)}] = \frac{2i}{N} \Tr\bigl( R^{(n)}(1+B\CD) \Gamma_{IJ} \bigr) \, , \qquad \Gamma_{IJ} \equiv \CD^{-1} \CD_I \CD^{-1} \CD_J \, .
\end{align}
For $n=0$, this reproduces (\ref{BIAJ0com}).
For $n=\infty$, it formally reproduces the exact algebra $[B_I,A_J^{(\infty)}] = \frac{2i}{N} \Tr \Gamma_{IJ} = \frac{2i}{N} G_{IJ}$. However a more careful analysis shows again that the limit $n \to \infty$ does not converge, but rather reaches an optimal approximation point at some value of $n$ of order $N$. To see this, note that substitution of (\ref{truncatedRseries}) in (\ref{BIAJnn}) gives
\begin{align} \label{BIAJnGIJ}
 [B_I,A_{J}^{(n)}] = \frac{2i}{N} \left( G_{IJ} +  (-1)^n \delta_{IJ}^{(n)} \right) \, , \qquad  \delta_{IJ}^{(n)} \equiv \Tr\bigl( (B\CD)^{n+1} \Gamma_{IJ} \bigr) \, .
\end{align}
Since $B \propto \frac{1}{\sqrt{N}}$ at large $N$ in the Gaussian approximation, we might naively expect the error term $\delta_{IJ}^{(n)}$ to become arbitrarily small when $n$ increases, but just as in (\ref{asymptoticbnvec}), this simple-minded estimate fails when $n$ becomes of order $N$.  
Consider first the vacuum expectation value of $\delta_{IJ}^{(1)}$, i.e.\ $\langle \delta_{IJ}^{(1)} \rangle = \Tr \bigl( \bigl\langle (B\CD)^2 \bigr\rangle \, \Gamma_{IJ} \bigr)$.
Denoting ${B_x}^y \equiv B_{xz} \CD^{zy}$, $Q^x = \CD^{xy} Q_y$ to reduce cluttering, we have
\begin{align}
 \langle {(B\CD B\CD)_x}^y \rangle = \langle {B_x}^z {B_z}^y \rangle = \frac{1}{N^2} \bigl\langle :\!Q_x Q^z\!: :\!Q_z Q^y\!: \bigr\rangle = \frac{K+1}{2 N} \, \delta_x^y \, .
\end{align}
Therefore we find for $\langle \delta_{IJ}^{(1)}\rangle$, and by a similar but somewhat longer computation for $\langle \delta_{IJ}^{(2)}\rangle$:
\begin{align} \label{lowerordererror}
 \langle \delta_{IJ}^{(1)} \rangle = \frac{K+1}{2N} \, G_{IJ} \, ,  \qquad 
 \langle \delta_{IJ}^{(2)} \rangle = \frac{K^2+3K+4}{(2N)^2}  \, G_{IJ} \, .
\end{align}
More generally we have
\begin{align}
 \langle \delta_{IJ}^{(n)} \rangle = \frac{1}{K} \bigl\langle \Tr (B\CD)^{n+1} \bigr\rangle \, G_{IJ} \, ,
\end{align}
where $\bigl\langle \Tr (B\CD)^n \bigr\rangle$ can be computed by a matrix integral independent of $\CD$:
\begin{align}
 \bigl\langle \Tr (B\CD)^n \bigr\rangle &= \frac{1}{Z} \int [d\tH] \, \Tr(\tH-1)^n \, (\det \tH)^N \, e^{-N \Tr \tH} \, \label{WishartModel} \\
 & = \frac{1}{Z'} \int d\tQ \, \Tr(\tQ\tQ-1)^n \, e^{-\Tr(\tQ\tQ)} \, .
\end{align}
The first matrix integral can be written in terms of the eigenvalues $\lambda_x>0$ of $H$ as
\begin{align}
 \delta^{(n)} \equiv \frac{1}{K} \bigl\langle \Tr (B\CD)^n \bigr\rangle = \frac{1}{Z} \int d \lambda \, \prod_{x<y} |\lambda_x - \lambda_y| \,  (\lambda_K - 1)^n \, \prod_x \lambda_x^{-\frac{K+1}{2} + N} \, e^{-N \sum_x \lambda_x} \, ,
\end{align}
where we used the eigenvalue permutation symmetry to rewrite the trace expectation value in terms of a single eigenvalue $\lambda_K$: $\langle \Tr(\tH-1)^n \rangle = \sum_{x=1}^K \langle (\lambda_x-1)^n \rangle = K \langle (\lambda_K-1)^n \rangle$. To find the minimal error in large $N$ saddle point approximation, we have to extremize the integrand with respect to the eigenvalues $\lambda_x$ and with respect to $n$. Extremization with respect to $n$ simply gives $\lambda_K = 2$.  Thus, the minimal value of $\delta^{(n)}$ in a large $N$ saddle point approximation is obtained as $\delta^{(n)}|_{\rm min} \approx P(\lambda_K=2)$, where $P(\lambda_K=2)$ is the probability density for the eigenvalue $\lambda_K$ to equal 2. (The optimal value of $n$ can in principle be found from the extremization equation with respect to variations of $\lambda_K$.) Notice that the perturbative expansion (\ref{Rseries}), i.e.\ $\tH^{-1}  =(1+\tB)^{-1} = \sum_k (-\tB)^k$  also breaks down exactly when one of the eigenvalues $\lambda$ of $\tH$ exceeds $\lambda \geq 2$. When $N \gg K$, all eigenvalues clump near $\lambda=1$. When $K$ grows, their spread increases due to the eigenvalue repulsion induced by the Vandermonde measure factor. Thus when $K$ gets too large, the eigenvalues will reach $\lambda \sim 2$ with probability essentially equal to 1. Then the minimal error will be of order 1 and the Heisenberg algebra cannot be approximately realized following this construction. Before we enter this regime though, when $K$ is still sufficiently small, the probability of $\lambda_K$ making it has high up as $\lambda=2$ is exponentially small at large $N$. In fact it will be essentially equal to the probability of the largest eigenvalue exceeding $2$, which in the limit
\begin{align} \label{KNlargelimi}
 K,N \to \infty, \qquad \kappa \equiv \frac{K}{2N} \mbox{ fixed},
\end{align}
is given by the Tracy-Widom distribution \cite{Tracy:1995xi}. Applied to the specific (Wishart) matrix ensemble (\ref{WishartModel}) of interest the probability density for the largest eigenvalue $\lambda$ is given in this limit by \cite{Chiani}
\begin{align}
 P(\lambda) = f_1\left(\tfrac{2N \lambda - \mu}{\sigma}\right) \, , 
\end{align} 
where $f_1(x)$ is the $\beta=1$ Tracy-Wishart probability density function  and
\begin{align}
 \mu = \bigl((N-\tfrac{1}{2})^{\frac{1}{2}} + (K-\tfrac{1}{2})^{\frac{1}{2}} \bigr)^2 \, , \qquad
 \sigma = \mu^{\frac{1}{2}} \bigl((N-\tfrac{1}{2})^{-\frac{1}{2}} + (K-\tfrac{1}{2})^{-\frac{1}{2}} \bigr)^{\frac{1}{3}} \, .
\end{align}
The relevant asymptotics of $f_1(x)$ are the $x \to \infty$ asymptotics, which can be found e.g.\ in \cite{Dominguez}: $f_1(x) \sim e^{-\frac{2}{3} x^{3/2}}$. In the limit (\ref{KNlargelimi}), we thus get for the minimal Heisenberg error
\begin{align} \label{deltaminHER}
 \delta_{\rm min} \sim  P(\lambda \geq 2) \sim e^{-g(\kappa) N} \, , \qquad g(\kappa) = \frac{4}{3} \,\frac{\kappa^{\frac{1}{4}} \left(1-2\sqrt{\kappa}-\kappa \right)^{\frac{3}{2}}}{\left( 1 + \sqrt{\kappa} \right)^2} \, , \qquad \kappa = \frac{K}{2N} \, .
\end{align}
A plot of the coefficient $g(\kappa)$ is given in fig.\ \ref{fig:gkappa}. Note that $g(\kappa)$ is positive only on the interval $(0,3-2\sqrt{2}) \approx (0,0.171573)$. Beyond this range, $P(\lambda>2)$ becomes of order 1 in this asymptotic limit, due to eigenvalue repulsion effects, and the perturbative Heisenberg algebra error cannot be made small.  Physically, the reason is that in this regime, some fluctuation takes us out of the perturbative regime with probability essentially 1. 
\begin{figure} 
 \begin{center}
   \includegraphics[width=0.5\textwidth]{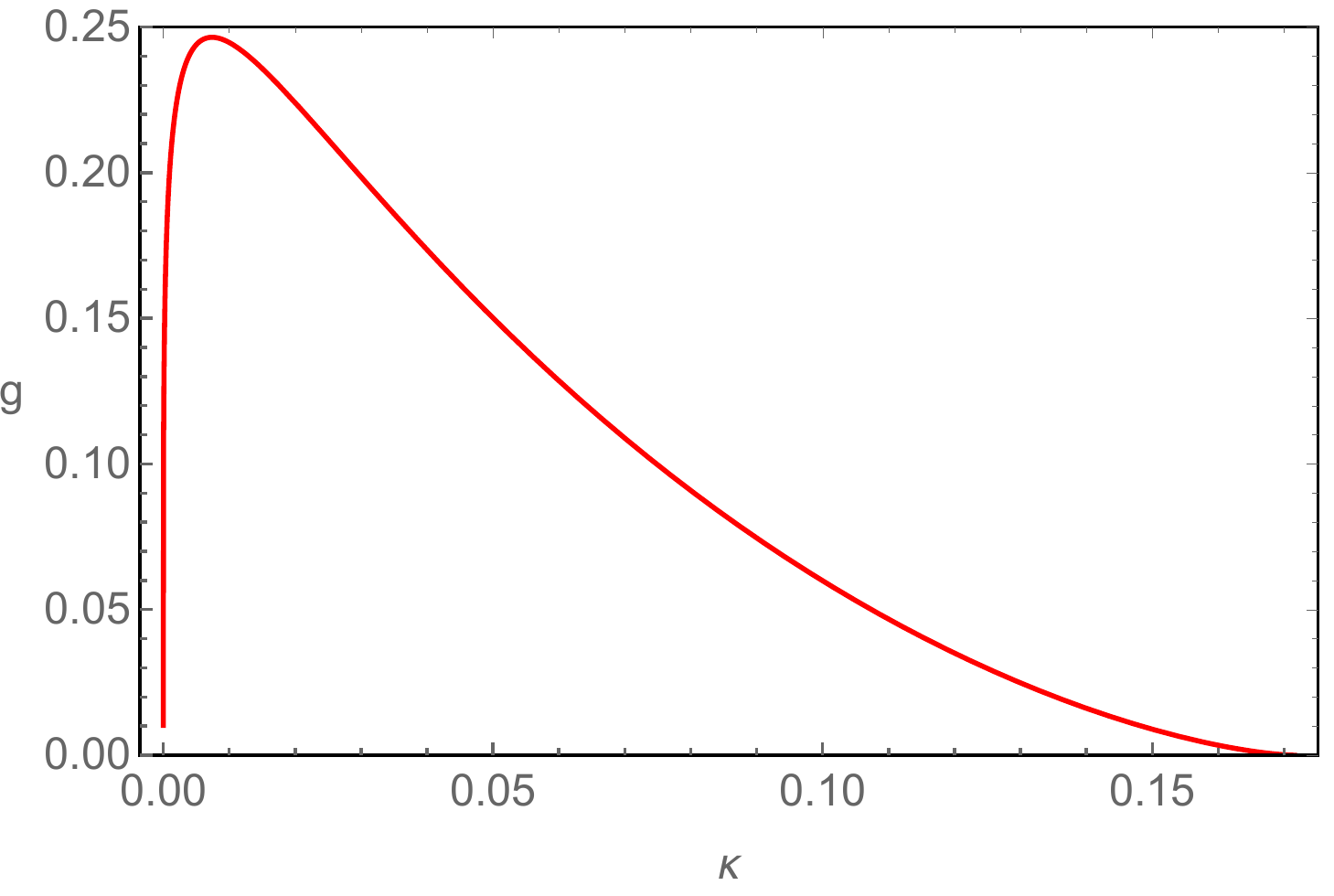}
 \end{center}
 \caption{Coefficient $g(\kappa)$ in $\delta_{\rm min} \sim e^{-g(\kappa) N}$, in limit $K,N \to \infty$, $\kappa = \frac{K}{2N}$ fixed. \label{fig:gkappa}}
\end{figure}

\subsubsection{$K>2N$ and continuum limit} \label{sec:contlimcoarsegr}

\begin{figure} 
 \begin{center}
   \includegraphics[width=\textwidth]{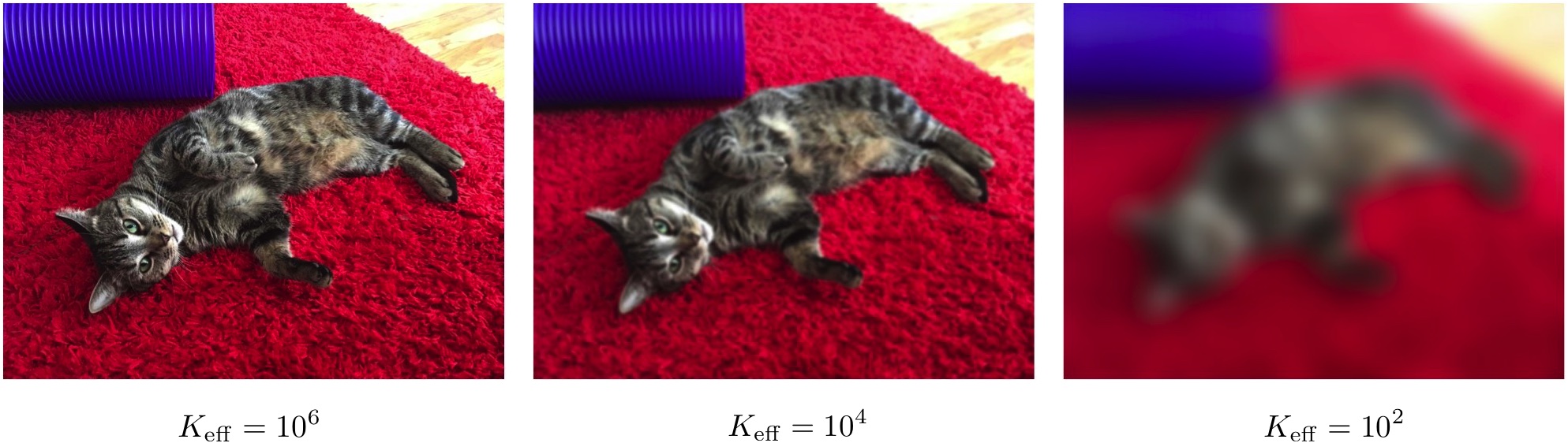}
 \end{center}
 \caption{Schr\"odinger's cat, coarse grained with a Gaussian kernel $W_\ell(x) \propto e^{-x^2/2\ell^2}$, for $\ell = 1,10,100$, in an area of size $V=10^7$. In two dimensions, this corresponds  to an effective number of pixels equal to $K_{\rm eff} = \frac{V}{4\pi\ell^2} \approx 10^6, 10^4, 10^2$. Realizing the perturbative QFT Heisenberg algebra at a resolution $K_{\rm eff}$ requires $N \gtrsim 3 K_{\rm eff}$ \label{fig:cats-lres}}
\end{figure}


It is clear already from the low order error terms (\ref{lowerordererror}) that the above construction of an approximate Heisenberg algebra breaks down when $K>2N$, so most definitely also in the continuum limit $K=\infty$. Our asymptotic analysis confirms this, and in fact indicated a breakdown well before this point, namely for $\frac{K}{2N} > 3-2\sqrt{2} \approx 0.171573$.   The basic reason for this was that when $K$ gets large, matrix eigenvalue repulsion effects push $B$ outside the radius of convergence of the perturbative expansion (\ref{Rseries}): the probability that $H=\frac{1}{N} Q Q^T$ has an eigenvalue larger than 2 becomes essentially equal to 1 in this regime. 

When $K>2N$, the impossibility of realizing the Heisenberg algebra also follows from a much simpler observation, namely the fact that $\Pi_{xy}$, the formal canonical conjugate to $\CH^{xy}$ defined in (\ref{Pixydef}), no longer exists, because $\CH=(\CD Q)(\CD Q)^T$ becomes non-invertible (as it has reduced rank $2N<K$). Similarly, in the continuum limit, one cannot expect to be able to construct the perturbative QFT Heisenberg algebra, because the full collection of perturbative higher spin fields yields an infinite number of degrees of freedom per spatial point, whereas the fundamental variables $Q$ only constitute $2N$ degrees of freedom per spatial point. This means in particular that the full collection of $QQ$-bilinears $B_I$ is not an independent set of variables, making it manifestly impossible to construct a Heisenberg algebra of the form (\ref{Heisa}).

The above observations suggest that we can at most hope to construct an approximate Heisenberg algebra for a much smaller set of degrees of freedom than naively suggested by the perturbative bulk QFT. Here we will consider an idea realizing this intuition. We will define a ``coarse grained'' version of the operators $A_I$, $B_I$, in such way that the {\it effective} number of coarse grained spatial ``pixels'' $K_{\rm eff}$ becomes less than the order $N$ bound suggested by the finite $K$ models, so we can effectively use the construction of section \ref{sec:pcctH}. The fundamental theory itself remains the same; only the operators we consider get coarse grained. To this end, we define coarse grained fundamental fields $\bar Q$ as follows: 
\begin{align}
 \bar Q_x^\alpha \equiv \int d^3 u \, W_\ell(u) \, Q_{x+u}^\alpha \, , \qquad W_\ell(u) \equiv \frac{1}{\ell^3} W(u/\ell) \, ,
\end{align} 
where $W(u)$ is a ``window'' function satisfying $\int d^3 u \, W(u) = 1$ (so $\int d^3u \, W_\ell(u) = 1$ as well). A convenient choice for practical computations is a Gaussian
\begin{align} \label{GaussianW}
 W(u) = \frac{1}{(2\pi)^{3/2}} \, e^{-u^2/2} \, .
\end{align}
Moreover we restrict the range of $x$ to some region of volume $V \gg \ell^3$. The coarse graining has the effect of reducing the effective spatial resolution to pixels of size $\sim \ell^3$, so we may expect to get an effective number of degrees of freedom of order $K_{\rm eff} \sim V/\ell^3$. The vacuum 2-point function of $\bar Q$ is
\begin{align}
 \langle \bar Q_x^\alpha \bar Q_y^\beta \rangle = \int d^3 u \int d^3 v \, W_\ell(u) W_\ell(v) \, \frac{\delta^{\alpha\beta}}{8\pi} \frac{1}{|x-y+u-v|} \, .
\end{align}
For the Gaussian window function (\ref{GaussianW}) this can be explicitly computed by first writing
\begin{align}
 \frac{1}{|x-y+u-v|} = \frac{1}{\sqrt{2\pi}} \int \frac{dt}{\sqrt{t}} \, e^{-\frac{1}{2} t(x-y+u-v)^2} \, ,
\end{align}
then performing all Gaussian integrals, and finally doing the integral over $t$. The result is \def\Erf{{\rm Erf}}
\begin{align}
 \langle \bar Q_x^\alpha \bar Q_y^\beta \rangle = \frac{1}{2} \,  g_{\ell, xy} \, \delta^{\alpha\beta} \, , \qquad g_{\ell, xy} \equiv \frac{\Erf\bigl(\frac{|x-y|}{2\ell}\bigr)}{4 \pi |x-y|} \, ,
\end{align}
where $\Erf(z)$ is the error function, $\Erf(z) = \frac{2}{\sqrt{\pi}} \int_0^z dx \, e^{-x^2}$, which satisfies $\Erf(z) \approx 1-\frac{e^{-z^2}}{\sqrt{\pi} z}$ for $z \gg 1$ and $\Erf(z) \approx \frac{2z}{\sqrt{\pi}}$ for $z \ll 1$. Thus, as was to be expected, the $\bar Q \bar Q$ 2-point function is essentially indistinguishable from the original $QQ$ 2-point function when $|x-y| \gg \ell$, while it gets regularized for $|x-y| \ll \ell$, capping off smoothly in the $x=y$ coincidence limit at a finite value $\frac{1}{4 \pi^{3/2} \ell}$. This is illustrated in fig.\ \ref{fig:coarsegr} on the left. 

\begin{figure} 
 \begin{center}
   \includegraphics[width=\textwidth]{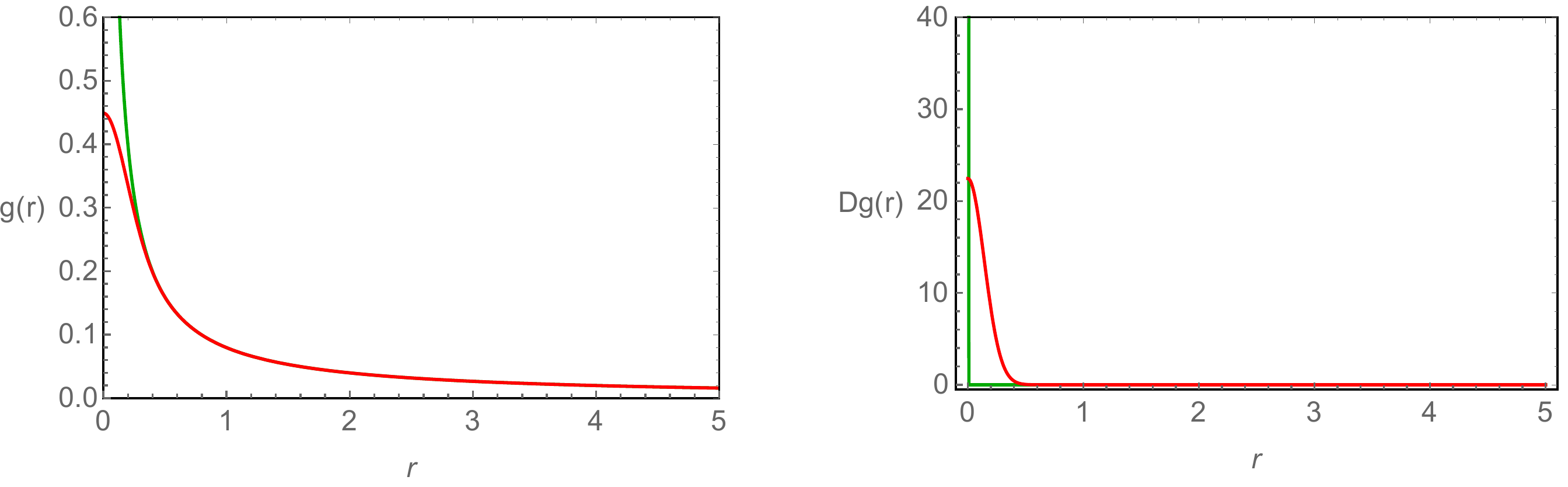}
 \end{center}
 \caption{Left: Coarse-grained two-point function $g_\ell(r)$ for $\ell=0.1$ (red) compared to original $g_0=\CD^{-1}$ (green). Right: $\delta_\ell = \CD g_\ell$ (red) vs $\delta = \delta_0 = \CD g_0$ (green).  \label{fig:coarsegr}}
\end{figure}

For the original 2-point function $(\CD^{-1})_{xy} = \frac{1}{4\pi|x-y|}$, we had the property $\CD^{xz}(\CD^{-1})_{zy} = \delta_y^x$. When acting with $\CD$ on the coarse-grained 2-point function on the other hand, we get a regularization of the delta function instead:
\begin{align} \label{regdelta}
 {\delta_{\ell\,}}_y^x \, \equiv \, \CD^{xz} g_{\ell,zy} = \frac{e^{-\frac{|x-y|^2}{4\ell^2}}}{8 \pi^{3/2} \ell^3} \, .
\end{align}
This is illustrated in fig.\ \ref{fig:coarsegr} on the right.
Taking the trace of this (i.e.\ the analog of computing $\delta^x_x = K = \infty$ in the non-coarse grained case) now gives
\begin{align}
  {\delta_{\ell\,}}_x^x  = \frac{V}{8 \pi^{3/2} \ell^3}  \equiv K_{\rm eff} \, ,
\end{align} 
which is finite as anticipated.
We next define coarse-grained versions of $B_{xy}$ and $B_I$ simply by replacing $Q$ everywhere by $\bar Q$, but keeping everything else unchanged:
\begin{align}
 \bar B_{xy} \equiv \frac{1}{N} \! :\! \bar Q_x^\alpha \bar Q_y^\alpha \! : \, , \qquad \bar B_I \equiv \Tr(\CD_I \bar B) \, . 
\end{align}
This renders formerly UV-divergent quantities finite. Recall for example that in the discretized, finite $K$ model, we had $\langle \Tr(\CD  B)^2 \rangle = \frac{K}{2N}(K+1)$, which diverges in the continuum limit $K=\infty$. In the coarse grained case we get instead
\begin{align}
 \langle \Tr(\CD \bar B)^2 \rangle &= \frac{1}{2N} \bigl( \CD^{xy} g_{\ell,yx} \CD^{x'y'} g_{\ell,y'x'} + \CD^{xy} g_{\ell,yy'} \CD^{y'x'} g_{\ell,x'x}  \bigr)  \nn \\
 &= \frac{1}{2N} \bigl( {\delta_{\ell\,}}_x^x \, {\delta_{\ell\,}}_{x'}^{x'} +  {\delta_{\ell\,}}_{y'}^x \, {\delta_{\ell\,}}_{y'}^{x}  \bigr)= \frac{K_{\rm eff}}{2N} \Bigl( K_{\rm eff} + \frac{1}{\sqrt{8}}  \Bigr) \approx  \frac{K_{\rm eff}^2}{2N} \, ,
\end{align}
where we used the fact that $V \gg \ell^3$, so $K_{\rm eff} \gg 1$. 

The vacuum 2-point functions of the bilinears $\bar B_I$ are
\begin{align}
 \langle \bar B_I \bar B_J \rangle = \frac{1}{N} \, \bar G_{IJ} \, , \qquad \bar G_{IJ} \equiv \Tr(g_\ell \CD_I g_\ell \CD_J) \, .
\end{align}
For example the scalar 2-point function is $\langle \bar B(x) \bar B(y) \rangle = \frac{1}{N} \bar G_{xy} = \frac{1}{N} g_{\ell,xy}^2$. Note that  $\bar G_{xy}$ is essentially indistinguishable from the original $G_{xy}=\frac{1}{(4\pi |x-y|)^2}$ if $|x-y| \gg \ell$, since in this regime $\bar G_{xy} = G_{xy} + O(e^{-|x-y|^2/4\ell^2})$. This remains true for the spin $s>0$ bilinears $B_I$ as well, as long as $s$ does not get too large. When $s$ increases, the range over which the deviation $\bar G_{IJ} - G_{IJ}$ is appreciably different from zero increases, due to the action of derivatives on the Gaussian window function. This is similar to the way in which the harmonic oscillator ground state wave function broadens into excited states when acting on it with derivatives. The effective width of the $s$-th excited state of the harmonic oscillator scales as $\sqrt{s}$, so we may expect the effective width to scale with the number of derivatives $s$ as $\sqrt
{s}$ in the case at hand as well. Hence we expect that for $I=(x,i_1,\ldots,i_s)$, $J=(y,j_1,\ldots,j_s)$, we have
\begin{align} \label{barGapproxG}
 \bar G_{IJ} = G_{IJ} + O(e^{-|x-y|^2/4\ell^2}) \quad \mbox{provided} \quad |x-y| \gg \sqrt{s} \, \ell \, .
\end{align} 
Since we necessarily have $|x-y| \lesssim L$ in the finite volume $V = L^3$ under consideration, we need $\sqrt{s} \, \ell \ll L$ for the above condition to be realizable, which puts a restriction on the range of spins for which the coarse grained setup can in any way be a good approximation to the original setup, to wit
\begin{align} \label{spinlimit}
 s \ll \frac{L^2}{\ell^2} \sim K_{\rm eff}^{2/3} \, .
\end{align} 
With this in mind, we are ready to repeat the construction of section \ref{sec:pcctH} to define approximate canonical conjugates $A_I$ to the $B_I$, at least within this range of spins. To this end, we define a coarse grained version of $P_x^\alpha$ defined in (\ref{freeapproxA0}),
\begin{align}
 \bar P_x^\alpha \equiv \int d^3 u \, W_\ell(u) \, P_{x+u}^\alpha \qquad \Rightarrow \qquad [\bar Q_x^\alpha,\bar P_y^\beta] = i \, \delta^{\alpha\beta} \, g_{\ell,xy} \, ,
\end{align}
and a coarse grained version of the $A_I^{(n)}$ defined in the paragraph below (\ref{AIQspace}) , 
\begin{align} \label{AIQspaceCG}
 \bar A_{I}^{(n)} \equiv \frac{1}{2N} \bigl( {{\bar R}_{x|n}}^z :\! \bar Q_z D_I^{xy} \bar P_y \! : + :\! \bar Q_z D_I^{xy} \bar P_y \! : {\bar R_{x|n}}^z \bigr)  \, , 
 \qquad  \bar R^{(n)} = \sum_{k=0}^n (-\bar B\CD)^k  \, .
\end{align}
Working out the commutator $[\bar B_I,\bar A_{J}^{(n)}]$, we now get a somewhat more complicated expression than (\ref{BIAJnGIJ}), because we no longer get exact cancelations in the sum over $k$:
\begin{align} \label{ABcommCG}
 [\bar B_I,\bar A_J] = \frac{2i}{N} \bigl( \bar G_{IJ} + (-1)^n \bar \delta_{IJ}^{(n)} + \Delta_{IJ}^{(n)} \bigr) \, , 
\end{align}
where
\begin{align}
 \bar\delta_{IJ}^{(n)} & =  \Tr\bigl( (\bar B \CD)^{n+1} g_\ell \CD_I g_\ell \CD_J \bigr) \\
 \Delta_{IJ}^{(n)} & = \sum_{k=0}^{n} (-1)^{k+1} \Tr\bigl[ (\bar B \CD)^{k+1} \bigl(g_\ell-g_0 \bigr) \CD_I g_\ell \CD_J \bigr] \, .
\end{align}
The first term in (\ref{ABcommCG}) reproduces the continuum Heisenberg algebra (\ref{Heisa}) to the extent that (\ref{barGapproxG}) is satisfied. Up to subleading terms in the large $K_{\rm eff}$ expansion, the vacuum expectation values of the error terms are
\begin{align}
 \bigl\langle \bar\delta_{IJ}^{(n)}  \bigr\rangle &\approx \Bigl( \frac{K_{\rm eff}}{2N} \Bigr)^n \, \bar G_{IJ} \, , \qquad \\
 \bigl\langle \bar\Delta_{IJ}^{(n)}  \bigr\rangle &\approx \sum_{k=0}^{n} (-1)^{k+1} \Bigl( \frac{K_{\rm eff}}{2N} \Bigr)^k  \Tr\bigl[\bigl(\delta_\ell - \delta_0 \bigr) \CD_I g_\ell \CD_J g_\ell \bigr] \, .
\end{align}
The first expression is just the coarse grained version of the analogous error term appearing in the finite $K$ model. The terms in the second expression are essentially zero in the regime (\ref{barGapproxG}). To see why, consider for example the $s=0$ case $\CD_I^{uv} = \delta_x^u \delta_x^v$, $\CD_J^{uv} = \delta_y^u \delta_y^v$. Then
\begin{align} \label{Trdelxy}
 \Tr\bigl[\bigl(\delta_\ell - \delta_0 \bigr) \CD_I g_\ell \CD_J g_\ell \bigr] = g_{\ell,xy} \int_z  g_{\ell,yz} (\delta_\ell - \delta_0)_x^z \, .
\end{align}
When $|x-y| \gg \ell$, we can put $g_{\ell,xy} \approx \frac{1}{4\pi|x-y|}$, $g_{\ell,yz} \approx \frac{1}{4\pi|y-z|}$, up to corrections of order $e^{-|x-y|^2/4\ell^2}$. But then the integral over $z$ equals the electrostatic potential at the point $y$ of a spherically symmetric charge density $\rho(z) = (\delta_\ell - \delta_0)_x^z$ centered at $x$, essentially concentrated within a region $|z-x| \lesssim O(\ell)$, with zero total charge. In other words it is the electrostatic potential of a neutral, spherically symmetric atom, at distances much larger than the radius of the atom. This vanishes, so (\ref{Trdelxy}) vanishes, up to terms of order $e^{-|x-y|^2/4\ell^2}$. Similar considerations hold for higher spins $s$, as long as (\ref{barGapproxG}) is satisfied, as well as for the $1/K_{\rm eff}$ corrections. We conclude that, at least when $n$ is not too large, the error term $\Delta_{IJ}$ is negligible in the regime of interest. Since the remaining $\delta_{IJ}$ error term is similar to the $\delta_{IJ}$ error term in the finite $K$ model, we may expect minimal error estimates similar to (\ref{deltaminHER}) to hold in the present case as well.  

Thus we conclude that in the regime (\ref{barGapproxG}), as long as
\begin{align} 
  K_{\rm eff} \sim \frac{V}{\ell^3} \lesssim N \, ,
\end{align}
we can realize the perturbative QFT Heisenberg algebra, up to errors of order $e^{-\tilde g(\kappa) N}$, where $\kappa = \frac{K_{\rm eff}}{2N}$ and $\tilde g(\kappa)$ is some O(1) function similar to $g(\kappa)$ in (\ref{deltaminHER}).

This is adequate for physical observables that can be computed accurately with these coarse-grained operators, such as suitable cosmological correlation functions at a coordinate distance scale $r$ such that $\ell \ll r \ll L$. The fundamental limitation in this construction is that the UV and IR cutoffs must be chosen such that the number of distinguishable spatial pixels $K_{\rm eff} \sim \frac{L^3}{\ell^3}$ is smaller than $N$. Moreover, by (\ref{spinlimit}), there is a bound on the spin $s$ that can be resolved in this way, requiring $s \ll N^{2/3}$. Although other constructions might be possible circumventing these limitations, this does hint at a  breakdown of perturbative bulk quantum field theory when trying to go beyond these bounds. Since the horizon entropy of Vasiliev de Sitter is of order $N$, this would be in line with expectation from the holographic principle (see \cite{Goheer:2002vf,Albrecht:2002xs,Banks:2003pt,ArkaniHamed:2007ky} for related discussions), although realized in a novel way. We leave a more thorough investigation of these observations to future work.

\section{Physical Hilbert space} \label{sec:GIaPHS}

As we already briefly discussed in section \ref{sec:conclandHphys}, the Hilbert space $\CCH$ of O(2N)-invariant wave functions $\psi(Q)$ with inner product $\langle \psi_1|\psi_2\rangle = \int dQ \, \psi_1(Q)^* \psi_2(Q)$ is not yet the physical Hilbert space $\CCH_{\rm phys}$. It should be viewed as the Hilbert space to which the perturbative bulk QFT Hilbert space $\CCH_{\rm Fock}$ is an approximation, while $\CCH_{\rm phys}$ should be viewed as the Hilbert space to which the physical perturbative bulk QFT Hilbert space $\CCH_{\rm Fock,\,phys}$ is an approximation. In both cases the difference between these Hilbert spaces lies in whether or not we require invariance under the group $\CG$ of residual spacetime gauge symmetries.

In this section we will take a closer look at these gauge symmetries and the construction of $\CCH_{\rm phys}$. We first point out that the choice of operator $\CD$ appearing in $\psi_0(Q)$ and in the Sp(N) model can be thought of as a partial gauge choice, and that this choice leaves a residual gauge group $\CG$ which can be thought of as the higher spin symmetry group, or a generalization thereof. This means that $\CCH_{\rm phys}$ can be defined, formally at least, as the $\CG$-invariant subspace of $\CCH$. We discuss some of the subtleties which quite generally arise when carrying out such a program in practice, and how they are  resolved in our framework.

\subsection{Gauge invariance}

For most of our concrete computations in the continuum limit, we took the operator $\CD$ appearing in $\psi_0(Q) \, \propto \, e^{-\frac{1}{2}\int Q\CD Q}$ to be the flat Laplacian on $\IR^3$, which corresponds to the future boundary of de Sitter in planar coordinates. In section \ref{sec:probandcorr}, we considered instead the conformal Laplacian on the round $S^3$, which corresponds to the future boundary of de Sitter in global coordinates.  More generally, upon performing asymptotic spacetime diffeomorphisms in the bulk, the late time metric could be taken to be any metric diffeomorphically and conformally equivalent to the original reference boundary metric, and the corresponding $\CD$ would be the conformal Laplacian for this metric.  
In the Sp(N) model living on the future boundary, these asymptotic spacetime diffeomorphisms are realized as spatial diffeomorphisms and local Weyl rescalings of the fundamental fields $\chi^a_x$. In the  $Q$-model, they are similarly realized as diffeomorphisms and Weyl rescalings of the $Q^\alpha_x$. 

In higher spin gravity there are infinitely many more bulk gauge transformations beyond diffeomorphisms, and therefore infinitely many more ways of transforming $\CD$ to a different but gauge-equivalent operator. All of these transformations still share the property that they act as linear field redefinitions $\chi_x \to {(R^{-1})_x}^y \chi_y$ on the fundamental fields in the Sp(N) model with action $S=\int \chi \CD \chi$, while mapping $\CD \to  \CD' = R^T \CD R$, keeping $S$  unchanged. 
Similarly, in the $Q$-model, mapping $Q \to R^{-1} Q$, $\CD \to \CD' = R^T \CD R$ keeps the wave function $\psi_0(Q) \propto e^{-\frac{1}{2} \int Q \CD Q}$ unchanged. Thus, (barring anomalies) vacuum correlation functions and other expectation values computed using $\CD$ or $\CD'=R^T \CD R$ are obtained from each other by a simple linear field redefinition $Q \to R^{-1} Q$, and operators $\CA(\CD,Q)$ invariant under gauge transformations $Q \to R^{-1} Q$, $\CD \to R^T \CD R$ have vacuum expectation values independent of the choice of $\CD$. In this sense, the choice of $\CD$ is tautologically a gauge choice. (From the bulk Vasiliev theory point of view, this can be thought of as a choice of the asymptotic form of the flat connection $W$.)

This choice does not completely fix the gauge, however. For example, there is a subgroup SO(1,4) of spatial diffeomorphisms and Weyl rescalings $R$ leaving the conformal Laplacian $\CD$ invariant, i.e.\ satisfying $R^T \CD R = \CD$. This is the conformal group from the point of view of the boundary, and the de Sitter isometry group from the point of view of the bulk. This group must be viewed as part of the gauge group of de Sitter quantum gravity even in perturbation theory \cite{Higuchi:1991tk,Higuchi:1991tm}; that is to say, physical states must be defined to be invariant under SO(1,4). Similarly, the higher spin extension of SO(1,4) is the subgroup $\CG$ of higher spin gauge transformations $R$ leaving $\CD$ invariant, i.e.\ maps $R$ satisfying $R^T \CD R = \CD$.  This condition defines the global higher spin symmetry group \cite{Eastwood:2002su,Mikhailov:2002bp,Joung:2014qya,Segal:2002gd}. The physical Hilbert space $\CCH_{\rm phys}$ is then defined as the subspace of $\CCH$ invariant under the residual gauge group $\CG$. 

Depending on restrictions one may wish to place on the set of admissible field redefinitions $R$, there may be subtleties in making the above discussions more precise. In conventional definitions of the higher spin symmetry group, one works at the level of the Lie algebra rather than at the level of finite group transformations, or equivalently at the level of infinitesimal group transformations $R=1+ \epsilon L$. The conventional (real) higher spin Lie algebra is then roughly speaking defined as the space of finite-order differential operators $L$ satisfying $L^T \CD + \CD L = 0$ \cite{Eastwood:2002su}. For example with $\CD=-\partial^2$, this is satisfied for translations $L=\partial_i$ as well as higher order $L=\partial_{i_1} \cdots \partial_{i_n}$ with $n$ odd, taking into account that $\partial_i^T = -\partial_i$. It is however a subtle matter whether these generators can be exponentiated to finite group transformations in a given representation of the algebra \cite{Monnier:2014tfa}. On the other hand, we can implement the residual gauge invariance constraints already at the level of the Lie algebra, by requiring physical states to be annihilated by the Lie algebra generators, so this does not affect our ability to define $\CCH_{\rm phys}$. Alternatively, one could drop the restriction to transformations $R$ generated by finite differential operators $L$, and simply allow ${\it all}$ continuous invertible linear transformations $R$, including nonlocal transformations \cite{Segal:2002gd}. In fact considering such nonlocal field redefinitions is quite natural in a dS context, since locality constraints arising from e.g.\ radial quantization of CFTs dual to AdS do not apply here. An example of such a nonlocal transformation (albeit not one that leaves the form of $\CD$ unchanged, so not part of $\CG$) is the Fourier transform, which is indeed often the preferred ``gauge'' for computation of cosmological correlation functions in dS, as illustrated for example in our computations in section \ref{sec:cosmocor}. 

All of this can be cast in the form of a formal gauge fixing procedure starting from an extended Hilbert space $\CCH_{\rm ext}$ of wave functions $\Psi(\CD,Q)$ with inner product $\langle \Psi_1|\Psi_2\rangle = \int [d\CD] (\det\CD)^N \int dQ \, \Psi_1(\CD,Q)^* \Psi_2(\CD,Q)$, where $[d\CD]$ is a  gauge invariant measure on the appropriate space of positive definite $\CD$, and the gauge transformation of the factor $(\det\CD)^N$ cancels the gauge transformation of the measure $dQ$. This inner product is divergent on gauge invariant wave functions such as $\Psi(\CD,Q)=e^{-\frac{1}{2} \int Q \CD Q}$ (already for discretized toy models with $K=1$, for which $[d\CD]=\frac{d\CD}{\CD}$), and must be gauge fixed. Gauge fixing is trivial in this setup, and consists simply of picking out a particular representative $\CD$. The gauge fixed inner product is then $\langle \psi_1|\psi_2\rangle = (\det \CD)^N \int dQ \, \psi_1(Q)^* \psi_2(Q)$, which reduces to our earlier definition upon absorbing a factor $(\det \CD)^{\frac{N}{2}}$ in the normalization of the wave functions $\psi(Q)$.

\def\rH{{\rm H}}
\def\CP{{\cal P}}

\subsection{Physical Hilbert space at finite $K$}

The construction of gauge invariant operators and states in $\CCH_{\rm phys}$ is most straightforward in  the finite-$K$ discretized models, for which $\psi_0(Q) \, \propto \, e^{-\frac{1}{2} \Tr(Q^T \CD Q)}$, with $\CD^{xy}$ a positive definite symmetric $K \times K$ matrix. The full gauge group is GL(K), acting as $Q \to R^{-1} Q$, $\CD \to R^T \CD R$, with the analog of the residual ``higher spin'' symmetry group being the subgroup $\CG$ satisfying $R^T\CD R = \CD$. The group $\CG$ is isomorphic to O(K). 

The operator algebra of $\CCH$ consists of linear combinations of O(2N)-invariant products of the operators $Q_x^\alpha$ and $P^x_\alpha = -i\partial_{Q_x^\alpha}$, such as $Q^\alpha_x Q^\alpha_y$, $P^x_\alpha P^y_\alpha$, $Q^\alpha_x P_\alpha^y + P^\alpha_y Q^\alpha_x$ and products thereof. The states of $\CCH$ are generated by acting with these operators on the vacuum state $|0\rangle$ with wave function $\psi_0(Q)$. Making use of $[Q_x^\alpha,P^y_\beta]=i \delta^\alpha_\beta \delta_x^y$ and $P^x_\alpha |0\rangle = i \CD^{xy} Q_y^\alpha |0\rangle$, it is clear that all states can be produced by acting with functions $f(\rH)$ of the O(2N)-invariant $QQ$ bilinears 
\begin{align} \label{hxydef}
  \rH_{xy} \equiv \frac{1}{\sqrt{N}} \,Q_x^\alpha Q_y^\alpha  
\end{align}
on $|0\rangle$, and that the computation of vacuum expectation values of arbitrary operators on $\CCH$ can be reduced to the computation of expectation values of operators of the form $f(h)$. Note that the normalization of $\rH_{xy}$ differs by a factor of $\sqrt{N}$ from that of $H_{xy}$ defined in (\ref{Hxydefi}).
 It is convenient to consider normal-ordered operators such as $:\! \rH_{x_1 y_1} \rH_{x_2 y_2} \rH_{x_3 y_3} \!:$, which are defined as usual by removing all vacuum self-contractions, or equivalently by expressing the operators $Q_x^\alpha$ in terms of creation and annihilation operators and moving all annihilation operators to the right. (Here the relevant annihilation operator is $a_x^\alpha=\frac{1}{\sqrt{2}}(Q_x^\alpha + i P_x^\alpha)$, where $P_x^\alpha \equiv (\CD^{-1})_{xy} P^y_\alpha$, since this satisfies $a_x^\alpha|0\rangle = 0$.) Then we may define ``$n$-particle'' states in $\CCH$ as
\begin{align} \label{HHHbasis}
 |\rH_{x_1 x_1'} \cdots \rH_{x_n x_n'} \rangle \, \equiv \, \, :\! \rH_{x_1 x_1'} \cdots \rH_{x_n x_n'} \! : |0\rangle \, .
\end{align}
The normal ordering ensures that states with different $n$ are automatically orthogonal to each other. With these definitions, we have for example, denoting $g_{xy} \equiv (\CD^{-1})_{xy}$,
\begin{align} \label{HHinenenen}
 \langle \rH_{x x'}|\rH_{yy'} \rangle = \frac{1}{2} \bigl( g_{x y} g_{x' y'} + g_{x y'} g_{x' y} \bigr) \, .
\end{align}
The normalization in (\ref{hxydef}) was chosen such that there is no $N$-dependence in this $n=1$ expression, and such that for general $n$, the leading term at large $N$ is of order $N^0$. 

The operator algebra of $\CCH_{\rm phys}$ consists of combinations of $Q_x^\alpha$ and $P^x_\alpha$ invariant under both O(2N) and $\CG$. For the construction of states in $\CCH_{\rm phys}$, it suffices to consider $\CG$-invariant operators $f(\rH)$ acting on $|0\rangle$. Such operators are generated by traces: denoting $Q^x_\alpha \equiv \CD^{xy} Q^\alpha_y$, let us define
\begin{align} \label{Tndefinition}
 T_n \, \equiv \,\, {\frac{1}{K^{\frac{n}{2}}}} \Tr (\rH \CD)^n  \, = {\frac{1}{(NK)^{\frac{n}{2}}}} \, Q_{x_1}^{\alpha_1} Q_{\alpha_1}^{x_2} \, Q_{x_2}^{\alpha_2} Q_{\alpha_2}^{x_3} \cdots Q_{x_n}^{\alpha_n} Q_{\alpha_n}^{x_1} \, .
\end{align}
$\CG$-invariant states are then produced by acting with normal-ordered products of such operators on the vacuum $|0\rangle$:
\begin{align} \label{TTTTstate}
 |T_{n_1} \cdots T_{n_k} \rangle \, \equiv \, :\! T_{n_1} \cdots T_{n_k} \! : |0\rangle \, .
\end{align}
This state can be interpreted as a gauge-invariant $n$-particle state, where $n=n_1 + \cdots + n_k$. For low $n$, it is easy to construct a basis of invariant $n$-particle states in $\CCH_{\rm phys}$:
\begin{align}
 n=0: & \quad |0\rangle \nn \\
 n=1: & \quad |T_1\rangle \nn \\
 n=2: & \quad |T_2\rangle \, , \quad |T_1^2\rangle \nn \\
 n=3: & \quad |T_3\rangle \, , \quad |T_2 T_1 \rangle \, , \quad |T_1^3 \rangle \nn \\
 n=4: & \quad |T_4\rangle \, , \quad |T_3 T_1 \rangle \, , \quad |T_2^2 \rangle \, , \quad |T_2 T_1^2 \rangle \, , \quad |T_1^4\rangle \, .  \label{basisdef}
\end{align}
Sectors with different $n$ are orthogonal due to the normal ordering in (\ref{TTTTstate}), and the inner products for a given $n$ are readily computed:
\begin{align}
 &\langle T_1|T_1\rangle = 1 \label{T1T1prod} \\
 &\langle T_2|T_2\rangle = 1 + \tfrac{1}{K} + \tfrac{1}{2N} + \tfrac{3}{2KN} \, , \quad \langle T_1^2|T_1^2\rangle = 2 + \tfrac{2}{KN} \, , \quad \langle T_2|T_1^2\rangle = \tfrac{1}{N} + \tfrac{2}{K} + \tfrac{1}{KN} \, . \label{T2T2prod}
\end{align}
The normalization in (\ref{Tndefinition}) was chosen such that these remain order 1 in the large-$N$ or large-$K$ limits. Note that these inner products are independent of $\CD$, consistent with our earlier observation that the choice of $\CD$ is a gauge choice.

Denoting this basis of $\CCH_{\rm phys}$ by $|r\rangle$, $r=0,1,2,3,\ldots$, i.e.\ $|0\rangle=|0\rangle$, $|1\rangle = |T_1\rangle$, $|2\rangle = |T_2\rangle$, $|3 \rangle = |T_1^2\rangle$ and so on, and defining
\begin{align}
 C_{rs} \, \equiv \langle r|s\rangle  \, ,
\end{align}
we can write the decomposition of unity of $\CCH_{\rm phys}$ as
\begin{align} \label{physproj}
 1_{\rm phys} = \sum_{rs} (C^{-1})^{rs} |r\rangle \langle s| \, .
\end{align}
Interpreted as an operator on $\CH$, $1_{\rm phys}$ acts as a projection operator onto $\CH_{\rm phys}$. We can use this to map arbitrary states $|\psi_1\rangle$ in $\CCH$ to $\CG$-invariant states $|\psi)$ in $\CCH_{\rm phys}$:
\begin{align} \label{physprojection}
 |\psi) \, \equiv 1_{\rm phys} |\psi\rangle = \sum_{rs} (C^{-1})^{rs} |r\rangle \langle s| \psi \rangle \, .
\end{align}
The inner product between two such projected states is
\begin{align} \label{projinnerprod}
 (\psi_1|\psi_2) = \langle \psi_1|1_{\rm phys}|\psi_2\rangle = \sum_{rs} (C^{-1})^{rs} \langle \psi_1|r\rangle \langle s| \psi_2 \rangle \, .
\end{align}
One can alternatively think of the projection operator $1_{\rm phys}$ as being the result of a ``group averaging'' procedure over the residual gauge group $\CG={\rm O(K)}$:
\begin{align}
 1_{\rm phys} = \int_{\CG} [dg] \, U(g) \, ,
\end{align}
where $U(g)$ denotes the unitary representation of $\CG$ on $\CCH$, and $[dg]$ is the Haar measure normalized such that $\int_{\CG} [dg] = 1$. Thus the projected inner product (\ref{projinnerprod}) is the finite-$K$ analog of the ``group averaged inner product'' for perturbative quantum gravity in de Sitter space as defined in \cite{Higuchi:1991tm} (see also \cite{Giddings:2007nu,Marolf:2008hg} and references therein). We avoid the need for any explicit integrations here due to the fact that we know in advance the complete list of invariant states at each level $n$, and we avoid the complications of having a noncompact group (at finite $K$) because $\CG = {\rm O(K)}$ is compact. We will discuss the continuum limit in section \ref{sec:Hphyscontlim}

To compute the $\CCH_{\rm phys}$-projected/group-averaged inner product (\ref{projinnerprod}) explicitly for the basis elements of $\CH$ defined in (\ref{HHHbasis}), we need to compute overlaps of these with the invariant basis of $\CH_{\rm phys}$. This is straightforward but somewhat tedious (though easily automated in Mathematica). For example,   
\begin{align} \label{HHTToverlaps}
 \langle \rH_{x x'} | T_1 \rangle &= \frac{1}{\sqrt{K}} \, g_{xx'} \nn \\
  \langle \rH_{x x'} \rH_{y y'} | T_1^2 \rangle &= \frac{1}{K} 
 \Bigl( 2 g_{xx'} g_{yy'} + \frac{1}{N} \bigl(g_{xx'} g_{yy'} + g_{xy} g_{x'y'} \bigr) \Bigr) \nn \\
 \langle \rH_{x x'} \rH_{y y'} | T_2 \rangle &= \frac{1}{K} 
 \Bigl( g_{xy} g_{x'y'} + g_{xy'} g_{yx'} + \frac{1}{2N} \bigl( g_{xy'} g_{yx'} + g_{xy} g_{x'y'} + 2 g_{xx'} g_{yy'} \bigr) \Bigr) \, .
\end{align}
Thus, 
\begin{align}
 (\rH_{xx'}|\rH_{yy'}) &= \langle\rH_{xx'}|1_{\rm phys}|\rH_{yy'}\rangle = \frac{1}{K} \, g_{xx'} g_{yy'} \\
 ( \rH_{..} \rH_{..}|\rH_{..} \rH_{..}) &= \langle \rH_{..} \rH_{..}|1_{\rm phys}|\rH_{..} \rH_{..}\rangle = \frac{1}{K^2} \bigl( g_{..} g_{..} g_{..} g_{..} + \tfrac{\cdots}{N} + \tfrac{\cdots}{K} \bigr) \, ,
\end{align}
where the second expression is schematic and meant to convey the nature of the $K$- and $N$-dependence. Note that the structure of these $\CCH_{\rm phys}$-projected  inner products are very different from the original inner products on $\CCH$ such as (\ref{HHinenenen}). 

Using these results, we can construct group-averaged states of unit norm; for example
\begin{align} \label{normalizedsingleparticlestate}
 |xx') \equiv \frac{\sqrt{K}}{g_{xx'}} \, 1_{\rm phys} |\rH_{xx'}\rangle  \qquad \Rightarrow \qquad (xx'|xx') = 1 \, .
\end{align}
Then we have more generally also $(xx'|yy') = 1$, for any choice of $x,x',y,y'$, expressing the fact that all single-particle states in $\CCH$ are gauge equivalent to each other in this theory. 

\subsection{Reduction to $2N \times 2N$ matrix model when $K \geq 2N$}

From the above it is clear that the computation of any gauge invariant quantity, e.g.\ overlaps of invariant states such as (\ref{TTTTstate}) or vacuum expectation values of gauge invariant operators, can be reduced to the computation of the vacuum expectation value of products of trace operators $T_n \, \propto \, \Tr(Q\CD Q^T)^n$ as defined in (\ref{Tndefinition}). Note that since the final result will be independent of $\CD$, we might as well set $\CD \equiv 1$ from the start. Thus, all gauge invariant quantities can be reduced to the computation of  Gaussian $K \times 2N$ matrix model integrals of the form
\begin{align} \label{Ik1k2k3}
 I_{k_1,k_2,k_3,\ldots}  = \int dQ \, e^{-\Tr(Q^T Q)} \, \bigl(\Tr(QQ^T) \bigr)^{k_1} \bigl(\Tr(QQ^T)^2 \bigr)^{k_2} \bigl(\Tr(QQ^T)^3 \bigr)^{k_3} \cdots   \, .
\end{align}
Here and below the normalization of the integration measure is chosen such that $I_{0,0,0,\ldots} = 1$.

When $K \leq 2N$, we can follow the reasoning of section \ref{sec:equivforKlessN} and change variables to the $K \times K$ symmetric positive definite matrix $H \equiv Q Q^T$ to rewrite this in terms of a Wishart matrix model,
\begin{align}
 I_{k_1,k_2,k_3,\ldots}  = \int [dH] \, (\det H)^N \, e^{-\Tr H} \, (\Tr H)^{k_1} (\Tr H^2)^{k_2} (\Tr H^3)^{k_3} \cdots \, ,
\end{align}
where $[dH]$ is the GL(K)-invariant measure $[dH]=\frac{dH}{(\det H)^{(K+1)/2}}$.

When $K>2N$, the $H$-matrix model becomes singular. However, in this case, we can follow a similar reasoning exchanging $K \leftrightarrow 2N$, based on the simple observation  
\begin{align} \label{MQQTdef}
 \Tr H^n = \Tr(QQ^T)^n = \Tr(Q^T Q)^n = \Tr M^n \, , \qquad M \equiv Q^T Q \, .
\end{align}
The point to note here is that $H=QQ^T$ is a $K \times K$ matrix, while $M = Q^T Q$ is a $(2N) \times (2N)$ matrix. Thus, when $K \geq 2N$ we can instead change variables to $M$ and repeat the same reasoning to rewrite (\ref{Ik1k2k3}) in terms of a $2N \times 2N$ Wishart matrix model:
\begin{align} \label{WishartMmodel}
 I_{k_1,k_2,k_3,\ldots}  = \int [dM] \, (\det M)^\frac{K}{2} \, e^{-\Tr M} \, (\Tr M)^{k_1} (\Tr M^2)^{k_2} (\Tr M^3)^{k_3} \cdots \, ,
\end{align}
where $[dM]$ is the GL(2N)-invariant measure $[dM] = \frac{dM}{(\det M)^{(2N+1)/2}}$. The O(2N)-symmetry of this matrix model is gauged, so the physical degrees of freedom are reduced in this way to the $2N$ eigenvalues of $M$. 

In particular, the $2N \times 2N$ matrix model computes all gauge invariant quantities in the limit $K \to \infty$ with $N$ fixed. In this limit, the above integral for finite values of the $k_i$ can be computed in a Gaussian saddle point approximation. To this end we write $(\det M)^\frac{K}{2} e^{-\Tr M} = e^{-F(M)}$ where $F(M)=-\frac{K}{2} \Tr \log M + \Tr M$. The saddle point equation $\partial_M F(M) = 0$ is solved by $M=\frac{K}{2} \, 1$. Gaussian fluctuations scale as $\delta M \sim \sqrt{K}$, so we expand $M=\frac{K}{2} + \sqrt{\frac{K}{2}} \, m$ to obtain
\begin{align}
 F(M) = NK(1- \log \tfrac{K}{2}) + \frac{1}{2} \Tr m^2  + O(\tfrac{1}{\sqrt{K}} m^3) \, ,
\end{align}
which in the limit $K \to \infty$ reduces the computation of any finite linear combination of the integrals $I_{k_1,k_2,k_3,\ldots}$ to a $2N \times 2N$ real symmetric Gaussian matrix integral. To get such finite linear combinations in the limit $K\to \infty$, it is more convenient to consider normal-ordered trace operators normalized with a factor $\frac{1}{\sqrt{K}}$ like the $T_n$ we defined in (\ref{Tndefinition}). For example $:\! T_1 \!: = T_1 -  \sqrt{NK} = \frac{1}{\sqrt{NK}} \Tr (M - \frac{K}{2}) = \frac{1}{\sqrt{2N}} \Tr m$, and
\begin{align}
 \bigl\langle (:\! T_1 \!:)^k \bigr\rangle = \frac{1}{(2N)^{k/2}} \int dm \, e^{-\frac{1}{2} \Tr m^2} \, (\Tr m)^k \, \qquad (K \to \infty) \, ,
\end{align}
where $dm$ is the flat measure on the space $2N \times 2N$ real symmetric matrices, normalized such that the above integral equals 1 for $k=0$. As a check, note that according to this formula, $\bigl\langle (:\! T_1 \!:)^2 \bigr\rangle = \frac{1}{2N} \partial_\lambda^2 \int dm \, e^{-\frac{1}{2} \Tr m^2 + \lambda \Tr m}|_{\lambda=0} = \frac{1}{2N} \partial_\lambda^2 e^{N \lambda^2}|_{\lambda=0} = 1$ in agreement with (\ref{T1T1prod}).

\subsection{Physical Hilbert space in continuum limit} \label{sec:Hphyscontlim}

In this section we will consider the construction of $\CCH_{\rm phys}$ in the continuum limit. We will define the continuum limit here as the limit $K \to \infty$ with $\CD$ limiting to, say, the Laplacian on flat $\IR^3$, or the conformal Laplacian on the round $S^3$. That is to say, we think of the finite $K$ model as a lattice version of the continuum theory with $K$ lattice points, and $\CD$ the discrete Laplacian on the lattice. The key question then is whether we obtain finite quantities in the continuum limit $K \to \infty$. In QFT, this typically involves delicate renormalization prescriptions. As we will see, the theory under consideration behaves in a much simpler way: once we impose standard normalization of states, all UV divergences disappear. This is essentially due to the fact that there exist only a finite number of $n$-particle states for each $n$. 

As we have seen above, it is certainly possible to define a basis of $\CCH_{\rm phys}$ of invariant states such as (\ref{basisdef}) whose inner products remain finite in the limit $K \to \infty$. For example from (\ref{T1T1prod})-(\ref{T2T2prod}) we get in this limit
\begin{align}
 \langle T_1|T_1\rangle = 1 \, , \qquad 
 \langle T_2|T_2\rangle = 1 + \frac{1}{2N}  \, , \quad \langle T_1^2|T_1^2\rangle = 2  \, , \quad \langle T_2|T_1^2\rangle = \frac{1}{N}  \, . \label{T2T2prodcont}
\end{align}
The physical interpretation of these quantities is obscure, however. The situation is somewhat better for invariant states $|\psi) = 1_{\rm phys} |\psi\rangle$ obtained by group averaging of non-invariant states $|\psi\rangle$, or equivalently projection onto $\CCH_{\rm phys}$, as defined in (\ref{physprojection}). For example if we take $|\psi\rangle$ to correspond to some ordinary $n$-particle state in $\CCH$, then we can think of as $|\psi)$ as the state symmetrized with respect to the residual gauge symmetry group $\CG$. The group $\CG$ includes for example rotations of the $S^3$, so the group-average includes averaging over the center of mass position of the $n$ particles on the sphere as well as their orientation, expressing there is no invariant physical meaning to the absolute position and orientation of $n$ particles on an $S^3$ spatial slice of de Sitter space. 

If we use properly normalized group-averaged states, i.e.\ unit normalized states such as (\ref{normalizedsingleparticlestate}), factors of $K$ cancel out and inner products between such states are finite even in the limit $K \to \infty$. To illustrate this, let us work out the inner products of group averaged 2-particle states to leading order at large $N$. In this limit, we have $\langle T_2|T_2\rangle = 1$, $\langle T_1^2|T_1^2 \rangle = 2$, $\langle T_2|T_1^2\rangle = 0$, hence, using (\ref{HHTToverlaps}) also to leading order at large $N$,
\begin{align}
 ( \rH_{xx'} \rH_{yy'} | \rH_{vv'} \rH_{ww'} ) &=  \langle \rH_{xx'} \rH_{yy'} |T_2\rangle \langle T_2| \rH_{vv'} \rH_{ww'} \rangle + \frac{1}{2} \langle \rH_{xx'} \rH_{yy'} |T_1^2\rangle \langle T_1^2| \rH_{vv'} \rH_{ww'} \rangle \nn \\
 &= \frac{1}{K^2} \bigl( 4 \,
   g_{x(y} g_{y')x'} \, g_{v(w} g_{w')v'}   + 2 \,  g_{xx'} g_{yy'} g_{vv'} g_{ww'} 
 \bigr), 
\end{align}
where $g_{xy} = \frac{1}{4 \pi |x-y|}$ in the continuum limit.
The corresponding unit-normalized states are
\begin{align}
 |xx',yy') \equiv \frac{K}{\sqrt{ 
   4  (g_{x(y} g_{y')x'})^2 + 2  (g_{xx'} g_{yy'})^2}} \, |\rH_{xx'} \rH_{yy'} ) \, .
\end{align}
The inner product between two unit-normalized states is 
\begin{align}
 (xx',yy'|vv',ww') = \frac{2 \, g_{x(y} g_{y')x'} \, g_{v(w} g_{w')v'} +  g_{xx'} g_{yy'} g_{vv'} g_{ww'}}{\sqrt{ 
   2(g_{x(y} g_{y')x'})^2 + (g_{xx'} g_{yy'})^2}
   \sqrt{ 2(g_{v(w} g_{w')v'})^2 +  (g_{vv'} g_{ww'})^2}} \, , 
\end{align}
which is independent of $K$ and thus remains finite in the continuum limit. To compute more specifically the inner product of unit-normalized group-averaged states corresponding to scalar field insertions $|\tilde\beta(x) \tilde\beta(y)\rangle$ and $|\tilde\beta(v) \tilde\beta(w) \rangle$ where $\tilde\beta(x)$ is the  shadow transform of the scalar field mode $\beta(x)$, we need to consider the coincident point limit $x'=x$, $y'=y$, $v'=v$, $w'=w$:
\begin{align} \label{xxyyvvww}
 (xx,yy|vv,ww) = \frac{2 g_{xy}^2 g_{vw}^2 + g_{xx} g_{yy} g_{vv} g_{ww}}{\sqrt{ 
   2 g_{xy}^4 +  g_{xx}^2 g_{yy}^2}
   \sqrt{ 2 g_{vw}^4 +  g_{vv}^2 g_{ww}^2}} \, .
\end{align}
Taking into account that $\lim_{x' \to x} g_{xx'} = \lim_{x' \to x} \frac{1}{4\pi|x'-x|}$ diverges in the continuum limit, this actually collapses to
\begin{align}
 (xx,yy|vv,ww) = 1 \, , 
\end{align}
indicating that all 2-particle scalar states are gauge equivalent. It can be checked that this final result remains the same at finite $N$. 

Defining $\rH_I \equiv \CD_I^{xy} \rH_{xy}$, using the notation originally introduced in (\ref{COIdef}), we can similarly construct 2-particle states $|\rH_I \rH_J\rangle$ with particles of arbitrary even spin, their group-averaged counterparts $|\rH_I \rH_J)$ and their unit-normalized versions which we  denote by $|IJ)$. To leading order at large $N$, the overlap between such states is 
\begin{align} \label{IJKLIP}
 (IJ|KL) = \frac{2 G_{IJ} G_{KL} + G_I G_J G_K G_L}{\sqrt{2 G_{IJ}^2 + G_I^2 G_J^2} \sqrt{2 G_{KL}^2 + G_K^2 G_L^2} } \, ,
\end{align}
where $G_{IJ} = \Tr(\CD^{-1} \CD_I \CD^{-1} \CD_J)$, $G_I = \Tr(\CD^{-1} \CD_I)$. This reproduces (\ref{xxyyvvww}) if $I,J,K,L$ all label scalars. Note that the $G_I$ are position-independent but UV-sensitive quantities. For spin $s=0$, $G_x=(\CD^{-1})_{xx}=g_{xx} \equiv \Lambda$, which diverges in the continuum limit. By adding the appropriate contact terms (i.e.\ terms involving factors of the Laplacian $\CD$, which do not affect the 2-point function $G_{IJ}$ for non-coincident points in the continuum limit), one can pick the $\CD_I$ such that for $s>0$, we have $G_I=0$. Thus, in general, we  set
\begin{align}
 G_I = \Lambda \delta_{s_I,0} \, , \qquad \Lambda \to \infty \, ,
\end{align}
where $s_I$ is the spin of the single-particle state labeled by $I$. If both $|IJ)$ and $|KL)$ have at least one higher-spin particle, then (\ref{IJKLIP}) reduces to $(IJ|KL) = \frac{G_{IJ} G_{KL}}{|G_{IJ}| |G_{KL}|} = \pm 1$, while if $|IJ)$ has at least one higher spin particle while $|KL)$ has two spin-0 particles, it reduces to $(IJ|KL)=\frac{2 G_{IJ} G_{KL}}{\sqrt{2}|G_{IJ}| \Lambda^2} = 0$ in the continuum limit. Thus, consistent with the fact that the physical Hilbert space of two-particle states is two-dimensional, all inner products of this form collapse to either 0 or $\pm 1$, depending on the spin content of the states:
\begin{align}
 (IJ|KL) = \left\{ 
 \begin{array}{ll}
   1 & \mbox{if }  s_I + s_J = 0 \mbox{ and } s_K + s_L = 0 \\
   \pm 1 & \mbox{if } s_I + s_J > 0 \mbox{ and } s_K + s_L > 0 \\
   0 & \mbox{otherwise} 
 \end{array}
 \right.
\end{align}
A similar but longer computation shows that at finite $N$, (\ref{IJKLIP}) becomes
\begin{align}
 (IJ|KL) = \frac{(2\!+\!\frac{1}{N}) G_{IJ} G_{KL} + G_I G_J G_K G_L + \frac{1}{N}(G_I G_J G_{KL} + G_{IJ} G_K G_L)}{\sqrt{(2\!+\!\frac{1}{N}) G_{IJ}^2 + G_I^2 G_J^2 + \frac{2}{N} G_I G_J G_{IJ}} \, \sqrt{(2\!+\!\frac{1}{N}) G_{KL}^2 + G_K^2 G_L^2 + \frac{2}{N} G_K G_L G_{KL}}  } \, . \nn
\end{align}
At finite $N$ there is some mixing between the two sectors which we identified above as orthogonal in the $N \to \infty$ limit: if $|IJ)$ has two spin-0 particles and $|KL)$ has at least one higher spin particle, we now get 
\begin{align}
 (IJ|KL) &= \lim_{\Lambda \to \infty} \frac{(2\!+\!\frac{1}{N}) G_{IJ} G_{KL} + \frac{1}{N} \Lambda^2 G_{KL}}{\sqrt{(2\!+\!\frac{1}{N}) G_{IJ}^2 + \Lambda^4 + \frac{2}{N} \Lambda^2 G_{IJ}} \, \sqrt{(2\!+\!\frac{1}{N}) G_{KL}^2}} \, = \, \pm \frac{1}{N \sqrt{2\!+\!\frac{1}{N}}} \, ,
\end{align}
where $\pm$ is the sign of $G_{KL}$. 

Finally, we can likewise construct physical operators $A_{\rm phys}$ acting on $\CCH_{\rm phys}$ from operators $A$ acting on $\CCH$ by defining 
\begin{align}
 A_{\rm phys} \equiv Z \, 1_{\rm phys} \, A \, 1_{\rm phys} \, ,
\end{align}
with $1_{\rm phys}$ again the projector (\ref{physproj}), and $Z$ an appropriate operator renormalization factor. Vacuum probability distributions $P(A_{\rm phys})$ for the physical observable described by this operator can be inferred e.g.\ from its moments
\begin{align}
 \langle 0|A_{\rm phys}^n|0\rangle = Z^n \, \langle 0|A \, 1_{\rm phys} \, A \, 1_{\rm phys} \, A \, 1_{\rm phys} \, \cdots \, A |0\rangle \, .
\end{align}
In general $P(A)$ computed on $\CCH$ will not coincide with $P(A_{\rm phys})$, of course.
Thus one could for example revisit the vacuum probability distribution $P(b^0)$ of the constant scalar deformation $b^0$, which we computed on the Hilbert space $\CCH$ in section \ref{sec:constscalarQspace}, and compute now instead the vacuum probability distribution $P(b^0_{\rm phys})$ on $\CCH_{\rm phys}$. We  leave this and other explorations of gauge-invariant observables to future work.

\subsection{Conclusions}

In this section we constructed the physical Hilbert space $\CCH_{\rm phys}$ as the subspace of $\CCH$ invariant under residual gauge transformations $\CG$, and started exploring its structure. We defined the continuum limit by first considering finite-$K$ models with residual gauge group $\CG = {\rm O(K)}$ and then taking the limit $K \to \infty$, in which $\CG$ becomes a version of the higher spin symmetry group. This allowed for a precise definition and computation of group-averaged states and their inner products. The resulting $\CCH_{\rm phys}$ has a topological flavor: sectors with definite ``particle number'' $n$ only have a finite number of physically distinct states, and all gauge invariant quantities are  computed by a $2N \times 2N$ matrix model. In particular this effectively eliminates the usual UV divergences appearing in the continuum limit of $\CCH$. Although we did not make this precise, it also suggests that the need for coarse graining to construct a perturbative Heisenberg algebra on $\CCH$, as discussed in section \ref{sec:contlimcoarsegr}, actually does not amount to a real loss of resolution from the point of view of $\CCH_{\rm phys}$, given the even greater coarseness of $\CCH_{\rm phys}$ at least at finite $n$. The full Hilbert space $\CCH_{\rm phys}$ of higher spin de Sitter space remains infinite-dimensional, because there is no bound on $n$. However it seems quite plausible to us that a suitable reduced density matrix obtained from the vacuum state, with a bulk interpretation of being the density matrix of a ``local observer'', will nevertheless have a finite entropy. Identifying this density matrix and computing its entropy would constitute a microscopic derivation of the de Sitter entropy in this framework.


\section{Outlook}

We have provided a precise microscopic definition of the Hilbert space $\CCH$ and $\CCH_{\rm phys}$ of minimal parity-even Vasiliev higher spin de Sitter gravity, its operator algebra and its vacuum state $|0\rangle$. $\CCH$ transforms unitarily under the higher spin symmetry group $\CG$ and $\CCH_{\rm phys}$ is its $\CG$-invariant subspace. While our construction answers many questions, it raises many more. We outline a few in what follows, and speculate on possible answers.

\vskip4mm \noindent {\bf I. Local bulk physics?} \vskip2mm

\noindent How do we identify observables accessible to a local observer in our framework, how does time emerge, and how do we reconstruct local dynamics? There are several possible approaches to these problems.

\begin{enumerate}

\item The most straightforward approach would seem to be a direct reconstruction of perturbative bulk quantum field theory analogous to \cite{Hamilton:2006az}. The appropriate setting for this is the Hilbert space $\CCH$. We discussed some steps this program in section \ref{breakdown}. The starting point is the reconstruction of the bulk QFT Heisenberg algebra of boundary fields. We have shown this is possible, but only up to operator-valued error terms whose vacuum expectation value is of order $e^{-c N}$, and only in a coarse grained sense, with the Heisenberg algebra effectively accessing no more than order $N \sim S_{\rm dS}$ spatial pixels on the boundary at future infinity. The error term is due to the existence of large quantum fluctuations exiting the radius of convergence of perturbation theory: such fluctuations are exponentially unlikely when restricting to observables accessible to the coarse-grained theory, but occur with probability 1 in the original fine-grained theory. This suggests a breakdown of bulk low energy effective field theory on time and length scales that are either too short or too long. For length scales this is fairly obvious, and for time scales this is suggested by the fact, roughly speaking, that by the usual boundary-bulk UV-IR correspondence, the UV coarse-graining cutoff $\ell$ corresponds to a late time cutoff in the bulk, while the IR coarse-graining cutoff $L$ corresponds to an early time cutoff. More coarsely, the necessity of coarse graining implies the nonexistence of a perturbative bulk QFT description carrying an exact representation of the full de Sitter isometry group. While all of these features are certainly reminiscent of limitations on effective field theories of inflationary universes inferred from consistency with holography and other considerations \cite{Dyson:2002pf,Goheer:2002vf,Albrecht:2002xs,Banks:2003pt,ArkaniHamed:2007ky}, clearly more work is needed to make this connection precise.  

\item To detect the emergence of time, it may however not be necessary to first reconstruct the perturbative bulk quantum fields and their canonical Heisenberg algebra. One could for example consider $n$-point functions of the $B_I$ such as those computed in section \ref{sec:cosmocor}, without any need to define canonically conjugate operators $A^I$, and infer bulk dynamical features directly from their analytic structure, or from their reinterpretation as amplitudes obtained by integrating bulk vertices over an emergent spacetime, along the lines of \cite{Arkani-Hamed:2017fdk}. Alternatively, as in \cite{Anninos:2011kh,Shaghoulian:2013qia}, one may be able to detect the emergence of time from the ultrametric, tree-like organization intrinsic to the probability distributions of the local boundary fields $B^I$, analogous to the intrinsic tree-like organization of species living at any given time. For species, this organization can be understood from evolutionary branching over time, due to accumulation of ``frozen-out'' fluctuations in DNA. In inflating spacetimes, this organization can be understood from branching of the wave function into an ensemble of effectively classical field profiles, due to accumulation of frozen-out quantum fluctuations. 

\item To define local observables, it may likewise not be necessary to first reconstruct the Heisenberg algebra, as one may be able to build observables such as local charge densities more directly from the microscopic operators representing such charge densities in the $Q$-model. For example the operator $:\! \partial_i Q \, P\! :$ can be thought of as a momentum density, in the sense that $\int\! :\! \partial_i Q \, P \! :$ is the generator of spatial translations. 

\item Finally, one could reasonably take the point of view that the only physical object is really $\CCH_{\rm phys}$, and that this should form the starting point for any discussion of the physics of the model, including local bulk physics. This would render the limitations on the constructability of the Heisenberg algebra on $\CCH$ less immediately relevant, although an effective bound of order $N$ on the number of accessible pixels reappears in a different guise in this context. As we have defined it in section \ref{sec:GIaPHS}, $\CCH_{\rm phys}$ is quasi-topological, in the sense that the subspace of gauge-invariant $n$-particle states is finite-dimensional for any finite $n$, and in the sense that all gauge-invariant quantities are computable as correlation functions of O(2N)-invariant traces in a $2N \times 2N$ matrix model. The question then arises how the familiar locally propagating field degrees of freedom are recovered in this framework. Conceivably this requires some form of spontaneous breaking of the residual higher spin gauge symmetry group $\CG$, perhaps by the branching of the wave function into frozen-out effectively classical field profiles, producing something akin to the background of ``fixed stars'' of Mach's principle. 
The gauge symmetry can also be (partially) broken explicitly by picking a local observer, or a coordinate patch with a choice of boundary conditions. Identifying the proper group of asymptotic (gauge vs physical) symmetries is quite subtle in general, and sensitive to the choice of boundary conditions, which in turn depend on the physics questions of interest \cite{Strominger:2017zoo,Anninos:2010zf,Anninos:2011jp,Ashtekar:2017dlf}.

\end{enumerate}

\vskip5mm \noindent {\bf II. Entropy?} \vskip2mm

\noindent The most obvious question left open in this work is probably the identification and microscopic computation of the de Sitter horizon entropy \cite{Gibbons:1977mu} in our framework. From the bulk point of view, the de Sitter entropy is naturally thought of as the entropy of the reduced density matrix of the static patch obtained from the global Hartle-Hawking vacuum state. In view of the highly nonlocal nature of the relation between bulk and boundary fields, and the obstruction to reconstructing a bulk effective field theory probing more than $N$ pixels, it is not immediately obvious how to translate this bulk definition to a natural quantity in the boundary $Q$-model. One could try to go the other way, and consider natural quantities in the $Q$-model with an entropic interpretation. The simplest one would be just the vacuum entanglement entropy of a region $\CR$ on the future boundary, computed naively in the Gaussian state $\psi_0(Q) = e^{-\frac{1}{2} \int Q \CD Q}$ on $\CCH$. The local nature of this wave function (as opposed to the nonlocal wave functions typically arising as ground states of free field theories) leads to an intriguing simplification, allowing to express the entanglement entropy purely in terms of a ``wave function'' $\Psi_\Sigma(q) = \int dQ_{\CR}|_q \, e^{-\int_{\CR} Q \CD Q}$ living on the region's boundary $\Sigma = \partial \CR$. Note that $\Psi_\Sigma(q)$ is the vacuum wave function of a radially quantized free CFT, living on a 2d surface, not to be confused with the Hartle-Hawking state $\psi_0(Q)$ living on the 3d spatial slice. This notion of entanglement entropy still leads to the usual UV-divergent area law, because $\CCH$ still has infinitely many short-distance degrees of freedom. A physically better motivated computation would impose the constraints arising from the infinite-dimensional residual gauge group, reducing $\CCH$ to a quasi-topological physical Hilbert space. A complicating factor is that picking a region explicitly breaks the global residual gauge group $\CG$ to a subgroup $\CG_{\rm in} \times \CG_{\rm out}$. It is conceivable that this reduces the computation of the physical entanglement entropy of $\CR$ to a matrix model larger than (\ref{WishartMmodel}), involving not just the matrix $M = Q^T Q$ of (\ref{MQQTdef}), but matrices $M_{\rm in , in} = Q_{\rm in}^T Q_{\rm in}$, $M_{\rm out,out} =  Q_{\rm out}^T Q_{\rm out}$ and $M_{\rm in,out} = Q_{\rm in}^T Q_{\rm out} = M_{\rm out,in}^T$. Quite plausibly this would lead to a finite result of order $N$. However, its interpretation as the dS entropy would still be far from clear. It would seem more akin to the dS entanglement entropy considered in \cite{Maldacena:2012xp}, but even this interpretation is not obvious, in view of the fact that the map from the local $QQ$ bilinears $B_I$ to the local boundary fields $\beta^I$ involves a nonlocal shadow transform, $B_I = G_{IJ} \beta^J$.

\vskip5mm \noindent {\bf III. Generalizations?} \vskip2mm

Although we motivated our constructions by comparison to results that were derived with a dS-CFT framework in mind, the structure we ended up with differs from this framework in important ways. Most importantly, what we construct is not a CFT with a Hilbert space living on 2d spatial slices, organized in highest-weight representations of SO(2,3), but a Hilbert space $\CCH$ living on 3d spatial slices, organized in unitary representations of SO(1,4) (as well as its higher spin extension $\CG$), and its $\CG$-invariant subspace $\CCH_{\rm phys}$. While correlation functions of the operators $B_I=\frac{1}{N} \! :\! \Tr(Q\CD_I Q) \! :$ do take the form of CFT correlation functions of the O(2N) vector model with fundamental fields $Q$, we also have canonically conjugate operators $P$ which have no counterpart in the O(2N) CFT. Without these, it is impossible to reproduce bulk quantum mechanics, as was clear already by considering the free bulk 2-point functions of both dynamical field modes $\alpha$ and $\beta$, as pointed out under (\ref{scalartwoptfunctionsvac}). Relatedly, although the wave function $\tilde \psi_0(\CB)$ in the O(2N)-invariant basis can be interpreted as the partition function of an Sp(N) CFT with sources $\CB$, the operators of the Sp(N) model are not  identified with operators acting on $\CCH$. 

Indeed, although several objects appearing in our construction have conventional local CFT interpretations, this may to some extent be a coincidence, made possible by the very special nature of higher spin theories. More general theories of quantum gravity in universes with a positive vacuum energy density might not share this property. Of course, the Hartle-Hawking wave function $\psi_{\rm HH}(\beta)$ in an asymptotically dS universe is necessarily invariant under the Euclidean conformal group SO(1,4), but this does not mean it is necessarily related to a local conformal field theory. In general, the unitary representations of SO(2,3) and SO(1,4) are very different, with only a small range of overlapping values for the allowed conformal dimensions $\Delta$. It just so happens that the set of primary single trace operators in the free O(N) CFT generates representations that are UIRs of both SO(2,3) and SO(1,4), characterized by a spin $s$ and conformal dimension $\Delta = \frac{1}{2} + s$. For SO(1,4) these correspond for $s=0$ to a representation in the complementary series, and for $s>0$ to a representation in the discrete series. The latter constitute isolated points in the space of possible values of $\Delta$. Another important very special ingredient is the infinite-dimensional higher spin symmetry group $\CG$, crucial in the construction of the physical Hilbert $\CCH_{\rm phys}$ as the $\CG$-invariant subspace of $\CCH$. If $\CG$ had been finite-dimensional, e.g.\ just the conformal group SO(1,4), the number of degrees of freedom of $\CCH_{\rm phys}$ would have been UV divergent and would moreover have scaled extensively with the volume, in considerable tension with expectations based on the holographic principle. Instead what happens is that the full $\CG$ in our model reduces the number of physical degrees of freedom to such extent that $\CCH_{\rm phys}$ becomes quasi-topological. Thus, in theories without such a large symmetry group, things may have to be set up quite differently in order to be consistent with holography. This indicates it will not necessarily be straightforward to broadly generalize the model considered in this paper without significant new ingredients. 

\def\CS{{\cal S}}

Certainly it should be possible though to generalize it to non-minimal or parity-odd Vasiliev gravity in four dimensions, presumably leading to a scalar bosonic U(2N) $Q$-model and a fermionic spinor $Q$-model, respectively. 
We could also consider the theory formulated on late-time slices $\CS$ of more general topologies, as was done in \cite{Banerjee:2013mca} for the original Sp(N) model. In fact, although we usually had the examples $\CS = \IR^3$ or $\CS = S^3$ in mind, nothing prevents us from considering more general cases in our setup, such as $\CS = S^1 \times S^2$. This will no longer have the full conformal group exactly realized as a subgroup of $\CG$ acting on $\CCH$ in the continuum limit, but will still have the same conformal correlation functions of the $B_I$ at scales much smaller than the size of $\CS$. It will moreover lead to exactly the same invariant states and inner products on $\CCH_{\rm phys}$, at least when the continuum limit is defined as a $K \to \infty$ limit of discretized models as in section \ref{sec:Hphyscontlim}. Note that we never couple the theory to a Chern-Simons gauge field, thus avoiding the disastrous divergences for nontrivial topologies noted in \cite{Banerjee:2013mca}. For local CFTs with a gauged symmetry such as O(N), it is necessary to couple the theory to a gauge field to ensure the path integral on a general 3-manifold $\CS$, cut open on a 2-dimensional slice $\Sigma$, can be interpreted as a sum over gauge invariant intermediate states of the conformal field theory Hilbert space living on $\Sigma$. However, in our setup, there is no reason for the path integral of the $Q$-model on $\CS$ to have a representation as a sum over O(2N)-invariant intermediate states living on a 2-dimensional cut surface $\Sigma$. The Hilbert space $\CCH$ we construct does consist of O(2N)-invariant states, but they live on $\CS$, not on $\Sigma$, and they are not states in a CFT Hilbert space.\footnote{This is not to say that there is {\it no} bulk interpretation for the CFT state space living on $\Sigma$. It has been shown in \cite{Ng:2012xp,Anninos:2010gh,Jafferis:2013qia} that such states are related to quasinormal modes in de Sitter.} Therefore, there is no need to couple the theory to a dynamical O(2N) gauge field. It may nevertheless be {\it possible} to do so, and this may lead to other interesting theories such as the family of parity-violating Vasiliev theories in dS$_4$. Such theories might also admit non-trivial large $N$ expansions.  
\newline\newline
Perhaps further study of the model presented in this work and generalizations thereof will suggest a set of physical consistency conditions akin to the unitarity and crossing symmetry constraints of the CFT bootstrap program. Solutions to these consistency conditions would correspond to consistent microscopic models of universes with a positive vacuum energy, including our own. Deriving all of known physics from a set of bootstrap equations would certainly be rather pleasing, but the obstacles to get there might still be quite insuperable.

\section*{Acknowledgements}

{We have benefited from discussions with Tarek Anous, Nima Arkani-Hamed, Tom Banks,  Sumit Das, Mike Douglas, Monica Guica, Sean Hartnoll, Daniel Harlow, Simeon Hellerman, Thomas Hertog, Diego Hofman, Daniel Jafferis, Antal Jevicki, George Konstantinidis, Albert Law, David Lowe, Alexandru Lupsasca, Juan Maldacena, Alex Maloney, Noah Miller, Greg Moore, Rachel Rosen, Nati Seiberg, David Simmons-Duffin, Edgar Shaghoulian, Douglas Stanford, Steve Shenker,  Misha Vasiliev, Erik Verlinde, Herman Verlinde, Edward Witten, and especially Adam Bzowski and Andy Strominger. This work was supported by the U.S.\ Department of Energy grant de-sc0011941, by FWO Flanders, the NSF, the AMIAS, the ANR grant Black-dS-String, by the CEA Enhanced Eurotalents project ``De Sitter Holography" and by the $\Delta$ Institute for Theoretical Physics.}

\appendix

\section{Shadow Transform}\label{shadowapp}

In this appendix we review the shadow transform \cite{Ferrara:1972uq,SimmonsDuffin:2012uy,Metsaev:2008fs,Metsaev:2009ym} in position and momentum space. 

\subsection{Position space}

The basic idea is as follows: given an object $\CO_{l,\bar\Delta}(x)$ transforming as a primary field of spin $l$ and scaling dimension $\bar\Delta$ under the $d$-dimensional Euclidean conformal group SO(1,d+1), we can construct a ``conjugate'' primary field 
\begin{align} \label{generalshadowtransf}
 \tilde{\CO}_{l,\Delta}(x)=\int d^dy \, G_{l,\Delta}(x-y) \, \CO_{l,\bar\Delta}(y)
\end{align}
called the {\it shadow transform} of $\CO_{l,\bar\Delta}$, with spin $l$ and conjugate scaling dimension 
\begin{align}
 \Delta = d- \bar\Delta \, .
\end{align}
The kernel $G_{l,\Delta}(x-y)$ takes the form of a 2-point function of spin $l$, dimension $\Delta$ operators in a $d$-dimensional CFT. Thus, for a scalar operator $\CO$, the shadow transform is
\begin{equation} \label{scalarshadowtr}
    \tilde{\CO}_\Delta(x)=\int d^d y \, \frac{c_{\Delta}}{|x-y|^{2\Delta}} \, \CO_{\bar\Delta}(y) \, .
\end{equation}
The constant $c_\Delta$ is a normalization factor, which we will leave arbitrary here. What is special about this particular choice of integration kernel --- as opposed to say a kernel of the form $|x-y|^{-2\Delta'}$ for some generic $\Delta'$ --- is that the resulting object $\tilde\CO(x)$ transforms again as a local primary field under the conformal group. For generic $\Delta'$ the resulting object would transform in a non-local way. The inverse of a shadow transform is again a shadow transform. For example the scalar relation (\ref{scalarshadowtr}) may be inverted as
\begin{align}
 \CO_{\bar \Delta}(x) = \int d^dy \,  \frac{\tilde c_{\bar\Delta}}{|x-y|^{2\Delta}} \, \tilde\CO_{\Delta}(y) \, ,
\end{align}
for a suitable constant $\tilde c_{\bar\Delta}$. This will become obvious in the momentum space description discussed further down. 

More generally, we may think of the shadow transform as follows. Let us consider a basis of objects $\CO_{\alpha m}(x)$ transforming as primary fields under the conformal group SO(1,d+1), and an SO(1,d+1)-invariant pairing
\begin{align}
 \langle \CO_{\alpha m}(x) \CO_{\beta n}(y) \rangle = G_{\alpha m x;\beta n y} = G_{\beta n y;\alpha m x} \, .
\end{align}
A typical example is a CFT containing primary operators labeled by $\alpha$, with $x$ labeling the position and $m$ the spin components, with the pairing being the CFT 2-point function for these operators. More generally $\alpha$ can be thought of as labeling different irreducible representations of SO(1,d+1). 
More succinctly, we can collect the labels $(\alpha,m,x)$ into a single label $I$, and write this pairing as
\begin{align}
 \langle \CO_I \CO_J \rangle = G_{IJ} = G_{JI} \, .
\end{align}
Then we may define shadows $\tilde\CO^I$ through the following relation:
\begin{align}
 \CO_I = G_{IJ} \tilde\CO^J \, ,
\end{align}
or in more detail $\CO_{\alpha m}(x) = \sum_{\beta,n} \int d^3 y \, G_{\alpha m x;\beta n y} \, \tilde\CO^{\beta n}(y)$. If $G_{IJ}$ is invertible, with inverse $G^{IJ}$, we have equivalently
\begin{align}
 \tilde\CO^I = G^{IJ} \CO_J \, .
\end{align}
This implies in particular
\begin{align}
 \langle \CO_I \tilde\CO^J \rangle = \delta_I^J \, , \qquad \langle \tilde\CO^I \tilde \CO^J \rangle = G^{IJ} \, .
\end{align}
From this point of view, the shadow transform is just the map from ``index up'' to ``index down'' objects induced by the metric $G_{IJ}$, analogous to the map between tangent and cotangent vectors. If we choose the basis of objects $\CO_{\alpha m}(x)$ to be such that the pairing is diagonal in the $(\alpha,\beta)$ indices, i.e.\ $G_{\alpha m x;\beta n y} = g_{\alpha,mn}(x,y) \, \delta_{\alpha\beta}$, this reduces to the form of the shadow transform given earlier in (\ref{generalshadowtransf}). 

It is possible that $G_{IJ}$ is not invertible. This arises naturally in CFTs when the operators $\CO_{\alpha m}(x)$ are conserved currents. In that case the 2-point function $G_{IJ}$ has null directions, so its inverse is not unique. This non-uniqueness is closely related to the gauge redundancy that occurs when coupling sources to conserved currents.

To obtain a general and simple explicit expression for the shadow kernel in (\ref{generalshadowtransf}) for operators of arbitrary spin $l$, it is convenient to use the auxiliary null vector formalism (see for example \cite{Giombi:2009wh,Sleight:2016dba}). For a dimension $\bar\Delta$, spin $l$,  traceless symmetric operator $\CO_{i_1i_2...i_l}(x)$, we can define a $z$-dependent operator 
\begin{equation}
\CO_{l,\bar\Delta}(x,z)=\CO_{i_1i_2...i_l}(x)z^{i_1}...z^{i_l}
\end{equation}
where $z^i$ is an auxiliary null complex vector in $d$ dimensions. Up to normalization, the two-point function takes the form 
\begin{equation}
\langle \CO_{l,\bar\Delta}(x_1,z_1)\CO_{l,\bar\Delta}(x_2,z_2)\rangle=\frac{\left[z_1^iH_{ij}(x_{12})z_{2}^j\right]^l}{|x_1-x_2|^{2\bar\Delta}},\quad
 \,\,\, H_{ij}(x)=\delta_{ij}-\frac{2x_ix_j}{x^2} \, ,
\end{equation}
hence the shadow of $\CO_{l,\bar\Delta}(x,z)$ is 
\begin{equation}
    \tilde{\CO}_{l,\Delta}(x_1,z_1)=\int d^dx_2 \, \frac{c_{l, \Delta}}{|x_1-x_2|^{2\Delta}} \, (z_1\!\cdot\! H(x_{12})\!\cdot\! z_2)^l \, \CO_{l,\bar\Delta}(x_2,z_2) \, ,
\end{equation}
where $c_{l,\Delta}$ is a normalization constant, and
the repeated $z_2$ on the right hand side of this equation means the contraction of indices. That is, if $f(z)$ and $g(z)$ encode two symmetric traceless rank $l$ tensors $f_{i_1...i_l}$ and $g_{i_1...i_l}$, then $f(z)g(z)$ with repeated $z$ just means the complete contraction of the two tensors $f_{i_1...i_l}g_{i_1...i_l}$.


\subsection{Momentum space}

For our purposes it will be particularly useful to express the shadow transform in momentum space. Our conventions for the Fourier transform are given by (\ref{fourier1}) and (\ref{fourier2}).
The Fourier transform of a CFT 2-point function $|x|^{-2\Delta}$ is  
\begin{equation} \label{fouriertransformxk}
    \int d^d x \, x^{-2\Delta} \, e^{-ik \cdot x}=\frac{\pi^{d/2}2^{d-2\Delta}\Gamma[\frac{d-2\Delta}{2}]}{\Gamma[\Delta]} \, k^{2\Delta-d} \, .
\end{equation}
This integral should be viewed as defined by analytic continuation when it does not converge. The shadow transform of a scalar operator in momentum space is  obtained by Fourier transforming (\ref{scalarshadowtr}):
\begin{equation}
\tilde\CO(k)=c_{\Delta}\frac{\pi^{d/2}2^{d-2\Delta}\Gamma[\frac{d-2\Delta}{2}]}{\Gamma[\Delta]} \, k^{2\Delta-d} \, \CO(k) \, .
\end{equation}
As an example, the shadow of a scalar primary $\CO$ of scaling dimension $\bar\Delta=2$ in $d=3$ is
\begin{equation}
    \tilde{\CO}(k)=c_{\Delta} \, 2\pi^2  \, \frac{1}{k} \, \CO(k) \, .
\end{equation}
For general higher spin operators, the Fourier transform of the shadow kernel   is much more complicated due to the tensor structure from the term $(z_1\cdot H(x)\cdot z_2)^l$. Fortunately, it has been worked out in detail in \cite{Dobrev:1977qv} by using the method of harmonic analysis on SO(d):
\begin{equation}
G_{l,\Delta}(k)=\int d^dx e^{-ikx}G_{l,\Delta}(x)=  \frac{c_{l,\Delta}\pi^{d/2}2^{d-2\Delta}\Gamma[\frac{d-2\Delta}{2}]}{(l+\Delta-1)\Gamma[\Delta-1]}k^{2\Delta-d}\sum_{s=0}^l\frac{(\bar\Delta+s-1)_{l-s}}{(\Delta+s-1)_{l-s}}\Pi^{ls}(k;z_1,z_2)  
\end{equation}
where $(x)_n = \frac{\Gamma(x+n)}{\Gamma(x)}$ is the Pochhammer symbol and $\{\Pi^{ls}(p,z_1,z_2)\}_{0\le s\le l}$ is a complete set of orthonormal projection operators with an extra ``transverse'' property 
\begin{equation} \label{trans}
k^{i_s}...k^{i_l}\Pi^{ls}(k)_{i_1...i_l;j_1...j_l}=0~.
\end{equation}
{The above property informs us that the shadow of a conserved current behaves like a gauge field. This plays an important role throughout our main discussion.} 

For the case that we are interested in, i.e.\ higher spin conserved currents, the spin-$l$ current has scaling dimension $\Delta_l=d-2+l$ which leads to a great simplification of $G_{l,\Delta_l}(k)$:
\begin{equation}
G_{l,\Delta_l}(k)=\frac{c_{l,\Delta_l}\pi^{d/2}2^{d-2\Delta_l}\Gamma[\frac{d-2\Delta_l}{2}]}{(l+\Delta_l-1)\Gamma[\Delta_l-1]} \, k^{2\Delta_l-d} \, \Pi^{ll}(k;z_1,z_2)   \, .
\end{equation}
The explicit form of $\Pi^{ll}(k;z_1,z_2)$ is
\begin{align}
   \Pi^{ll}(k;z_1,z_2)&=(-)^ll!(d/2-1)_l\frac{d+2l-3}{(d-3)_{2l+1}}(2\hat{k}\cdot z_1\hat{k}\cdot z_2)^l C^{(d-3)/2}_l(\omega), \nn \\
   \omega&=1-\frac{k^2z_1\cdot z_2}{k\cdot z_1 k\cdot z_2} \, ,
\end{align}
where $\hat{k}$ is the unit vector in the $k$ direction and the $C_l^\alpha(\omega)$ are Gegenbauer polynomials.

The shadow transformation of higher spin operators in momentum space consists two parts: one rescales the original operator by an appropriate power of $k$ and the other projects the resulting operator into the transverse subspace of $\hat{k}$ due to equation (\ref{trans}). Taking the limit $d\to 3$, $G_{l,\Delta_l}(k)$ is of the following form
\begin{equation}
G_{l,1+l}(k)=\frac{c_{l,1+l}2^{1-l}\pi^2}{\Gamma[2l+1]}\left(\frac{z_1\cdot z_2}{1-\omega}\right)^l T_l(\omega) \, k^{2l-1}=\frac{c_{l,1+l}2^{1-l}\pi^2}{\Gamma[2l+1]}(\hat{k}\!\cdot\! z_1 \, \hat{k}\!\cdot\! z_2)^l \, T_l(\omega) \, k^{2l-1}
\end{equation}
where $T_l(x)$ is the Chebyshev T-function reduced from Gegenbauer polynomials by using
$\frac{2}{n}T_n(x)=\lim_{\alpha\to 0}\frac{1}{\alpha}C^{(\alpha)}_n(x)$. Here, we list some lower spin examples:
\begin{equation}
\begin{split}
l&=0:\, \, G_{l=0, \Delta=1}(k)=2\pi^2 \, c_{0,1}\, k^{-1}\\
l&=1:\,\, G_{l=1, \Delta=2}(k)=\frac{\pi^2}{2}\, c_{1,2}\, k\, (-z_1\cdot z_2+\hat{k}\cdot z_1\hat{k}\cdot z_2)\\
l&=2:\,\, G_{l=2,\Delta=3}(k)=\frac{\pi^2}{24}\, c_{2,3}\, k^3\left[(z_1\cdot z_2-\hat{k}\cdot z_1\hat{k}\cdot z_2)^2-\frac{1}{2}(\hat{k}\cdot z_1\hat{k}\cdot z_2)^2\right]
\end{split}
\end{equation}
or with explicit indices 
\begin{equation}\label{shadowmon}
\begin{split}
l&=0:\, \, G_{l=0, \Delta=1}(k)=2\pi^2c_{l=0, \Delta=1}k^{-1}\\
l&=1:\,\, G_{l=1, \Delta=2}(k)_{i,m}=\frac{\pi^2}{2}c_{l=1, \Delta=2}k\Pi_{im}(\hat{k})\\
l&=2:\,\, G_{l=2,\Delta=3}(k)_{ij,mn}=\frac{\pi^2}{24}c_{l=2, \Delta=3}k^3\Pi_{ij,mn}(\hat k)\\
&...\\
l&=s:\,\, G_{l=s,\Delta=s+1}(k)_{i_1...i_s,m_1...m_s}=\frac{(-)^s\pi^2}{(2s)!}c_{l=s, \Delta=s+1}k^{2s-1}\Pi_{i_1...i_s,m_1...m_s}(\hat k)\\
\end{split}
\end{equation}
where in the last line, we have used 
\begin{align}
(\hat k\cdot z_1 \hat k\cdot z_2)^sT_s(\omega)=(-)^s2^{s-1}\Pi_{i_1...i_s,m_1...m_s}(\hat k)z_1^{i_1}...z_1^{i_s}z_2^{m_1}...z_2^{m_s}, \quad s\ge 1~.
\end{align}

\section{Normalization of $\psi_0(H)$} \label{app:normalization}

In this appendix we calculate the noralization constant in (\ref{ceq}). The relevant integral is
\begin{equation}\label{c}
\frac{1}{c^2} = \int_+\frac{dH}{(\det H)^{(K+1)/2}}(\det H)^\N e^{-\N \Tr H}
\end{equation}
where $H$ is a $K\times K$ positive symmetric real matrix. 


The integral measure consists of two parts. One depends on the eigenvalues only and the other one only involves the O(K) part. As long as the integrand is O(K) invariant, the O(K) dependent measure will only give an overall constant and it suffices to integrate over the eigenvalues (see for example \cite{Marino:2012zq})
\begin{equation}\label{factor}
\int dH f(H)=\CN \int  \prod_{i=1}^K d\lambda_i \prod_{i<j}|\lambda_i-\lambda_j|f(\lambda)
\end{equation}
where $f(H)$ is an O(K)-invariant function, $\lambda_i$'s are eigenvalues of $H$ and $\CN$ is the constant `volume' of the O(K) group, which will be computed shortly. 

After certain rescalings, (\ref{c}) can be expressed as
\begin{equation}
\frac{1}{c^2}=\frac{\CN}{\N^{K\N}}\int^\infty_0 \prod_{i=1}^K (d\lambda_i \lambda_i^{\N-(K+1)/2}e^{-\lambda_i})\prod_{i<j}|\lambda_i-\lambda_j|
\end{equation} 
which is a special form of Selberg integral \cite{selberg}
\begin{equation}
\int^\infty_0 \prod_{i=1}^K d\lambda_i \lambda_i^{x-1}e^{-\lambda_i} \, \prod_{i<j}|\lambda_i-\lambda_j|^{2z}=\prod_{j=1}^K\frac{\Gamma[x+(j-1)z]\Gamma[jz+1]}{\Gamma[z+1]}~.
\end{equation}
In our case $x=\N-(K-1)/2$ and $z=1/2$. Thus we have
\begin{equation}\label{c'}
\frac{1}{c^{2}}=\frac{\CN}{\N^{K\N}}\prod_{j=1}^K\frac{\Gamma[\N-(K-j)/2]\Gamma[j/2+1]}{\Gamma[3/2]}
\end{equation}
Given that the Selberg integral requires $x>0$ our result holds for $2\N\ge K$.

What remains is to find the constant $\CN$. Consider the following Gaussian integral over symmetric real matrices
\begin{equation}
I=\int dG\, e^{-1/2 \Tr G^2}~.
\end{equation}
On the one hand, $I$ is the product of a collection of ordinary Gaussian integrals
\begin{equation}
I=\int \prod_idG_{ii}e^{-1/2G_{ii}}\prod_{i<j}dG_{ij}e^{- G_{ij}G_{ij}}=(2\pi)^{K/2}\pi^{\frac{K(K-1)}{4}}~.
\end{equation}
On the other hand, by noting than the integrand is O(K)-invariant, $I$ can be reduced to an integral over eigenvalues which are now valued in the whole real line:
\begin{equation}
I=\CN\int^\infty_{-\infty} \prod_{i=1}^K d\lambda_i e^{-\lambda_i^2/2}\prod_{i<j}|\lambda_i-\lambda_j|
\end{equation}
This is another type of Selberg integral
\begin{equation}
\int^\infty_{-\infty} \prod_{i=1}^K \frac{d\lambda_i}{\sqrt{2\pi}} e^{-\lambda_i^2/2}\prod_{i<j}|\lambda_i-\lambda_j|^{2z}=\prod_{j=1}^K\frac{\Gamma[jz+1]}{\Gamma[z+1]}
\end{equation}
Putting everything together leads to
\begin{equation}
\CN=\pi^{\frac{K(K-1)}{4}}\prod_{j=1}^K\frac{\Gamma[3/2]}{\Gamma[j/2+1]}~,
\end{equation}
and finally
\begin{equation}\label{final}
{\frac{1}{c^2}=\frac{\pi^{\frac{K(K-1)}{4}}}{\N^{\N K}}\prod_{j=1}^K\Gamma[\N-(K-j)/2]}~.
\end{equation}
One can check the above formula explicitly for low values of $K$ and $N$.

\end{document}